\documentclass[twocolumn,letterpaper,aps,prc,superscriptaddress,longbibliography,nofootinbib,floatfix]{revtex4-2}

\usepackage{graphicx}	

\usepackage{xspace}
\usepackage{amsmath}
\usepackage{appendix}

\newcommand{\sqs}{\mbox{$\sqrt{s}$}\xspace}
\newcommand{\sqsn}{\mbox{$\sqrt{s_{_{NN}}}$}\xspace}

\newcommand{\s}[1]{\mbox{$\sqrt{s}$ = #1\,GeV}}
\newcommand{\snn}[1]{\mbox{$\sqrt{s_{_{NN}}}$ = #1\,GeV}}
 
\newcommand{\AB}[2]{#1$+$#2}

\newcommand{\pp}{$p$$+$$p$\xspace}

\newcommand{\pbpb}{\mbox{Pb$+$Pb}\xspace}
\newcommand{\auau}{\mbox{Au$+$Au}\xspace}
\newcommand{\cucu}{\mbox{Cu$+$Cu}\xspace}
\newcommand{\piz}{\mbox{$\pi^0$}\xspace}
\newcommand{\gammahadr}{\mbox{$\gamma^{\rm hadr}$}\xspace}
\newcommand{\gammapiz}{\mbox{$\gamma^{\rm \pi^0}$}\xspace}
\newcommand{\gammaincl}{\mbox{$\gamma^{\rm incl}$}\xspace}
\newcommand{\gammadir}{\mbox{$\gamma^{\rm dir}$}\xspace}
\newcommand{\mhbd}{\mbox{$M_{\rm HBD}$}\xspace}
\newcommand{\mvtx}{\mbox{$M_{\rm vtx}$}\xspace}
\newcommand{\m}[1]{\rm{#1}}
\newcommand{\pt}{\mbox{$p_T$}\xspace}
\newcommand{\ptmin}{\mbox{$p_{\rm T,{\rm min}}$}\xspace}

\newcommand{\ee}{\mbox{$e^+e^-$}\xspace}

\newcommand{\Ncoll}{\mbox{$N_{\rm coll}$}\xspace}
\newcommand{\Nch}{\mbox{$N_{\rm ch}$}\xspace}
\newcommand{\dNch}{\mbox{$dN_{\rm ch}/d\eta$}\xspace}

\newcommand{\Rg}{\mbox{$R_\gamma$}\xspace}
\newcommand{\Nincl}{\mbox{$N_\gamma^{\rm incl}$}\xspace}
\newcommand{\Ntag} {\mbox{$N_\gamma^{\pi^0, \rm tag}$}\xspace}
\newcommand{\ef}{\mbox{$\langle\varepsilon_{\gamma} f \rangle$}\xspace}

\newcommand{\gevc}{\mbox{GeV/$c$}\xspace}
\newcommand{\gevcc}{\mbox{GeV/$c^2$}\xspace}

\newcommand{\Lb}{\left(}
\newcommand{\Rb}{\right)}
\def\m{{\boldsymbol m}}

\begin{document}

\title{Low-$p_T$ direct-photon production in Au$+$Au collisions 
at $\sqrt{s_{_{NN}}}=39$ and 62.4~GeV}

\newcommand{\abilene}{Abilene Christian University, Abilene, Texas 79699, USA}
\newcommand{\augie}{Department of Physics, Augustana University, Sioux Falls, South Dakota 57197, USA}
\newcommand{\banaras}{Department of Physics, Banaras Hindu University, Varanasi 221005, India}
\newcommand{\barc}{Bhabha Atomic Research Centre, Bombay 400 085, India}
\newcommand{\baruch}{Baruch College, City University of New York, New York, New York, 10010 USA}
\newcommand{\bnlcoll}{Collider-Accelerator Department, Brookhaven National Laboratory, Upton, New York 11973-5000, USA}
\newcommand{\bnlphys}{Physics Department, Brookhaven National Laboratory, Upton, New York 11973-5000, USA}
\newcommand{\caucr}{University of California-Riverside, Riverside, California 92521, USA}
\newcommand{\charlesczech}{Charles University, Faculty of Mathematics and Physics, 180 00 Troja, Prague, Czech Republic}
\newcommand{\cns}{Center for Nuclear Study, Graduate School of Science, University of Tokyo, 7-3-1 Hongo, Bunkyo, Tokyo 113-0033, Japan}
\newcommand{\colorado}{University of Colorado, Boulder, Colorado 80309, USA}
\newcommand{\columbia}{Columbia University, New York, New York 10027 and Nevis Laboratories, Irvington, New York 10533, USA}
\newcommand{\czechtech}{Czech Technical University, Zikova 4, 166 36 Prague 6, Czech Republic}
\newcommand{\dapnia}{Dapnia, CEA Saclay, F-91191, Gif-sur-Yvette, France}
\newcommand{\debrecen}{Debrecen University, H-4010 Debrecen, Egyetem t{\'e}r 1, Hungary}
\newcommand{\elte}{ELTE, E{\"o}tv{\"o}s Lor{\'a}nd University, H-1117 Budapest, P{\'a}zm{\'a}ny P.~s.~1/A, Hungary}
\newcommand{\ewha}{Ewha Womans University, Seoul 120-750, Korea}
\newcommand{\famu}{Florida A\&M University, Tallahassee, FL 32307, USA}
\newcommand{\fsu}{Florida State University, Tallahassee, Florida 32306, USA}
\newcommand{\gsu}{Georgia State University, Atlanta, Georgia 30303, USA}
\newcommand{\hanyang}{Hanyang University, Seoul 133-792, Korea}
\newcommand{\hiroshima}{Hiroshima University, Kagamiyama, Higashi-Hiroshima 739-8526, Japan}
\newcommand{\howard}{Department of Physics and Astronomy, Howard University, Washington, DC 20059, USA}
\newcommand{\ihepprot}{IHEP Protvino, State Research Center of Russian Federation, Institute for High Energy Physics, Protvino, 142281, Russia}
\newcommand{\illuiuc}{University of Illinois at Urbana-Champaign, Urbana, Illinois 61801, USA}
\newcommand{\inrras}{Institute for Nuclear Research of the Russian Academy of Sciences, prospekt 60-letiya Oktyabrya 7a, Moscow 117312, Russia}
\newcommand{\instpasczech}{Institute of Physics, Academy of Sciences of the Czech Republic, Na Slovance 2, 182 21 Prague 8, Czech Republic}
\newcommand{\isu}{Iowa State University, Ames, Iowa 50011, USA}
\newcommand{\jaea}{Advanced Science Research Center, Japan Atomic Energy Agency, 2-4 Shirakata Shirane, Tokai-mura, Naka-gun, Ibaraki-ken 319-1195, Japan}
\newcommand{\jeonbuk}{Jeonbuk National University, Jeonju, 54896, Korea}
\newcommand{\jyvaskyla}{Helsinki Institute of Physics and University of Jyv{\"a}skyl{\"a}, P.O.Box 35, FI-40014 Jyv{\"a}skyl{\"a}, Finland}
\newcommand{\kek}{KEK, High Energy Accelerator Research Organization, Tsukuba, Ibaraki 305-0801, Japan}
\newcommand{\korea}{Korea University, Seoul 02841, Korea}
\newcommand{\kurchatov}{National Research Center ``Kurchatov Institute", Moscow, 123098 Russia}
\newcommand{\kyoto}{Kyoto University, Kyoto 606-8502, Japan}
\newcommand{\labllr}{Laboratoire Leprince-Ringuet, Ecole Polytechnique, CNRS-IN2P3, Route de Saclay, F-91128, Palaiseau, France}
\newcommand{\lahorelums}{Physics Department, Lahore University of Management Sciences, Lahore 54792, Pakistan}
\newcommand{\lawllnl}{Lawrence Livermore National Laboratory, Livermore, California 94550, USA}
\newcommand{\losalamos}{Los Alamos National Laboratory, Los Alamos, New Mexico 87545, USA}
\newcommand{\lpc}{LPC, Universit{\'e} Blaise Pascal, CNRS-IN2P3, Clermont-Fd, 63177 Aubiere Cedex, France}
\newcommand{\lund}{Department of Physics, Lund University, Box 118, SE-221 00 Lund, Sweden}
\newcommand{\maryland}{University of Maryland, College Park, Maryland 20742, USA}
\newcommand{\mass}{Department of Physics, University of Massachusetts, Amherst, Massachusetts 01003-9337, USA}
\newcommand{\mate}{MATE, Laboratory of Femtoscopy, K\'aroly R\'obert Campus, H-3200 Gy\"ongy\"os, M\'atrai\'ut 36, Hungary}
\newcommand{\michigan}{Department of Physics, University of Michigan, Ann Arbor, Michigan 48109-1040, USA}
\newcommand{\miss}{Mississippi State University, Mississippi State, Mississippi 39762, USA}
\newcommand{\muhlenberg}{Muhlenberg College, Allentown, Pennsylvania 18104-5586, USA}
\newcommand{\myongji}{Myongji University, Yongin, Kyonggido 449-728, Korea}
\newcommand{\nagasaki}{Nagasaki Institute of Applied Science, Nagasaki-shi, Nagasaki 851-0193, Japan}
\newcommand{\nara}{Nara Women's University, Kita-uoya Nishi-machi Nara 630-8506, Japan}
\newcommand{\natmephi}{National Research Nuclear University, MEPhI, Moscow Engineering Physics Institute, Moscow, 115409, Russia}
\newcommand{\newmex}{University of New Mexico, Albuquerque, New Mexico 87131, USA}
\newcommand{\nmsu}{New Mexico State University, Las Cruces, New Mexico 88003, USA}
\newcommand{\northcg}{Physics and Astronomy Department, University of North Carolina at Greensboro, Greensboro, North Carolina 27412, USA}
\newcommand{\ohio}{Department of Physics and Astronomy, Ohio University, Athens, Ohio 45701, USA}
\newcommand{\ornl}{Oak Ridge National Laboratory, Oak Ridge, Tennessee 37831, USA}
\newcommand{\orsay}{IPN-Orsay, Univ.~Paris-Sud, CNRS/IN2P3, Universit\'e Paris-Saclay, BP1, F-91406, Orsay, France}
\newcommand{\pnpi}{PNPI, Petersburg Nuclear Physics Institute, Gatchina, Leningrad region, 188300, Russia}
\newcommand{\pusan}{Pusan National University, Pusan 46241, Korea}
\newcommand{\riken}{RIKEN Nishina Center for Accelerator-Based Science, Wako, Saitama 351-0198, Japan}
\newcommand{\rikjrbrc}{RIKEN BNL Research Center, Brookhaven National Laboratory, Upton, New York 11973-5000, USA}
\newcommand{\rikkyo}{Physics Department, Rikkyo University, 3-34-1 Nishi-Ikebukuro, Toshima, Tokyo 171-8501, Japan}
\newcommand{\saispbstu}{Saint Petersburg State Polytechnic University, St.~Petersburg, 195251 Russia}
\newcommand{\saopaulo}{Universidade de S{\~a}o Paulo, Instituto de F\'{\i}sica, Caixa Postal 66318, S{\~a}o Paulo CEP05315-970, Brazil}
\newcommand{\seoulnat}{Department of Physics and Astronomy, Seoul National University, Seoul 151-742, Korea}
\newcommand{\stonybrkc}{Chemistry Department, Stony Brook University, SUNY, Stony Brook, New York 11794-3400, USA}
\newcommand{\stonycrkp}{Department of Physics and Astronomy, Stony Brook University, SUNY, Stony Brook, New York 11794-3800, USA}
\newcommand{\sungskku}{Sungkyunkwan University, Suwon, 440-746, Korea}
\newcommand{\tenn}{University of Tennessee, Knoxville, Tennessee 37996, USA}
\newcommand{\texsu}{Texas Southern University, Houston, TX 77004, USA}
\newcommand{\titech}{Department of Physics, Tokyo Institute of Technology, Oh-okayama, Meguro, Tokyo 152-8551, Japan}
\newcommand{\tsukuba}{Tomonaga Center for the History of the Universe, University of Tsukuba, Tsukuba, Ibaraki 305, Japan}
\newcommand{\vandy}{Vanderbilt University, Nashville, Tennessee 37235, USA}
\newcommand{\weizmann}{Weizmann Institute, Rehovot 76100, Israel}
\newcommand{\wigner}{Institute for Particle and Nuclear Physics, Wigner Research Centre for Physics, Hungarian Academy of Sciences (Wigner RCP, RMKI) H-1525 Budapest 114, POBox 49, Budapest, Hungary}
\newcommand{\yonsei}{Yonsei University, IPAP, Seoul 120-749, Korea}
\newcommand{\zagreb}{Department of Physics, Faculty of Science, University of Zagreb, Bijeni\v{c}ka c.~32 HR-10002 Zagreb, Croatia}
\newcommand{\zambia}{Department of Physics, School of Natural Sciences, University of Zambia, Great East Road Campus, Box 32379, Lusaka, Zambia}
\affiliation{\abilene}
\affiliation{\augie}
\affiliation{\banaras}
\affiliation{\barc}
\affiliation{\baruch}
\affiliation{\bnlcoll}
\affiliation{\bnlphys}
\affiliation{\caucr}
\affiliation{\charlesczech}
\affiliation{\cns}
\affiliation{\colorado}
\affiliation{\columbia}
\affiliation{\czechtech}
\affiliation{\dapnia}
\affiliation{\debrecen}
\affiliation{\elte}
\affiliation{\ewha}
\affiliation{\famu}
\affiliation{\fsu}
\affiliation{\gsu}
\affiliation{\hanyang}
\affiliation{\hiroshima}
\affiliation{\howard}
\affiliation{\ihepprot}
\affiliation{\illuiuc}
\affiliation{\inrras}
\affiliation{\instpasczech}
\affiliation{\isu}
\affiliation{\jaea}
\affiliation{\jeonbuk}
\affiliation{\jyvaskyla}
\affiliation{\kek}
\affiliation{\korea}
\affiliation{\kurchatov}
\affiliation{\kyoto}
\affiliation{\labllr}
\affiliation{\lahorelums}
\affiliation{\lawllnl}
\affiliation{\losalamos}
\affiliation{\lpc}
\affiliation{\lund}
\affiliation{\maryland}
\affiliation{\mass}
\affiliation{\mate}
\affiliation{\michigan}
\affiliation{\miss}
\affiliation{\muhlenberg}
\affiliation{\myongji}
\affiliation{\nagasaki}
\affiliation{\nara}
\affiliation{\natmephi}
\affiliation{\newmex}
\affiliation{\nmsu}
\affiliation{\northcg}
\affiliation{\ohio}
\affiliation{\ornl}
\affiliation{\orsay}
\affiliation{\pnpi}
\affiliation{\pusan}
\affiliation{\riken}
\affiliation{\rikjrbrc}
\affiliation{\rikkyo}
\affiliation{\saispbstu}
\affiliation{\saopaulo}
\affiliation{\seoulnat}
\affiliation{\stonybrkc}
\affiliation{\stonycrkp}
\affiliation{\sungskku}
\affiliation{\tenn}
\affiliation{\texsu}
\affiliation{\titech}
\affiliation{\tsukuba}
\affiliation{\vandy}
\affiliation{\weizmann}
\affiliation{\wigner}
\affiliation{\yonsei}
\affiliation{\zagreb}
\affiliation{\zambia}
\author{N.J.~Abdulameer} \affiliation{\debrecen}
\author{U.~Acharya} \affiliation{\gsu} 
\author{A.~Adare} \affiliation{\colorado} 
\author{C.~Aidala} \affiliation{\losalamos} \affiliation{\michigan} 
\author{N.N.~Ajitanand} \altaffiliation{Deceased} \affiliation{\stonybrkc} 
\author{Y.~Akiba} \email[PHENIX Spokesperson: ]{akiba@rcf.rhic.bnl.gov} \affiliation{\riken} \affiliation{\rikjrbrc} 
\author{R.~Akimoto} \affiliation{\cns} 
\author{H.~Al-Ta'ani} \affiliation{\nmsu} 
\author{J.~Alexander} \affiliation{\stonybrkc} 
\author{M.~Alfred} \affiliation{\howard} 
% ...
\collaboration{PHENIX Collaboration}  \noaffiliation

\date{\today}


\begin{abstract}

The measurement of direct photons from Au$+$Au collisions at 
$\sqrt{s_{_{NN}}}=39$ and 62.4 GeV in the transverse-momentum range 
$0.4<p_T<3$ Gev/$c$ is presented by the PHENIX collaboration at the 
Relativistic Heavy Ion Collider. A significant direct-photon yield is 
observed in both collision systems. A universal scaling is observed when 
the direct-photon $p_T$ spectra for different center-of-mass energies 
and for different centrality selections at $\sqrt{s_{_{NN}}}=62.4$ GeV 
is scaled with $(dN_{\rm ch}/d\eta)^{\alpha}$ for 
$\alpha=1.21{\pm}0.04$. This scaling also holds true for direct-photon 
spectra from Au$+$Au collisions at $\sqrt{s_{_{NN}}}=200$ GeV measured 
earlier by PHENIX, as well as the spectra from Pb$+$Pb at 
$\sqrt{s_{_{NN}}}=2760$ GeV published by ALICE.  The scaling power 
$\alpha$ seems to be independent of $p_T$, center of mass energy, and 
collision centrality.  The spectra from different collision energies 
have a similar shape up to $p_T$ of 2 GeV/$c$. The spectra have a local 
inverse slope $T_{\rm eff}$ increasing with $p_T$ of $0.174\pm0.018$ 
GeV/$c$ in the range $0.4<p_T<1.3$ GeV/$c$ and increasing to 
$0.289\pm0.024$ GeV/$c$ for $0.9<p_T<2.1$ GeV/$c$. The observed 
similarity of low-$p_T$ direct-photon production from $\sqrt{s_{_{NN}}}= 
39$ to 2760 GeV suggests a common source of direct photons for the 
different collision energies and event centrality selections, and 
suggests a comparable space-time evolution of direct-photon emission.

\end{abstract}

\maketitle

\section{Introduction}
\label{sec1:Intro}

The measurement of direct-photon emission plays an important role in the 
study of collisions of heavy 
ions~\cite{Stankus:2005eq,David:2006sr,Linnyk:2015rco,David:2019wpt}. 
Due to their very small interaction cross section with the strongly 
interacting matter, photons are likely to escape the collision region 
with almost no final-state interactions. Thus, they carry information 
about the properties and dynamics of the environment in which they are 
produced, such as the energy density, temperature, and collective motion, 
integrated over space and time.

Direct photons with transverse momenta (\pt) of up to a few \gevc are 
expected to be dominantly of thermal origin, radiated from a thermalized 
hot ``fireball'' of quark-gluon plasma (QGP), throughout its expansion 
and transition to a gas of hadrons, until the hadrons cease to interact. 
In addition to the fireball, hard-scattering processes in the initial 
phase of the collision also emit photons. These prompt photons typically 
have larger \pt and dominate the direct-photon production at \pt above 
several \gevc.  Experimentally, direct photons are measured together 
with a much larger number of photons resulting from decays of unstable 
hadrons, such as \piz and $\eta$ decays. The contribution of these decay 
photons to the total number of photons needs to be removed with an 
accuracy of a few percent, which is the main experimental challenge.

The production of thermal photons has been extensively studied through a 
variety of models with different production processes and mechanisms, 
different photon rates, as well as a range of assumptions about the 
initial state of the matter and its space-time evolution. Some of the 
well-known examples include models developed with an 
``elliptic-fireball'' expansion approach~\cite{vanHees:2011vb, 
vanHees:2014ida}, hydrodynamic simulations of the ``fireball'' 
evolution~\cite{Dion:2011pp, Shen:2013vja, Shen:2015qba,Paquet:2015lta}, 
the parton-hadron-string dynamics transport 
approach~\cite{Bratkovskaya:2008iq,Bratkovskaya:2014mva,Linnyk:2013wma}, 
the thermalizing 
Glasma~\cite{Chiu:2012ij,McLerran:2014hza,McLerran:2015mda,Klein-Bosing:2014uaa} 
and the thermalizing Glasma plus bottom-up thermalization scenarios for 
calculations of the pre-equilibrium and equilibrium 
phases~\cite{Berges:2017eom,Khachatryan:2018ori}, reduced radiation from 
the QGP until the transition temperature is 
reached~\cite{monnai:2014kqa,Lee:2014pwa}, as well as calculations in 
the late hadron-gas phase using the spectral-function 
approach~\cite{Lee:2014pwa,Turbide:2003si,Dusling:2009ej,Kim:2016ylr,Heffernan:2014mla,Linnyk:2015tha}. 
The strong magnetic fields emerging in heavy ion collisions have been 
considered as an additional, significant source of 
photons~\cite{Basar:2012bp,Basar:2014swa,Muller:2013ila,Ayala:2016awt}.

The PHENIX experiment at the Relativistic Heavy Ion Collider (RHIC) was 
the first to detect a large yield of direct photons in heavy ion 
collisions at \snn{200}~\cite{PHENIX:2005yls}. Earlier evidence was 
presented by WA98~\cite{WA98:2000vxl,WA98:2003ukc} for \snn{17.3}, with 
mostly upper limits below 1.5 \gevc in \pt, except for two points 
obtained from interferometry in the 0.1--0.3\,\gevc \pt range, which is 
below our \pt threshold. Multiple subsequent publications from PHENIX 
established that at RHIC energies the direct-photon yield below 
transverse momenta of 2 \gevc exceeded what was expected from hard 
processes by a factor of $\approx$10~\cite{Adare:2008ab}, showed a 
stronger-than-linear increase with the collision 
volume~\cite{Adare:2014fwh}, and a large anisotropy with respect to the 
reaction plane~\cite{PHENIX:2011oxq,Adare:2015lcd}. The STAR 
collaboration also reported an enhanced yield of direct photons at low 
\pt in \auau collisions at \snn{200}~\cite{STAR:2016use}; for minimum 
bias (MB) events the yield measured by STAR is a factor of $\approx$3 
lower for \pt below 2~\gevc, while it is consistent at higher 
\pt\footnote{The persisting discrepancy between STAR and PHENIX 
measurements at low \pt is noted, but can not be resolved by PHENIX 
alone and thus is not further discussed in this paper.}. Observations 
consistent with the PHENIX \auau measurements at \snn{200} were made by 
the ALICE Collaboration at the Large Hadron Collider 
(LHC)~\cite{Adam:2015lda} in \pbpb collisions at \sqsn = 2.76 TeV and, 
more recently, by PHENIX at the lower energies of 39 and 62.4 
GeV~\cite{Adare:2018wgc}.  The low transverse-momentum yield, for \pt 
below 2 \gevc, shows a power-law dependence on $\dNch|_{\eta\approx 0}$ 
with a power $\alpha\approx 1.25$~\cite{Adare:2018wgc}. The power 
$\alpha$ is independent of centrality or collision 
energy\footnote{Throughout the rest of the paper the subscript 
$\eta{\approx}0$ will be dropped and \dNch will always imply density at 
midrapidity.}. These experimental findings are qualitatively consistent 
with thermal radiation being emitted from a rapidly expanding and 
cooling fireball. However, it is challenging for theoretical models to 
describe all data quantitatively.

To further constrain the sources of low-momentum direct photons, PHENIX 
continues its program on such measurements in large- and small-system 
collisions. This paper extends a previous publication on Au$+$Au 
collisions at $\sqrt{s_{_{NN}}}=39$ and 62.4 GeV~\cite{Adare:2018wgc} 
and provides more detail about the measurement and the universal 
features exhibited by direct photons emitted from heavy ion collisions 
from RHIC to LHC energies, including inverse slopes and the scaling with 
\dNch, both as a function of \pt.

The paper is organized as follows:  Section~\ref{sec2:AuAu39-62} presents 
the measurement and the results of low-momentum direct-photon production 
in \auau at \sqsn = 39 and 62.4~GeV.  Section~\ref{sec:Comparisons} puts 
these results into context with other direct-photon measurements. 
Section~\ref{sec:Summary} gives the summary and conclusions.

\section{Low-momentum direct-photon production at \sqsn = 39 and 62.4~GeV}
\label{sec2:AuAu39-62}

\subsection{Experimental method for measuring direct photons}
\label{sec2:method}

\begin{figure}[hbt]
\includegraphics[width=1.0\linewidth]{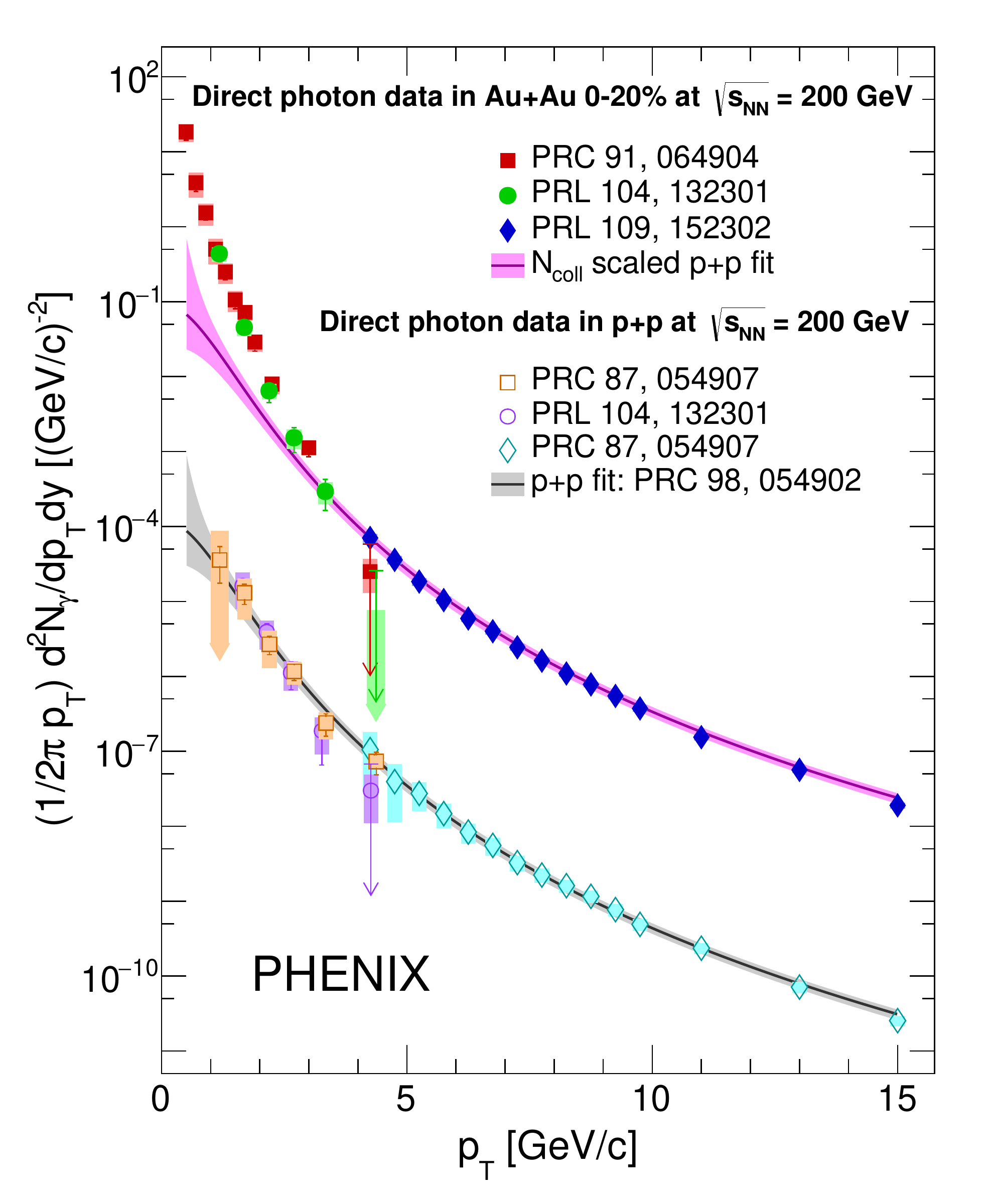}
\caption{The upper data points are the invariant yield of direct photons 
in \auau collisions in 0\%--20\% centrality bin at 200~GeV: the full 
square data are from an analysis based on external 
conversions~\cite{Adare:2014fwh}, the full circle data are from an 
analysis based on internal conversions~\cite{Adare:2008ab}, the full 
diamond data are from calorimeter measurements~\cite{Afanasiev:2012dg}. 
The lower data points are the invariant yield of direct photons in \pp 
collisions at 200~GeV: the open square and open circle data are from 
internal conversions~\cite{Adare:2012vn,Adare:2008ab}, the open diamond 
data are from calorimeter measurements~\cite{Adare:2012yt,Adler:2006yt}. 
The lower curve is a fit to the combined set of \pp data, extrapolated 
below 
1~GeV/$c$~\cite{Adare:2018wgc,Khachatryan:2018evz,Drees:2019ila,PHENIX:2018che}, 
and the upper curve is the $N_{\rm coll}$ scaled \pp fit with 
$N_{\rm coll} = 779.0$~\cite{Adare:2014fwh}.}
\label{fig:PHENIX}
\end{figure}

Figure~\ref{fig:PHENIX} presents the direct-photon $p_{T}$ spectra measured by 
PHENIX in \auau collisions in the 0\%--20\% centrality bin at \sqsn = 200~GeV, 
including data points from an analysis based on external 
conversions~\cite{Adare:2014fwh}, internal conversions~\cite{Adare:2008ab}, and 
from calorimeter measurements~\cite{Afanasiev:2012dg}.  Also shown are invariant 
yields of direct photons in \pp collisions at 200~GeV from internal 
conversions~\cite{Adare:2012vn,Adare:2008ab}, calorimeter 
measurements~\cite{Adare:2012yt,Adler:2006yt}, a fit to the combined set of \pp 
data, extrapolated below 
1~GeV/$c$~\cite{Adare:2018wgc,Khachatryan:2018evz,Drees:2019ila,PHENIX:2018che}, 
and an $N_{\rm coll}$-scaled \pp fit with 
$N_{\rm coll}=779.0$~\cite{Adare:2014fwh}.

The three techniques used for measuring direct photons deploy different detector 
systems within the PHENIX central arms\footnote{The PHENIX central arm acceptance 
is 0.7 units around midrapidity. Thus there is little difference between momentum 
and transverse momentum, so the terms will be used interchangeably in the 
following discussion.} (see Ref.~\cite{PHENIX:overview}) and various strategies 
to extract the direct photons from the decay-photon background include measuring:

\begin{itemize}

\item[(i)] photons that directly deposit energy into electromagnetic 
calorimeters. This is the method of choice to measure high momentum 
photons. At \pt below a few \gevc, the method suffers from significant 
background contamination from hadrons depositing energy in the 
calorimeter and the limited energy resolution~\cite{Afanasiev:2012dg}.

\item[(ii)] virtual photons that internally convert into \ee pairs and 
extrapolating their measured yield to zero mass. This technique was used 
for the original discovery of low-momentum direct photons at 
RHIC~\cite{Adare:2008ab}. The pairs are measured in the mass region 
above the \piz mass, which eliminates more than 90\% of the 
hadron-decay-photon background. The extrapolation to zero mass requires 
the pair mass to be much smaller than the pair momentum, and thus limits 
the measurement to \pt$>1$ \gevc.

\item[(iii)] photons that convert to \ee pairs in the detector material 
("external conversion method"). This method gives access to a nearly 
background-free sample of photons down to \pt below 1 
\gevc~\cite{Adare:2014fwh}.

\end{itemize}

The external-conversion method is used for the analysis presented here, 
which is the identical method used to analyze direct-photon production 
from 2010 \auau collisions at 
\snn{200}~\cite{Adare:2014fwh,Adare:2015lcd}.  Additional 
details can be found in Ref.~\cite{Khachatryan:2017gqp}. The analysis 
proceeds in multiple steps.  First established is \Nincl, which is a 
sample of conversion photons measured in the PHENIX-detector acceptance. 
This is done in bins of conversion photon \pt.  For a given \pt 
selection, the \Nincl sample relates to the true number of photons 
\gammaincl in that \pt range as follows:

\begin{equation}
\Nincl  = \varepsilon_{ee}a_{ee}\,p_{\rm conv}\,\gammaincl,
\label{eqn:incl}
\end{equation}

\noindent where $a_{ee}$ is the \ee pair acceptance, $\varepsilon_{ee}$ 
is the pair reconstruction efficiency, and $p_{\rm conv}$ is the conversion 
probability. In the next step a subsample \Ntag of \Nincl is tagged as 
\piz decay photons; details of how the \Ntag subsample is determined are 
described in Sec.~\ref{sec2:Ntag} below. Because \Ntag is a subset of 
\Nincl, it is related to the true number of \piz decay photons 
$\gamma^{\piz}$ among \gammaincl by:

\begin{equation}
\Ntag  = \varepsilon_{ee}a_{ee}\,p_{\rm conv}\,\ef \gamma^{\pi^{0}},
\label{eqn:pitag}
\end{equation}

\noindent with \ef being the average conditional probability of 
detecting the second photon in the PHENIX acceptance, given that one 
\piz decay photon converted and was reconstructed in the desired 
conversion photon \pt range. Here the average is taken over all possible 
\piz \pt. Taking the ratio of Eq.~\ref{eqn:incl} and 
Eq.~\ref{eqn:pitag} gives:

\begin{equation}
\frac{\gammaincl}{\gamma^{\pi^{0}}} = 
\Lb\ef\Rb_{\rm Sim} \Lb\frac{\Nincl}{\Ntag} \Rb_{\rm Data}.
\label{eqn:g2piratio}
\end{equation}

\noindent This ratio is constructed such that 
$\varepsilon_{ee}a_{ee}\,p_{\rm conv}$ explicitly cancels, eliminating 
the need to determine these quantities and the related systematic 
uncertainties. The only correction necessary is the conditional 
probability \ef, which is determined from a full Monte-Carlo simulation 
of the PHENIX detector indicated by the subscript $_{\rm Sim}$. The 
second factor is a ratio of directly measured quantities, indicated by 
$_{\rm Data}$. Finally, Eq.~\ref{eqn:g2piratio} can be divided by the 
fraction of hadron decay photons (\gammahadr) per \piz decay photon, 
which defines \Rg as a double ratio:

\begin{equation}
R_{\gamma} = \frac{\gammaincl}{\gammahadr} = 
\frac{\Lb \ef \Rb_{\rm Sim} 
\left( \Nincl / \Ntag \right)_{\rm Data}}{\left(\gammahadr / \gamma^{\pi^{0}}\right)_{\rm Gen}}.
\label{eqn:rgamma}
\end{equation}

\noindent where the ratio \gammahadr/$\gamma^{\pi^0}$ was determined with
a particle-decay generator, indicated by the subscript $_{\rm Gen}$.

If direct photons are emitted from the collision system in a particular 
\pt range, \Rg will be larger than unity. The denominator in 
Eq.~\ref{eqn:rgamma} can be obtained from the PHENIX hadron-decay 
generator {\sc exodus}~\cite{PHENIX:2009gyd}, based on the measured \piz 
spectra.  In the following sections, the determination of \Nincl, \Ntag, 
\ef, and \gammahadr/$\gamma^{\piz}$ will be discussed separately.

\subsection{Determining the inclusive photon sample {\Nincl} }
\label{sec2:Nincl}

The 2010 data samples of $7.79 \times10^7$ (at 39~GeV) and 
$2.12{\times}10^8$ (at 62.4~GeV) MB Au$+$Au collisions were recorded 
with the two PHENIX central-arm spectrometers, each of which has an 
acceptance of $\pi/2$ in azimuthal angle and $|\eta|<$ 0.35 in 
pseudorapidity. For both collision energies, the MB data sets cover a range 
of 0\%--86\% of the interaction cross section. The data sample for 62.4~GeV 
is large enough so that two centrality classes (0\%--20\% central 
collisions, 20\%--40\% midcentral collisions) could be analyzed separately.  
The event centrality is categorized by the charge measured in the PHENIX 
beam-beam counters~\cite{PHENIX:BBC}, which are located at a distance of 
144~cm from the nominal interaction point in both beam directions, 
covering the pseudorapidity range of $3.1<|\eta|<3.9$ and $2\pi$ in azimuth.

The PHENIX central-arm drift chambers and pad 
chambers~\cite{PHENIX:tracking}, located from 200 to 250~cm radially to 
the beam axis, are used to determine the trajectories and momenta of 
charged particles. The momenta are measured assuming the track 
originated at the event vertex (vtx) and traversed the full magnetic field. 
The tracks are identified as electrons or positrons by a combination of 
a minimum signal in the ring-imaging \v{C}erenkov (RICH) 
detector~\cite{PHENIX:PID} and a match of the track momentum with the 
energy measured in the electromagnetic calorimeter 
(EMCal)~\cite{PHENIX:EMCal}. The RICH cut requires that a minimum of 
three RICH phototubes be matched to the charged track within a radius 
interval of 3.4~cm $< r <$ 8.4~cm at the expected ring location. For 
each electron candidate an associated energy measurement in the EMCal is 
required, with an energy/momentum ratio, E/$p$, greater than 0.5. 
Electrons and positrons are combined to \ee pairs and further selection 
cuts are applied to establish a clean sample of photon conversions. Most 
photon conversions occur in the readout boards and electronics at the 
back plane of the hadron blind detector (HBD)~\cite{PHENIX:HBD}, located 
at a radius of $\approx$60~cm from the nominal beam axis. The relative 
thickness in terms of radiation length is equal to $X/X_{0} \approx 
2.5\%$; all other material between the beam axis and the drift chamber 
is significantly thinner. Electrons and positrons from these conversions 
do not traverse the full magnetic field\footnote{A special field 
configuration was used in 2010 for the operation of the HBD. In this 
configuration there is a nearly field free region around the beam axis 
out to 60~cm. Thus the field integral missed by tracks from photon 
conversions in the HBD back plane is rather small.}. Projecting the 
tracks back to the interaction point results in a small distortion 
of the reconstructed momenta, both in magnitude and in direction, which 
in turn results in an artificial opening angle of the \ee pair. This 
gives the pair an apparent mass (\mvtx), which depends monotonically on 
the radial location of the conversion point and is approximately 0.0125 
\gevcc for conversions in the HBD back plane.

\begin{figure}[hbt]
\includegraphics[width=1.0\linewidth]{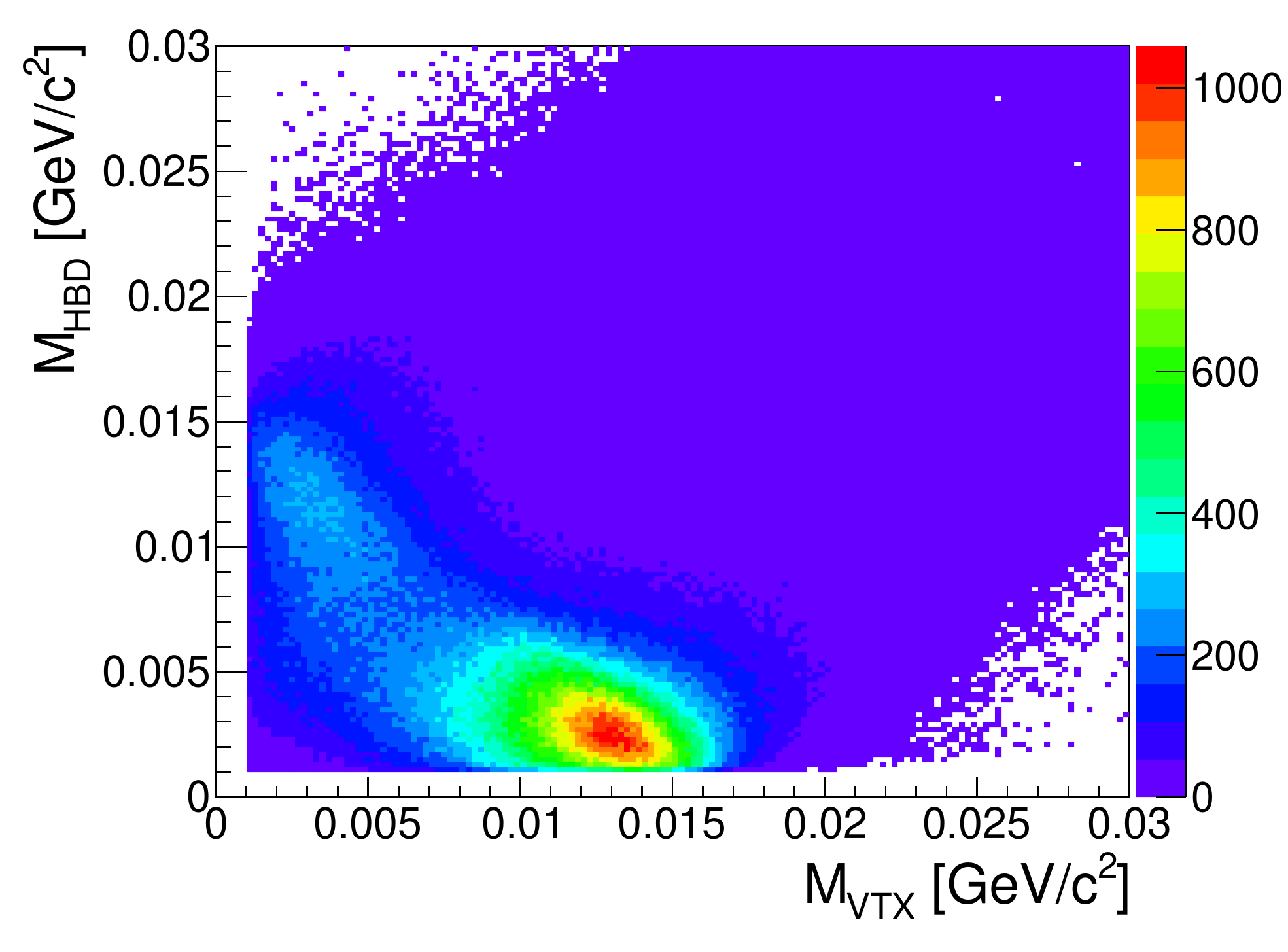}
\caption{Mass correlation of \ee pairs measured in \auau collisions at 
\sqsn = 62.4~GeV. Conversion photon \ee pairs are identified by the 
correlation between the mass calculated assuming the track originated at 
the interaction point (\mvtx) or at the HBD back plane (\mhbd). }
\label{fig:2DMass_agg}
\end{figure}

\begin{figure*}[hbt]
\begin{minipage}{0.48\linewidth}
\vspace{-0.5cm}
\includegraphics[width=0.99\linewidth]{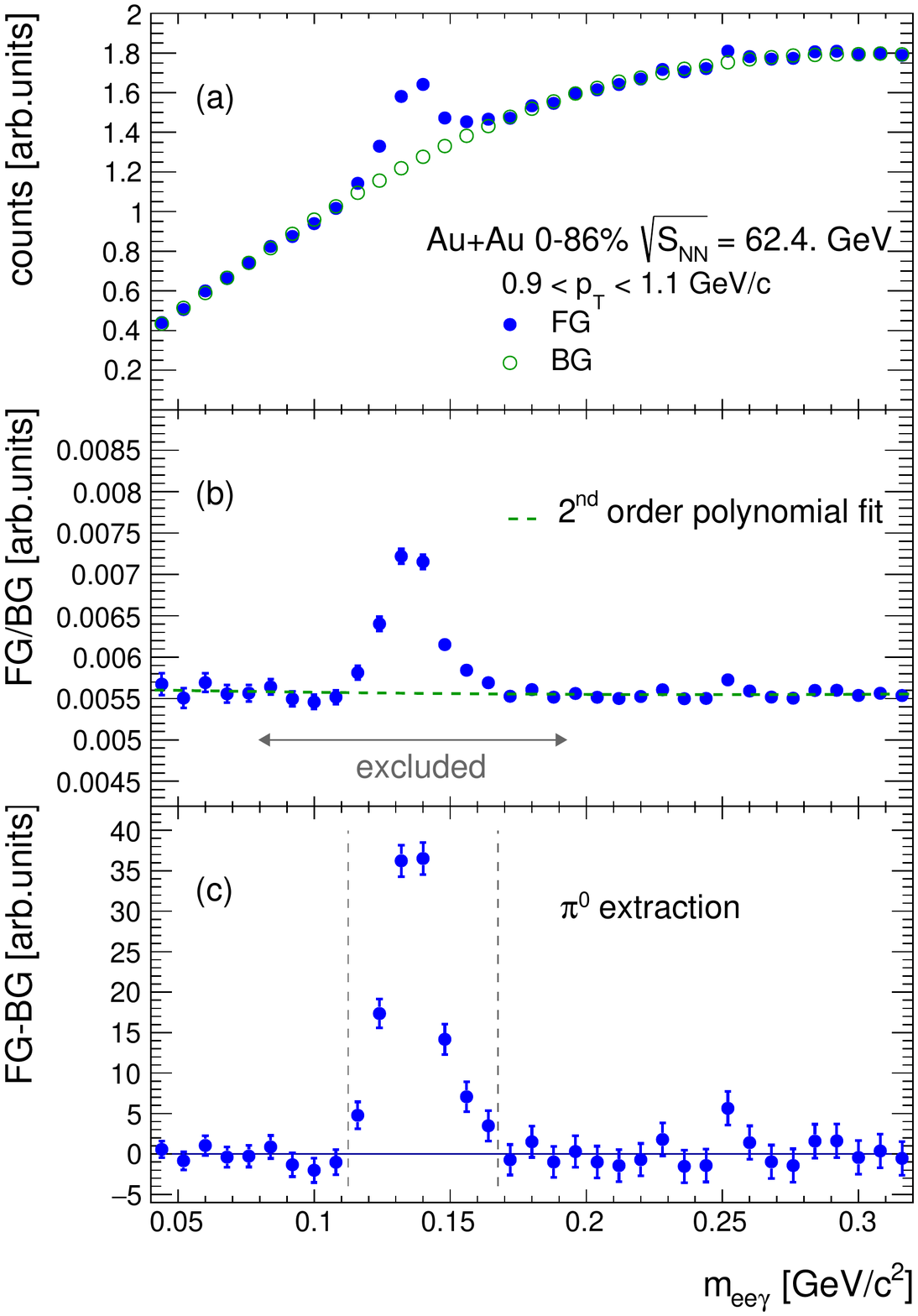}
\caption{Illustration of the \piz peak extraction method for one \pt bin 
from 0.9 to 1.1 \gevc in MB \auau  collisions at \snn{64}. Panel 
(a) shows the \ee$\gamma$ foreground (FG) and the normalized mixed-event 
background (BG). The middle panel (b) gives the ratio of foreground to 
background used to normalize the mixed event background. Panel (c) 
presents the counts after subtracting the normalized mixed-event 
background. }
\label{fig:pizExtraction}
\end{minipage}
\hspace{0.2cm}
\begin{minipage}{0.48\linewidth}
\vspace{-1.6cm}
\includegraphics[width=0.99\linewidth]{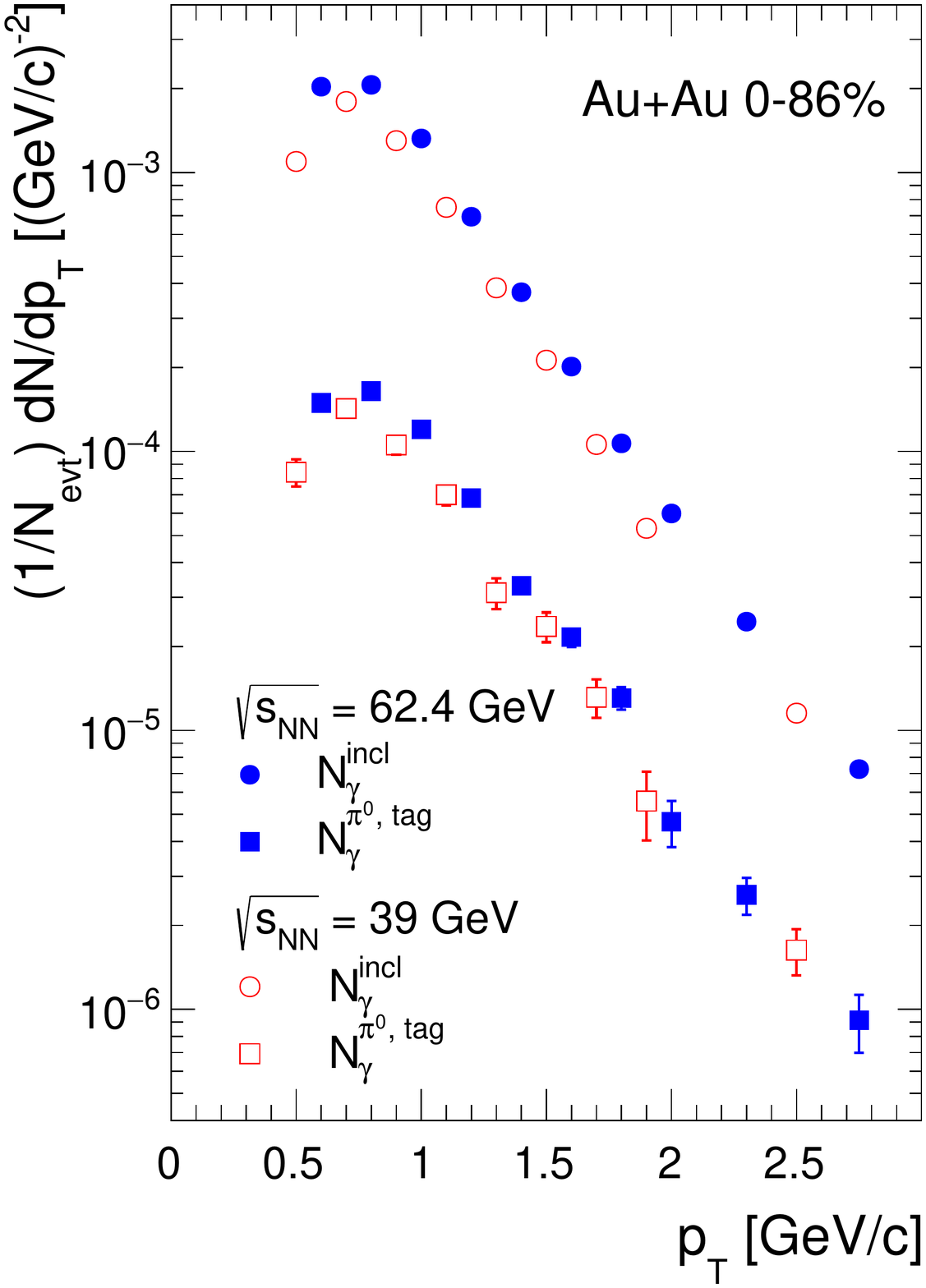}
\caption{Raw counts of \Nincl and its subsample \Ntag, which was tagged 
as photons from \piz decays.  Data for MB \auau collisions from 
39 and 62.4 GeV are given. }
\label{fig:NinclNtag}
\end{minipage}
\end{figure*}

To select photon conversions in the HBD back plane, the track momenta 
are re-evaluated assuming the tracks originated at the HBD back plane. 
For \ee pair from conversions in the HBD back plane, a mass (\mhbd) of 
below 0.005 \gevcc is calculated with a distribution expected for an \ee 
pair of zero mass measured with the PHENIX-detector resolution. 
Figure~\ref{fig:2DMass_agg} shows the correlation between the two 
different masses calculated for each pair. Photon conversions in the HBD 
back plane are clearly separated from \ee pairs from \piz Dalitz decays, 
$\pi^{0} \rightarrow \gamma e^{+}e^{-}$, which populate a region 
$\mvtx<0.005 \ \gevcc$ and \mhbd around 0.012 \gevcc. The region between 
the \ee pairs from Dalitz decays and conversion in the HBD back plane is 
populated by conversions at radii smaller than 60~cm. To select a clean 
sample of photon conversions in the HBD back plane, \Nincl, a two 
dimensional cut is applied: $\mhbd < 0.0045$ \ \gevcc and $0.01 < \mvtx 
< 0.015$ \gevcc. The purity of this photon sample was determined with a 
full Monte-Carlo simulation and is better than 99\%. The sample sizes 
are $9.42 \times10^4$ and $3.28 \times 10^5$, for 39 and 62.4 GeV, 
respectively.

\subsection{Tagging photons from {$\piz \rightarrow \gamma\gamma$} decays}
\label{sec2:Ntag}

Once the conversion-photon sample \Nincl is established, all \ee pairs 
in a given \pt bin are combined with showers reconstructed in the EMCal 
in the same event and then the invariant mass is calculated. A 
minimum-energy cut of 0.4 GeV is applied to remove charged particles 
that leave a minimum-ionizing signal in the EMCal and further reduce 
the hadron contamination by applying a shower-shape cut. 
Figure~\ref{fig:pizExtraction}(a) shows one example of the resulting 
mass distributions for a \pt bin around 1 \gevc from the 62.4-GeV MB 
data set. The \piz peak is clearly visible above a combinatorial 
background, which results from combining \ee pairs with all showers in 
the event, most of which are not correlated with the \ee pair.

A mixed-event technique is used to determine and subtract the mass 
distribution of these random combinations. In event mixing, all \ee 
pairs in a given event are combined with the EMCal showers from several 
other events. These other events are chosen to be in the same 10\% 
centrality selection and within 1~cm of the interaction point of the 
event with the \ee pair. The ratio of the measured (foreground) mass 
distribution and mixed event (background) mass distribution is fitted 
with a 2nd-order polynomial, excluding the mass range 
$0.08<m_{ee\gamma}<0.19$~\gevcc, around the \piz peak. 
Figure~\ref{fig:pizExtraction}(b) shows the ratio and the fit, which is 
used to normalize the mixed event background distribution over the full 
mass range; the result is included in Fig.~\ref{fig:pizExtraction}(b).

Figure~\ref{fig:pizExtraction}(c) depicts the counts remaining after the 
mixed event background distribution is subtracted from the foreground 
distribution. The raw yield of tagged \piz is calculated as the sum of 
all counts in mass window $0.11 < m_{ee\gamma} < 0.165\ \gevcc$. The 
counts in two side bands around the \piz peak are evaluated to account 
for any possible remaining mismatch of the shape of the combinatorial 
background from mixed events compared to the true shape. These side 
bands are $0.035<m_{ee\gamma}<0.110$~\gevcc and $0.165<m_{ee\gamma}< 
0.240$~\gevcc. The average counts per mass bin in the side-bands is 
subtracted from the raw tagged \piz counts, the resulting counts are the 
number of tagged \piz, \Ntag in the given \pt bin. 

\begin{figure}[t]
\begin{minipage}{0.99\linewidth}
\includegraphics[width=1.0\linewidth]{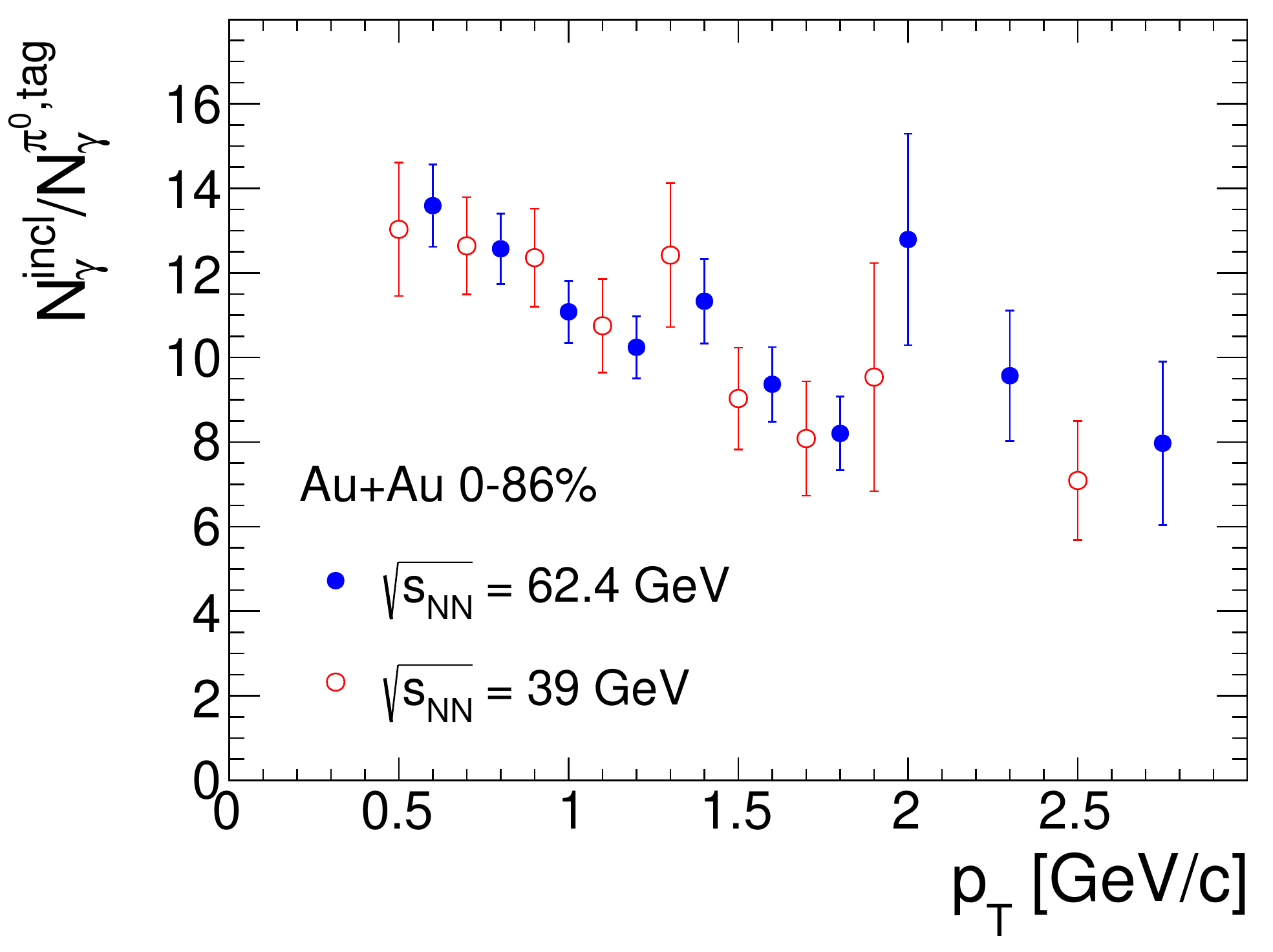}
\caption{The ratio of the measured inclusive photon yield \Nincl to the 
yield \Ntag of those photons tagged as \piz decay photons for MB 
data samples at \sqsn = 39 and 62.4~GeV.  The x-axis is the \pt
of the \ee pair.}
\label{fig:RatioNinclNtag}
\end{minipage}
\begin{minipage}{0.99\linewidth}
\includegraphics[width=1.0\linewidth]{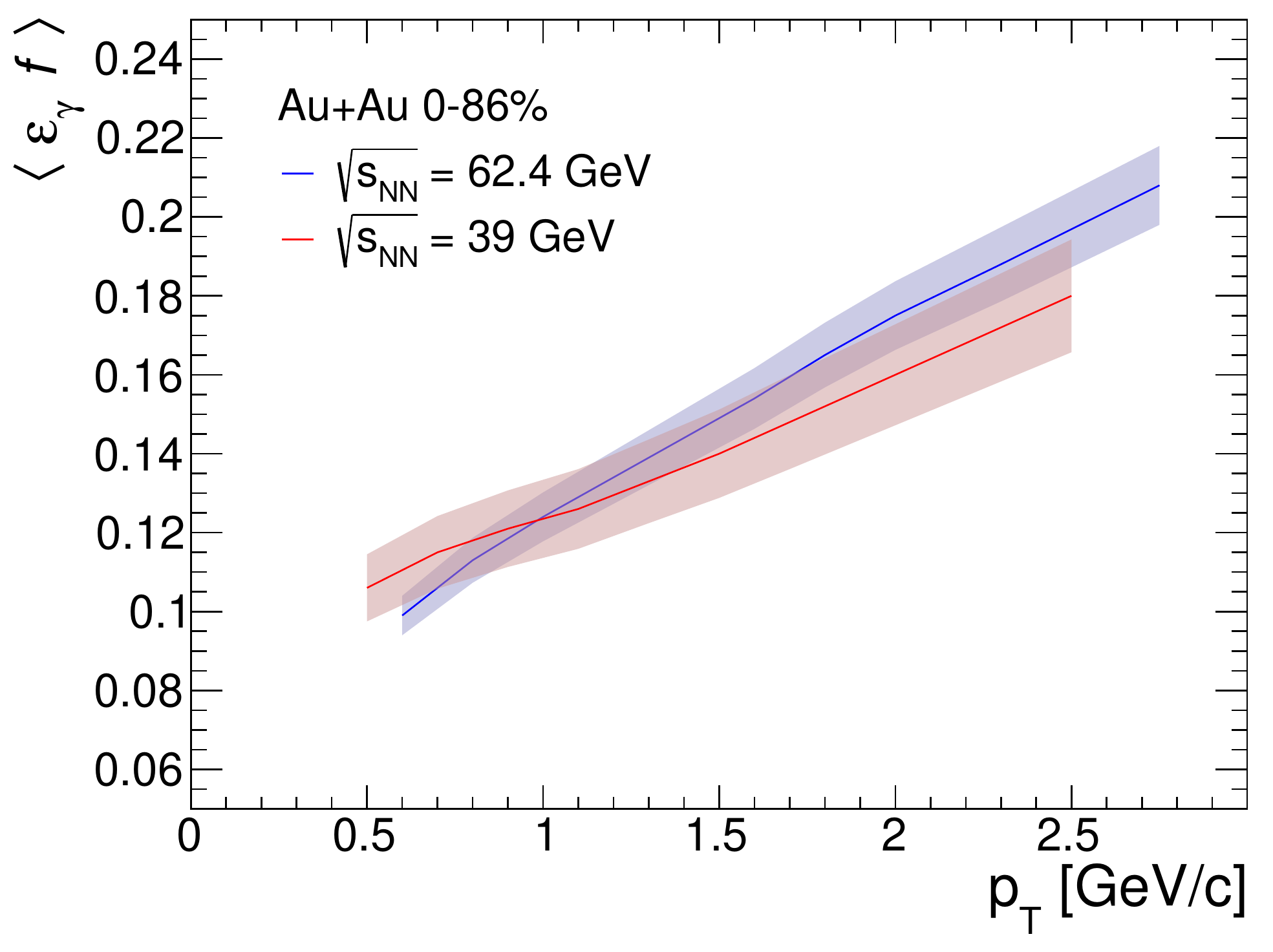}
\caption{Simulated conditional probability, \ef, to detect the second 
photon from a \piz decay in the MB data samples at \sqsn = 
39 and 62.4~GeV.  The x-axis is the \pt of the \ee pair.}
\label{fig:ef}
\end{minipage}
\end{figure}

Figure~\ref{fig:NinclNtag} shows both \Nincl and \Ntag for 39 and 62.4 
GeV MB \auau data. Figure~\ref{fig:RatioNinclNtag} gives the ratios, 
\Nincl/\Ntag.

The systematic uncertainties of the peak-extraction procedure were 
evaluated by choosing different-order polynomial function for the 
normalization and the various mass windows were varied in the procedure. 
It is found that \Ntag changes by less than 8\% and 5\% for 39 and 62.4 GeV 
data, respectively. These systematic uncertainties are mostly 
uncorrelated between \pt bins and thus are added in quadrature to the 
statistical uncertainties on \Ntag.

\subsection{The conditional {\piz} tagging probability}
\label{sec2:ef}

The conditional probability \ef, to tag an \ee pair that resulted from a 
conversion of a \piz decay photon with the second decay photon, depends 
on the parent \piz \pt spectrum, the \piz decay kinematics, the detector 
acceptance, and, the photon reconstruction efficiency. A Monte-Carlo 
method is used to calculate \ef. The method was developed for the 
direct-photon measurement from \auau collisions at \snn{200}, also 
recorded during 2010, as described in Ref.~\cite{Adare:2014fwh}. The 
calculation is done separately for MB and centrality selected \auau 
collisions at 39 and 62.4 GeV. Each calculation is based on an input 
\piz spectrum that was measured for the same data 
sample~\cite{PHENIX:2012oed}.

Figure~\ref{fig:ef} shows the results for MB collisions. The conditional 
probability \ef is small; it increases from approximately 10\% to 20\% 
over the \pt range from 0.8 to 2.5 \gevc. The visible difference between 
\ef for 39 and 62.4 GeV is due to the \sqs dependence of the \piz \pt 
spectra, which are much softer for the lower energies. Because \ef is 
evaluated for a fixed \pt range of the \ee pair, it is averaged over all 
possible \piz \pt. Thus the value of \ef at a fixed \ee pair \pt is 
sensitive to the parent \piz \pt spectrum.
 
The EMCal acceptance contributes a multiplicative factor of 0.35 to \ef 
at an \ee pair \pt = 0.8 \gevc, the factor increases to 0.45 at 2.5 
\gevc. This includes the geometrical dimension and the location of the 
EMCal sectors, the fiducial cuts around the sector boundaries and any 
dead areas in the EMCal. The minimum-energy cut of 0.4 GeV is the main 
contributor to the photon-reconstruction efficiency loss. This cut is 
equivalent to an asymmetry cut on the \piz decay photons; the effect 
being largest at the lowest \piz momenta that can contribute in a given 
\ee pair \pt bin. With additional, but small, contributions from the 
shower-shape cut and the conversion of the second photon, the 
reconstruction efficiency rises from $\approx$0.3 to 0.45 over the \pt 
range of 0.8 to 2.5 \gevc.

Figure~\ref{fig:ef} also shows the systematic uncertainties on \ef, 
which are 8\% and 5\% for 39 and 62.4 GeV, respectively. The 
uncertainty of the energy calibration and the accuracy of the \piz \pt 
spectra are the two dominant sources of systematic uncertainties. A 2\% 
change in the energy calibration, and with it a change of the actual 
energy cutoff, modifies \ef by 3\% to 4\%. For 62.4 GeV, the measured \piz 
\pt spectra agree in shape within $\pm$10\% with the charged-pion data 
from STAR~\cite{STAR:2007zea}. Possible shape variations within this 
range translate into an uncertainty of 3\% on \ef. 

For 39 GeV, STAR has published charged-pion data up to 2 
\gevc~\cite{STAR:2017sal}, these data agree in shape with the PHENIX 
\piz data within $\pm$10\%. However, due to the limited \pt range, the 
systematic uncertainties on the shape of the \piz \pt spectrum were 
determined from the systematic uncertainties of the PHENIX measurement 
alone, which is less restrictive and, thus, results in a larger 
uncertainty.

\begin{figure}[t]
\includegraphics[width=1.0\linewidth]{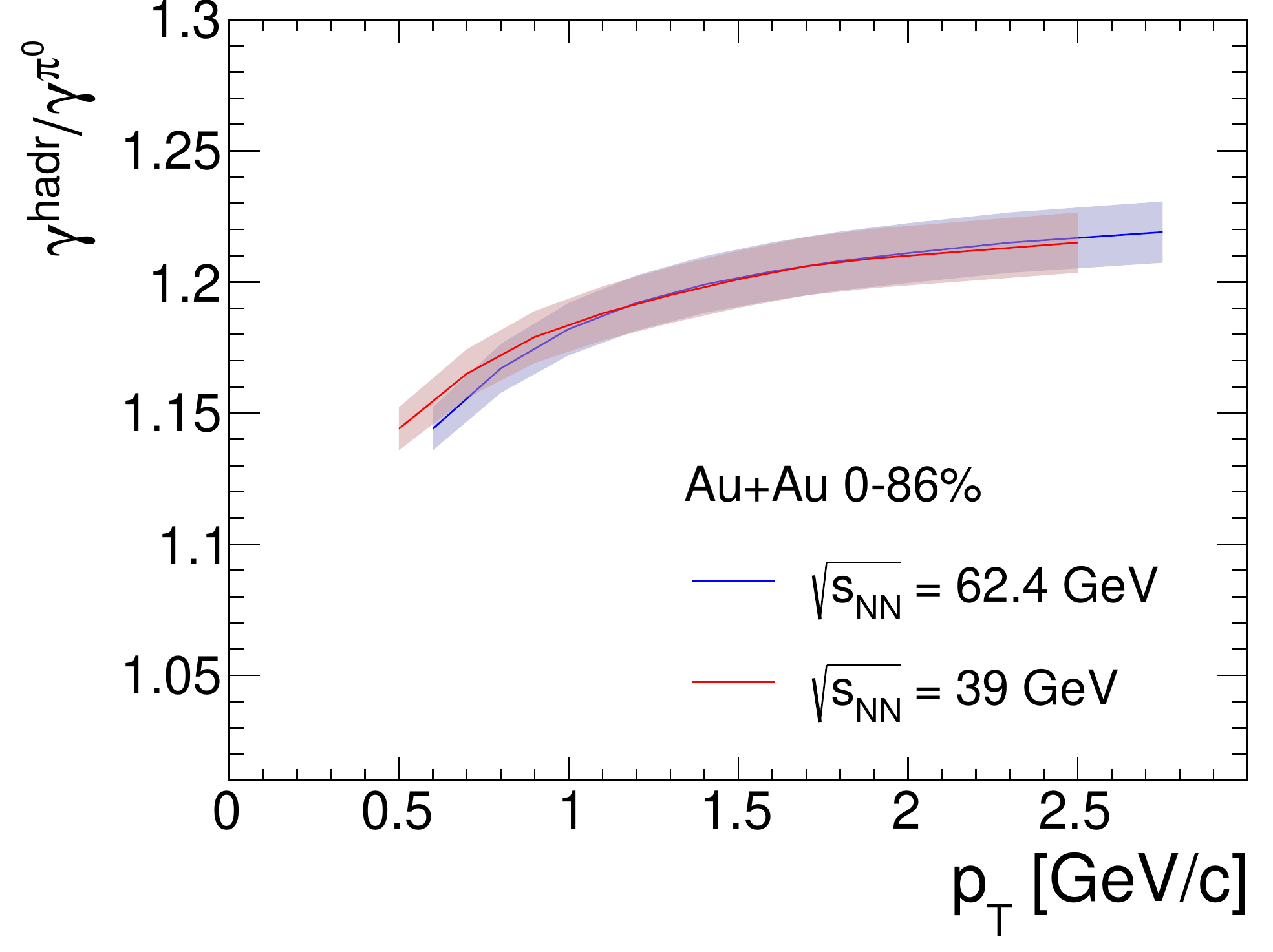}
\caption{Simulated ratio of photons from hadron decays to those from 
$\pi^{0}$ decays in the MB data samples at \sqsn = 39 and 62.4~GeV.
The x-axis is the \pt of the \ee pair.}
\label{fig:gh}
\end{figure}

\subsection{Decay photons form hadron decays}
\label{sec2:gh}

The ratio of all photons from hadron decays to those from \piz decays, 
$\gammahadr / \gamma^{\pi^{0}}$ in the denominator of 
Eq.~\ref{eqn:g2piratio}, is the final component that is needed to 
calculate \Rg. In addition to decays of \piz, decays of the $\eta$, 
$\omega$, and $\eta'$ mesons contribute to \gammahadr, with the $\eta$ 
decay being the largest contributor. Any other decays emit a negligible 
number of photons.

Photons from hadron decays are modeled based on the parent \pt distributions. For 
each centrality class, the measured \piz \pt spectrum is used to generate 
$\pi^{0}$s, which are subsequently decayed to photons using the known branching 
ratios and decay kinematics. The decay photons from $\eta$, $\omega$ and 
$\eta^{\prime}$ are modeled similarly, with a parent \pt distribution derived 
from the measured \piz \pt distributions, assuming $\m_T$ scaling 
(see~Refs.~\cite{Adare:2008ab,Ren:2021xbh} for 
details)\footnote{Ref.~\cite{Ren:2021xbh} recently noted that using $m_T$ scaling 
overestimates the $\eta$ meson yield in \pp collisions for \pt below 2 \gevc. The 
same work also shows that in \auau collisions at RHIC energies, this depletion is 
partially compensated by radial flow, which enhances the yield of $\eta$ in the 
same \pt region. For this analysis, removing the $m_T$ scaling assumption, while 
including the effect of radial flow, will reduce the number of photons from 
hadron decays by $\approx$2\% for $p_T{\approx}1$~\gevc, where the change is the 
largest. Correspondingly the direct-photon yield would increase by 2\%, which is 
within the systematic errors of 2.4\% quoted on the contribution of 
\gammahadr/\gammapiz to \Rg and much smaller than the overall statistical 
($>$7\%) and systematic ($>$5\%) uncertainties of the \Rg measurement at \pt of 
1\gevc.} The normalization of photons from $\eta$, $\omega$, and $\eta^{\prime}$ 
is set to $\eta/\pi^{0}=0.46{\pm}0.06$, $\omega/\pi^{0}=0.9{\pm}0.06$ and 
$\eta^{\prime}/\pi^{0}=0.25{\pm}0.075$ all at $p_{T}=5$~GeV/$c$.

Figure~\ref{fig:gh} shows the $\gammahadr / \gamma^{\pi^{0}}$ ratio. 
The ratio increases with \pt and saturates at high \pt between 1.22 and 
1.23.  There is no appreciable \sqs dependence of $\gammahadr / 
\gamma^{\pi^{0}}$. Following Ref.~\cite{Adare:2014fwh}, the systematic 
uncertainties from $\gammahadr/\gamma^{\pi^{0}}$ on \Rg are estimated 
to be 2.4\%.

\subsection{Direct-photon spectra} 
\label{sec2:directg}

\begin{figure}[hbt]
\includegraphics[width=1.0\linewidth]{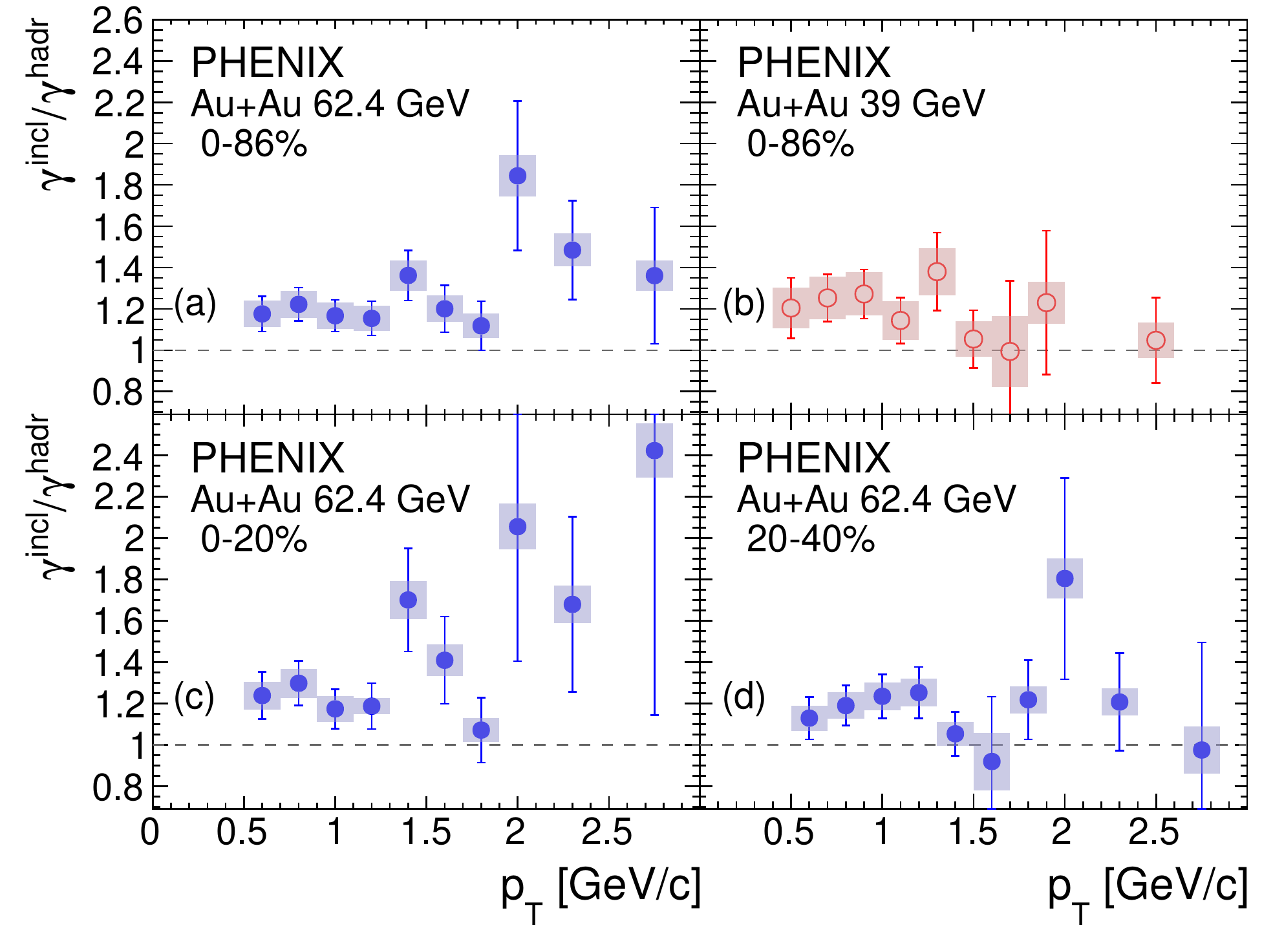}
\caption{$R_{\gamma}$ (\gammaincl/\gammahadr) for MB (0\%--86\%) 
\auau collision at \sqsn = (a) 62.4 and (b) 39~GeV.  Also shown for 
62.4~GeV are centrality bins (c) 0\%--20\% and (d) 20\%--40\%.  Data 
points are shown with statistical (bar) and systematic (box) 
uncertainties.  }
\label{fig:Rgamma}
\end{figure}

After each factor in Eq.~\ref{eqn:rgamma} is determined, \Rg can be 
calculated.  Figure~\ref{fig:Rgamma} shows the results for all centrality 
classes. Despite the significant statistical and systematic 
uncertainties, the majority of the data points are above unity at a 
value around $\Rg \approx 1.2$.  This indicates the presence of a 
direct-photon component of $\approx$20\% relative to hadron-decay photons 
in \auau collisions at 39 and 62.4 GeV. There is no obvious \pt 
dependence over the observed range; furthermore, the \sqs and centrality 
dependence, if any, must be small.

\begin{figure}[hbt]
   \includegraphics[width=1.0\linewidth]{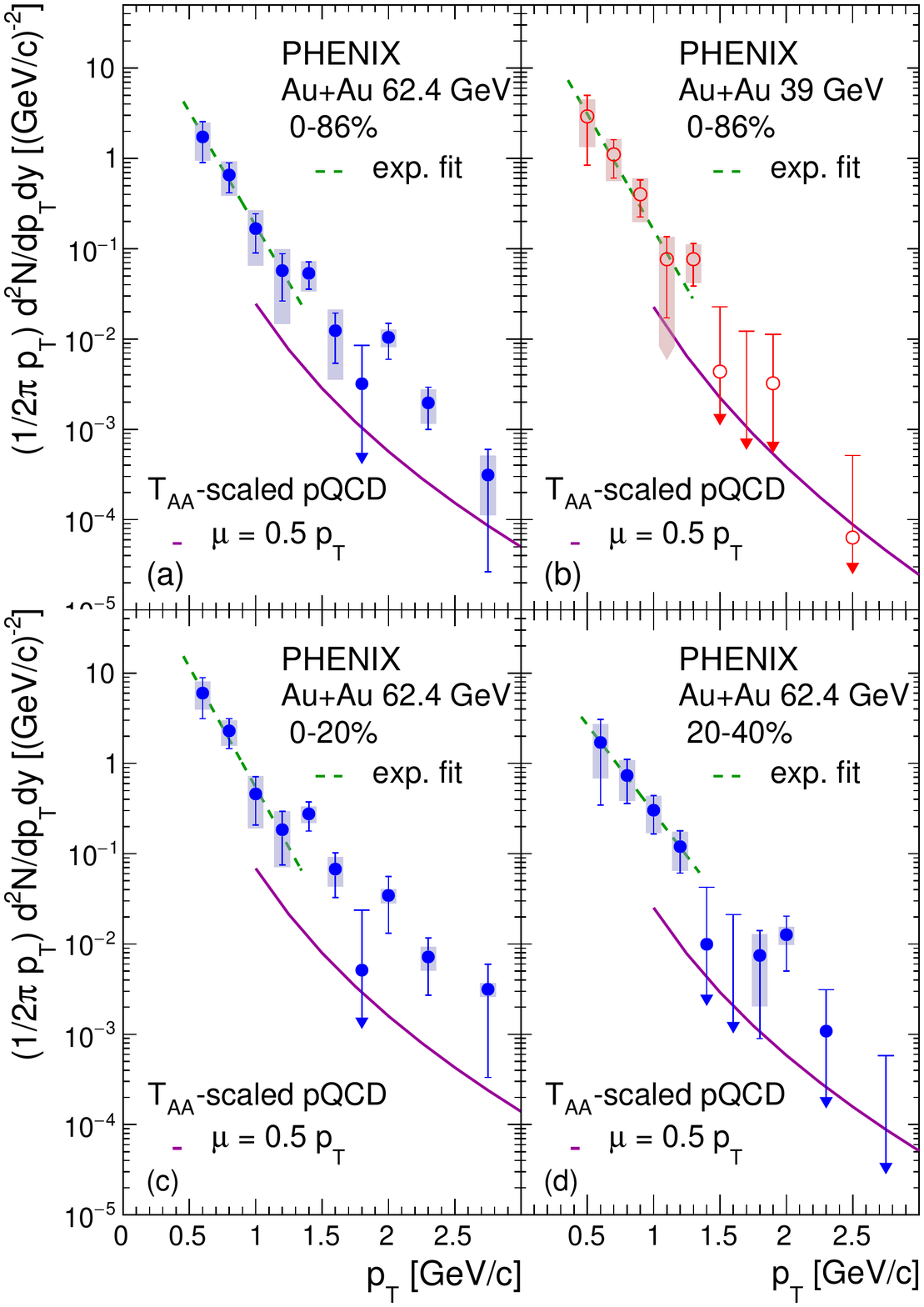}
\caption{Direct-photon $p_{T}$-spectra in MB (0\%--86\%) \auau 
collisions at \sqsn = (a) 62.4 and (b) 39~GeV. Also shown for 62.4~GeV 
are the centrality bins (c) 0\%--20\% and (d) 20\%--40\%.  Data points 
are shown with statistical (bar) and systematic (box) uncertainties, 
unless the central value is negative (arrows) or is consistent with zero 
within the statistical uncertainties (arrows with data point). In these 
cases the upper limits are given with confidence levels of 95\%.}
\label{fig:Spectra}
\end{figure}

To further analyze the data \Rg is converted to a direct-photon \pt 
spectrum $\gammadir$ using the hadron-decay-photon spectra calculated in 
Sec.~\ref{sec2:gh}:
\begin{equation}
\gammadir= \Lb R_{\gamma} - 1 \Rb \gammahadr.
\label{eqn_direct2}
\end{equation}

\begin{figure*}[hbt]
\includegraphics[width=0.99\linewidth]{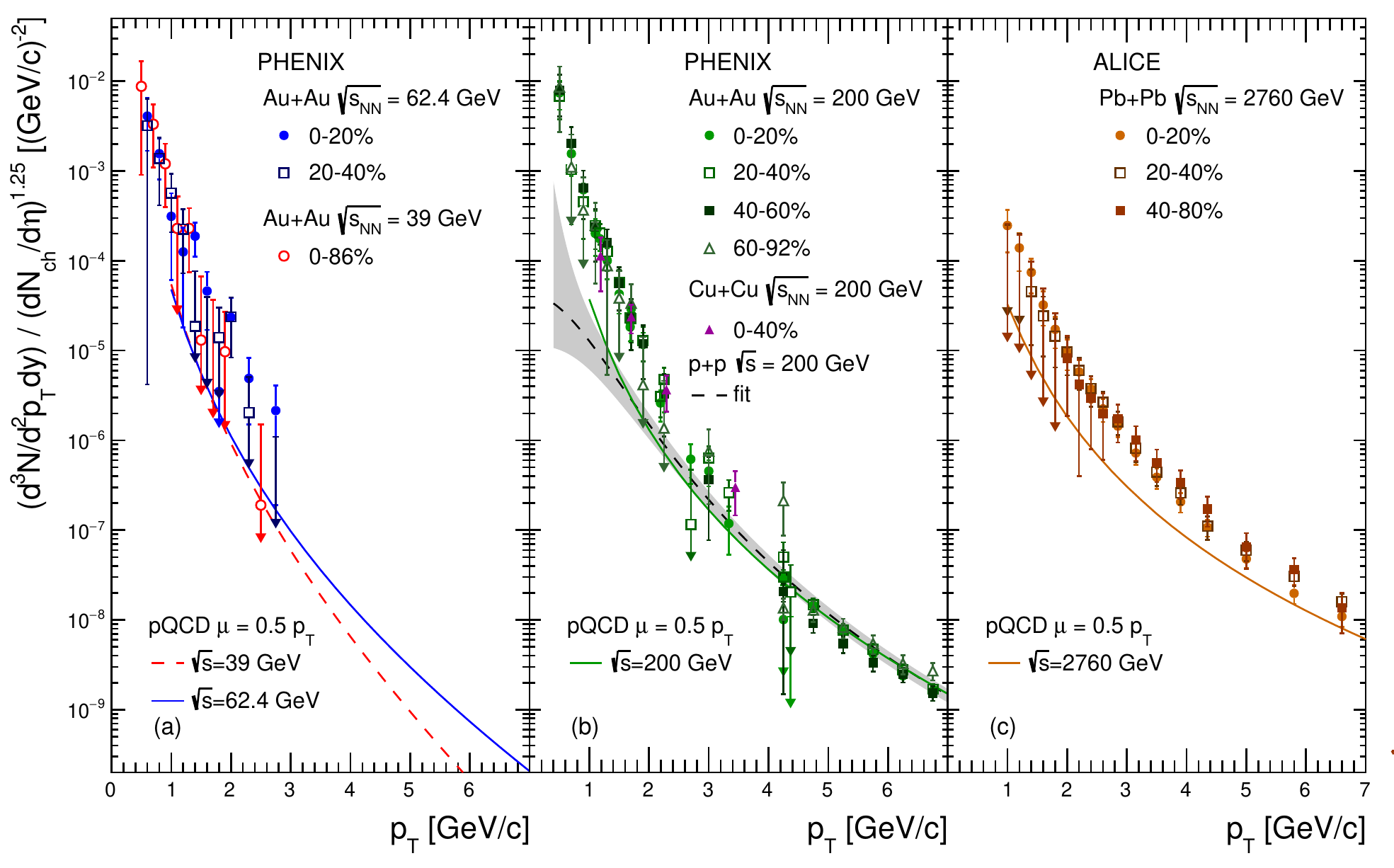}
\caption{Direct-photon $p_{T}$-spectra normalized by $(\dNch)^{1.25}$ 
for (a) the MB \auau 39 and centrality selected 62.4~GeV data sets from 
Fig.~\ref{fig:Spectra}, (b) various centrality selected 200~GeV 
\auau~\protect\cite{Afanasiev:2012dg,Adare:2008ab,Adare:2014fwh} and 
\cucu~\protect\cite{PHENIX:2018che} data sets, and (c) various 
centrality selected \pbpb 2760~GeV data 
sets~\protect\cite{Adam:2015lda}. Also shown in panel (c) is the \pp fit 
discussed in the text. The pQCD curves in the three panels are from 
Refs.~\protect\cite{Paquet:2015lta,Paquet:2017}.  The error bars shown 
are total uncertainties, i.e. the quadrature sum of statistical and 
systematic uncertainties.}
\label{fig:Norm_pTspectra_all}
\end{figure*}

Figure~\ref{fig:Spectra} presents the calculated direct-photon $p_{T}$ 
spectra.  In addition to the systematic uncertainty on \Rg, the 
hadron-decay-photon spectra contribute $\approx$10\% to the systematic 
uncertainties. These uncertainties cancel in $\gammahadr / 
\gamma^{\piz}$, but need to be considered here. Each centrality and 
energy selection is compared to the expected prompt-photon contribution 
from hard-scattering processes based on 
perturbative-quantum-chomodynamics (pQCD) calculations 
from~\cite{Paquet:2015lta,Paquet:2017}. Shown are the calculations at 
the scale $\mu=0.5\,p_{T}$, which were extrapolated down to 
$p_T=1$~GeV/$c$.  The scale was selected as it typically gives a good 
description of prompt-photon measurements in \pp collisions (see also 
Fig.~\ref{fig:Norm_pTspectra_all}). To represent hard scattering in 
\auau collisions, the calculation is multiplied with the nuclear-overlap 
function $T_{\rm AA}$ for the given event 
selection~\cite{Adare:2015bua}, assuming an inelastic \pp cross sections 
of $\sigma_{\rm inel}=33.8$~mb at 39~GeV $\sigma_{\rm inel}=35.61$~mb 
at 62.4~GeV.  Table~\ref{Tab:TAA} gives the values. 
Below 1.5 \gevc, there is a clear enhancement of the data above the 
scaled pQCD calculation, consistent with the expectation of a 
significant thermal contribution.

\begin{table}[htb!]
\caption{The values of $T_{\rm AA}$ obtained from Ref.~\cite{Adare:2015bua}.}
\label{Tab:TAA}
\begin{ruledtabular}  \begin{tabular}{ccc}
   \sqsn  &   Centrality-class   &  $T_{\rm AA}$    \\
   (GeV)  &     selection        &  (mb$^{-1}$)     \\
   \hline
   62.4   &   0\%--20\%          &  18.44 $\pm$ 2.49   \\
   62.4   &  20\%--40\%          &   6.77 $\pm$ 0.82   \\
   62.4   &   0\%--86\%          &   6.59 $\pm$ 0.89   \\
   39     &   0\%--86\%          &   6.76 $\pm$ 1.08   \\
   \end{tabular}   \end{ruledtabular}
\end{table}

To characterize the enhancement, the data is fitted with a falling 
exponential function given by
\begin{equation}
\frac{1}{2\pi}\frac{d^{2}N}{dp_{T}dy} \approx \exp{\! \Lb -\frac{p_{T}}{T_{\rm eff}} \Rb}.
\label{eqn_Teff}
\end{equation}

The data sets were fitted below a \pt of 1.3~\gevc, where statistics are 
sufficient.  Table~\ref{Tab:Teff} summarizes the results, which are also 
shown in Fig.~\ref{fig:Spectra}. Systematic uncertainties were obtained 
with the conservative assumption that the uncertainties are 
anticorrelatated over the observed \pt range. All values are consistent 
with a common inverse slope $T_{\rm eff}$ of $\approx$0.170~\gevc.  For the MB 
and 0\%--20\% centrality \auau sample at 62.4 GeV, the data in the range 
from 0.9 to 2.1 \gevc is also fitted. The values are slightly above 0.24 
\gevc and are larger than the value extracted for the lower-\pt range.  
A possible increase of $T_{\rm eff}$ with \pt is consistent with the values 
obtained from \auau at 200 GeV~\cite{Adare:2014fwh} and \pbpb at 2.76 
TeV~\cite{Adam:2015lda}, which were fitted in the higher-\pt range.  See 
a more detailed discussion in the next section.


\begin{table}[htb!]
\caption{Inverse slopes fitted to the direct-photon spectra in
    different \pt ranges.}
\label{Tab:Teff}
   \begin{ruledtabular} \begin{tabular}{ccccc}
   \pt      &  \sqsn  &   Centrality  &  $T_{\rm eff}$    &  $\rm{\chi^{2}/{\rm NDF}} $\\
 (\gevc)    &  (GeV)  &   class       &  (\gevc)      \\
\hline
$\pt<1.3$   &  62.4   &    0\%--20\%  &  0.163 $\pm$ 0.031 $\pm ^{0.016}_{0.009}$ & 0.44/2 \\
            &  62.4   &   20\%--40\%  &  0.224 $\pm$ 0.067 $\pm ^{0.034}_{0.018}$ & 0.01/2 \\
            &  62.4   &    0\%--86\%  &  0.172 $\pm$ 0.032 $\pm ^{0.022}_{0.011}$ & 0.16/2 \\
            &  39     &    0\%--86\%  &  0.169 $\pm$ 0.035 $\pm ^{0.020}_{0.011}$ & 0.41/2 \\
\\
$0.9<\pt<2.1$  &  62.4  &  0\%--20\%  &  0.241 $\pm$ 0.048 $\pm ^{0.024}_{0.012}$ & 6.96/4\\
               &  62.4  &  0\%--86\%  &  0.245 $\pm$ 0.046 $\pm ^{0.044}_{0.016}$ & 5.61/4\\
   \end{tabular}   \end{ruledtabular}
\end{table}


\section{Comparison to Direct-Photon Measurements from higher collision 
energies}
\label{sec:Comparisons}

In this section, the direct-photon results from \auau collisions at 39 
and 62.4 GeV are discussed in the context of other direct-photon 
measurements from heavy ion collisions at higher collision energies, 
specifically \auau collisions at 200 GeV from RHIC and \pbpb collisions 
at 2.76 TeV from LHC. The discussion is divided into three parts.  The 
first part recalls the already published scaling behavior of the direct 
photon yield with $(\dNch)^{\alpha}$~\cite{Adare:2018wgc}. The next part takes
a closer look at the \pt and \sqsn dependence of the inverse 
slope $T_{\rm eff}$ of the direct-photon \pt spectra. The last part 
investigates the dependence or independence of the scaling variable 
$\alpha$ on the \pt range.

\begin{table*}[hbt!]
\caption{Values for \dNch and \Ncoll obtained from published 
experimental data. The collaboration and Ref. numbers are indicated in 
column six. See text for explanation of the extrapolation used for the 
\pp collision data at 62.4~GeV. The same \dNch and \Ncoll were used for 
the corresponding pQCD curves in Figs.~10 and 13--15.
}
\label{Tab:dNch}
   \begin{ruledtabular}   \begin{tabular}{cccccc}
 Collision  system & \sqsn (GeV) & Centrality class & \dNch & \Ncoll & Collaboration [Ref.]
\\
   \hline
\pp   & 62.4  &  -         & $1.86 \pm 0.08$   &    1  &  UA5~\cite{Zyla:2020zbs,UA5:1982ygd,UA5:1986yef}   \\
      & 200   &  -         & $2.38 \pm 0.17$   &    1  &  PHENIX~\cite{Adare:2015bua}   \\
      & 2760  &  -         & $3.75 \pm 0.26$   &    1  &  ALICE~\cite{Adam:2015gka}   \\
\\
\cucu & 200   &  0\%--40\% & $109.3 \pm 7.8$  &   $108.2  \pm 12.0$  &  PHENIX~\cite{Adare:2015bua} \\
      & 200   &  0\%--94\% & $51.7 \pm 3.6$   &    $51.8  \pm 5.6$   &    "  \\
\\
\auau & 39    &  0\%--86\% & $104.3 \pm 8.9$   &   $228.4  \pm 36.5$  &  PHENIX~\cite{Adare:2015bua} \\
      & 62.4  &  0\%--86\% & $131.5 \pm 11.2$  &   $228.5  \pm 30.9$  &    " \\
      & 62.4  &  0\%--20\% & $341.2 \pm 29.3$  &   $656.6  \pm 88.7$  &   " \\
      & 62.4  & 20\%--40\% & $151.8 \pm 12.7$  &   $241.1  \pm 29.2$  &   " \\
      & 200   &  0\%--20\% & $519.2 \pm 26.3$  &   $770.6  \pm 79.9$  &   " \\
      & 200   & 20\%--40\% & $225.4 \pm 13.2$  &   $241.1  \pm 28.4$  &   " \\
      & 200   & 40\%--60\% & $85.5  \pm 8.0$   &   $82.6   \pm 9.3$   &   " \\
      & 200   & 60\%--92\% & $16.4  \pm 2.8$   &   $12.1   \pm 3.1$   &   " \\
\\
\pbpb & 2760  &  0\%--20\% & $1206.8 \pm 45.8$ &   $1210.9 \pm 132.5$ &  ALICE~\cite{Aamodt:2010cz} \\
      & 2760  & 20\%--40\% & $537.5 \pm 19.0$  &   $438.4  \pm 42.0$   &  " \\
      & 2760  & 40\%--80\% & $130.3 \pm 5.3$   &   $77.2   \pm 18.0$  &   " \\
   \end{tabular}   \end{ruledtabular}
\end{table*}

\subsection{Scaling of the direct-photon yield with {$(\dNch)^{\alpha}$}}
\label{sec3:scaledyield} 

It was shown in Ref.~\cite{Adare:2018wgc} that the direct-photon yield 
from heavy ion collisions is approximately proportional to 
$(\dNch)^{\alpha}$ with common power $\alpha \approx 1.25$ across 
collision energies, systems, and centrality. 
Figure~\ref{fig:Norm_pTspectra_all} presents the direct-photon yield 
normalized to $(\dNch)^{1.25}$ for a large range of data 
sets\footnote{The WA98 data are not shown here and in the following 
plots. The upper limits from WA98 for \pt$<1.5$~\gevc are consistent 
with the lower end of the uncertainties of the PHENIX 62.4 GeV and 39 
GeV data, but they do not significantly constrain the scaling behavior 
at low \pt. The STAR data are also not shown as the tension with the 
PHENIX data remains unresolved, while the multiple publications from 
PHENIX, based on different data sets and analysis methods, show self 
consistent results. If taken at face value, the STAR data do 
demonstrate a similar scaling behavior with \Nch for \pt$<2$~\gevc, but 
at a factor-3-lower direct-photon yield.}. Panel (a) shows the 
data sets that are derived from the \auau measurements at 39 and 
62.4~GeV shown in Fig.~\ref{fig:Spectra}. 
Panel (b) presents PHENIX measurements from 
\auau~\cite{Afanasiev:2012dg,Adare:2008ab,Adare:2014fwh} and
\cucu~\cite{PHENIX:2018che} collisions at \snn{200}. 
Panel (c) uses the ALICE measurement from \pbpb 
collisions at \snn{2760}~\cite{Adam:2015lda}. All panels show pQCD 
calculations for \pp collisions at the corresponding \sqs, extrapolated 
to \pt= 1 \gevc at the scale 
$\mu=0.5~p_{T}$~\cite{Paquet:2015lta,Paquet:2017}.  

Table~\ref{Tab:dNch} gives the \dNch and \Ncoll values, which are are 
used to normalize the integrated yields and are obtained from published 
experimental data. The values for \pp collisions at 62.4 are taken from 
Fig.~52.1 of Ref.~\cite{Zyla:2020zbs}, which was interpolated between 
UA5 data at \s{53}~\cite{UA5:1982ygd} and 200 GeV~\cite{UA5:1986yef}. 
The values for \pp and heavy ion collisions from \sqsn= 7.7~GeV to 
200~GeV are from PHENIX~\cite{Adare:2015bua}; the values for 2760 GeV 
\pp data are from ALICE~\cite{Adam:2015gka}; and the values for \pbpb 
collision data at 2760~GeV are also from ALICE~\cite{Aamodt:2010cz}.


Figure~\ref{fig:Norm_pTspectra_all}(b) also gives a fit to the \pp data 
at $\sqs = 200$ GeV~\cite{Adare:2018wgc,PHENIX:2018che} with the 
empirical form:
\begin{equation}
\frac{d^3N}{d^2\pt dy} =  \frac{A_{pp}}{(1 + (\frac{p_{T}}{p_{0}})^2 )^{n}},
\label{eqn:ppfit}
\end{equation}

\noindent where the parameters are 
$A_{pp}=1.60\!\cdot\!10^{-4}$~(GeV/$c$)$^{-2}$, $p_{0}=1.45$~GeV/$c$ 
and $n=3.3$. The band represents the uncertainty of the fit.

All three panels in Fig.~\ref{fig:Norm_pTspectra_all} show that at a 
given \sqsn the normalized direct-photon yield from \AB{A}{A} collisions 
is independent of the collision centrality. This is true both for low 
and high \pt. Comparing the yield at \pt below 3--4~\gevc across panels 
reveals that the yield is also remarkably independent of \sqsn. Above 
\pt of 4 to 5~\gevc the normalized yield does show the expected \sqsn 
dependence and is described by the pQCD calculations.

In the high-\pt range, hard-scattering processes dominate direct-photon 
production, and these direct-photon contributions are not altered 
significantly by final-state effects. Different centrality selections 
show the same normalized yield, which reflects that empirically $\Ncoll 
\propto \dNch^{1.25}$~\cite{Adare:2018wgc}. It remains surprising that 
within uncertainties the same scaling also holds at lower \pt where 
direct-photon emission should be dominated by thermal radiation from the 
fireball. In the following sections, the similarity of the low-\pt 
direct-photon spectra, both in shape and in normalized yield, is 
analyzed more quantitatively.

\begin{figure}[hbt]
 \includegraphics[width=1.0\linewidth]{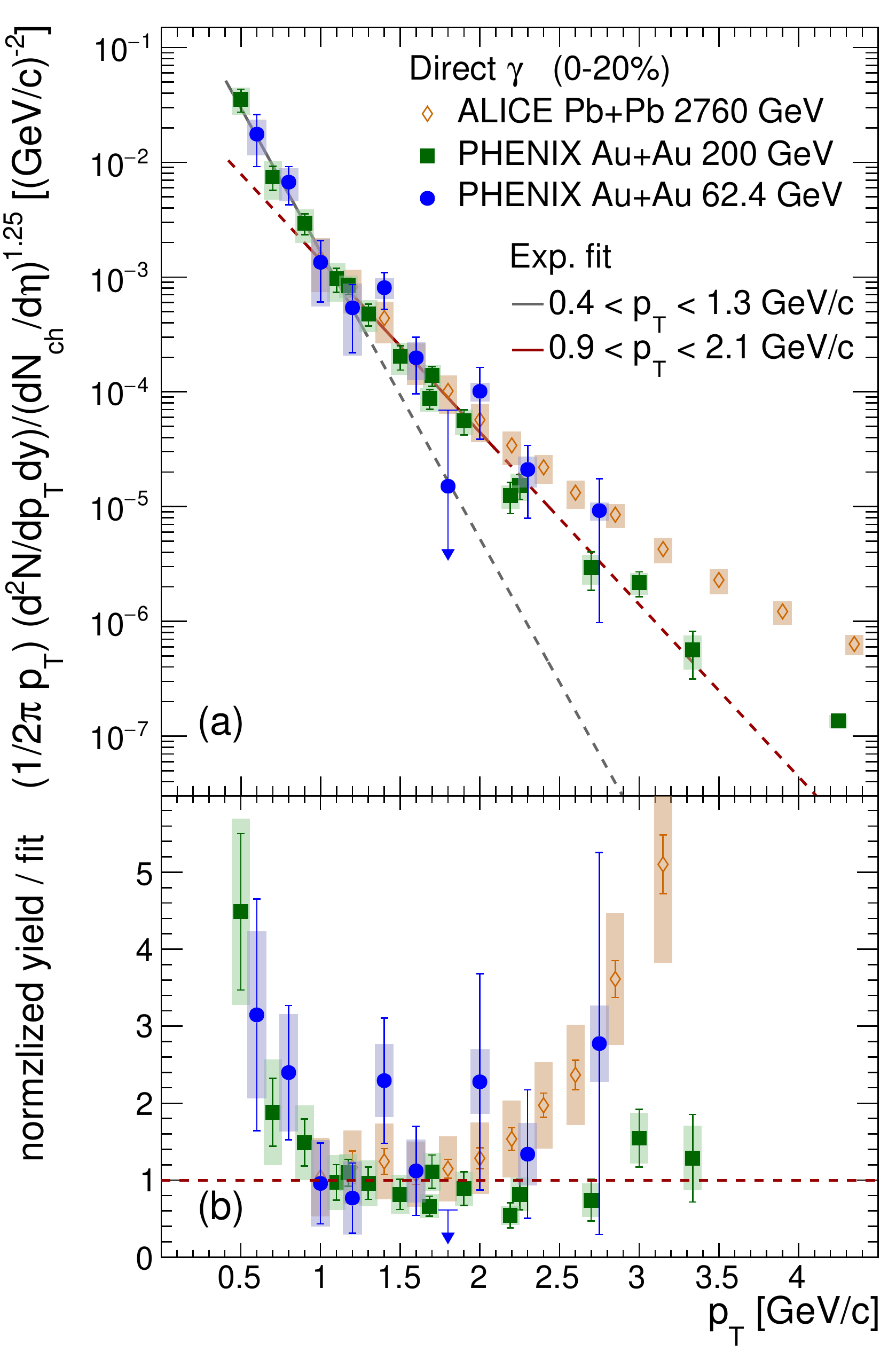}
\caption{Direct-photon yield normalized to $(\dNch)^{1.25}$ in the 
low-\pt region for 0\%--20\% centrality in \pbpb at 2760 GeV, \auau at 200 
GeV, and \auau at 62.4 GeV. Panel (a) gives the normalized yield and two 
exponential fits to the data in the \pt region below 1.3~\gevc and from 
0.9 to 2.1~\gevc. The dashed line extrapolates the fits beyond the fit 
ranges.  Panel (b) shows the ratio of the data sets to the fit in the 
range 0.9 to 2.1~\gevc range.}
\label{fig:LowPtSpectra}
\end{figure} 

\begin{figure}[hbt!]
 \includegraphics[width=1.0\linewidth]{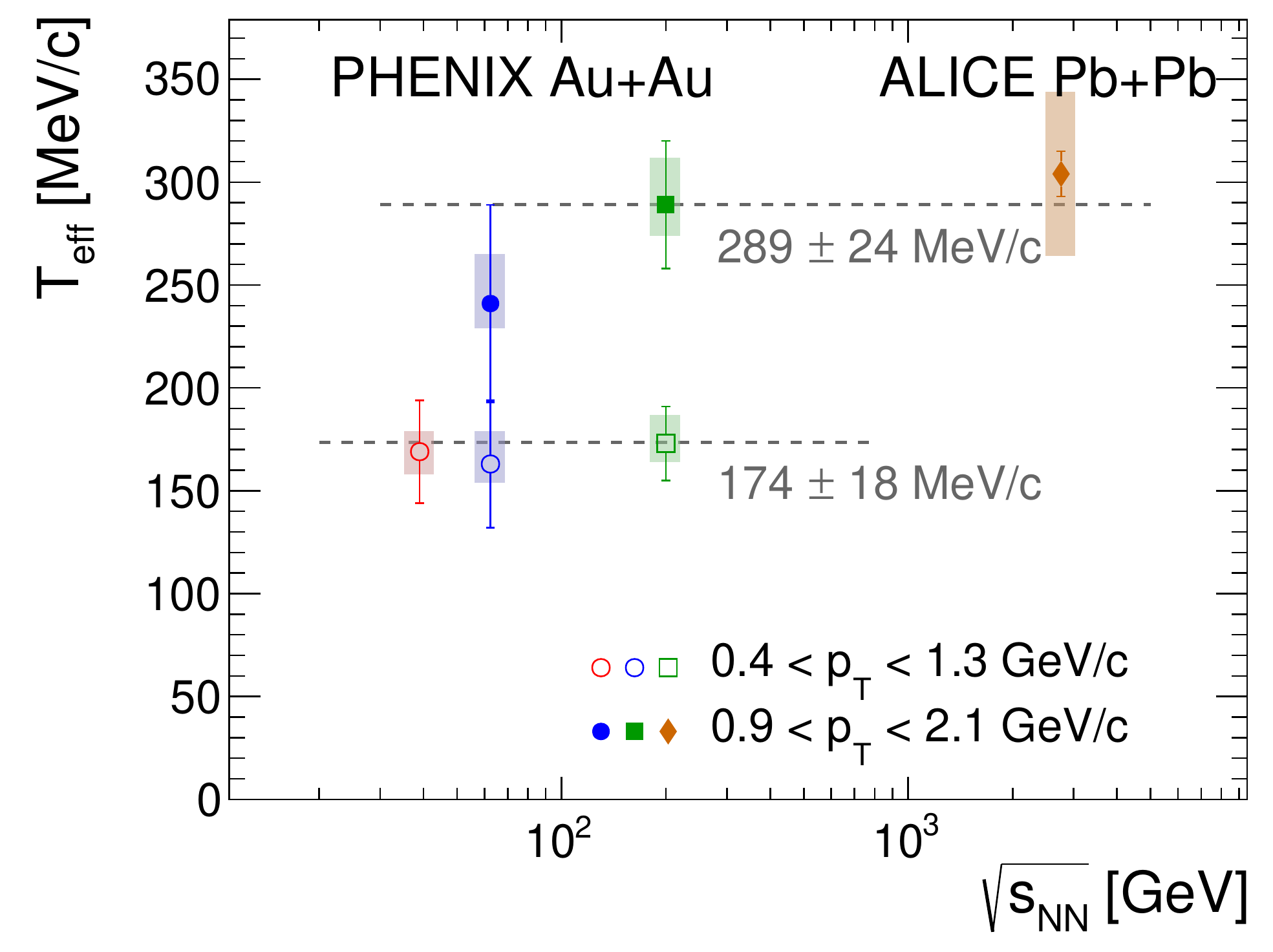}
\caption{Inverse slopes, $T_{\rm eff}$, obtained from fitting the combined 
data from central collisions shown in Fig.~\ref{fig:LowPtSpectra} is 
compared to the fit results of the individual data sets at 62.4, 
200, and 2760 GeV.  Also included is the value for \snn{39} obtained 
from fitting the MB data set in the lower-\pt range. }
\label{fig:Teff_low-pT}
\end{figure} 

\subsection{Direct-photon inverse slope {$T_{\rm eff}$}}
\label{sec3:inverseslope}

To better reveal the similarity of the low-\pt direct-photon spectra 
across \sqsn, the normalized yield from the most-central samples 
(0\%--20\%) for \pbpb at \snn{2760}, \auau at 200 GeV, and \auau at 62.4 
GeV are superimposed on Fig.~\ref{fig:LowPtSpectra}(a). Below 2.5 \gevc, 
the data agree very well, even though they span almost two orders of 
magnitude in \sqsn. As already suggested earlier by exponential fits to 
the 39 and 62.4 GeV data, the low-\pt direct-photon spectra cannot be 
described by a single inverse slope, but seem consistent with an inverse 
slope that increases with \pt. Fitting all data shown in the \pt range 
$\pt<1.3$~\gevc and $0.9<\pt<2.1$~\gevc results in inverse slopes of 
$T_{\rm eff}=0.174{\pm}0.018$~\gevc and $0.289{\pm}0.024$~\gevc, 
respectively. Here the statistical and systematic uncertainties were 
added in quadrature in the fitting procedure. The fits are also shown in 
Fig.~\ref{fig:LowPtSpectra}, where the dashed lines extrapolate the fits 
over the full \pt range.

Figure~\ref{fig:Teff_low-pT} compares the inverse slopes from the common 
fit to the fits of the individual data sets. For \snn{62.4}, the values 
are from Table~\ref{Tab:Teff}, for 200 GeV the 
data~\cite{Adare:2008ab,Adare:2014fwh} were fitted in the two \pt 
ranges, and for 2760 GeV the value published in~Ref.~\cite{Aamodt:2010cz} 
is shown. For the lower-\pt range a value for MB collisions at \snn{39} 
is also included.

Another way to illustrate the commonality of the spectra is to compare 
the ratio of the normalized yield divided by the extrapolated fit for 
$0.9 < \pt < 2.1$ \gevc. The result is shown in  
Fig.~\ref{fig:LowPtSpectra}(b).  Within the uncertainties the ratios are 
consistent with unity over the fit range for all three \sqsn. Below 1 
\gevc, where there is no data from \snn{2760}, the other two energies 
also agree very well.

The similarity of the spectra in the \pt range up to $\approx$2~\gevc 
indicates that the source that emits these photons must be very similar, 
independent of \sqsn, a finding that would be consistent with radiation 
from an expanding and cooling fireball evolving through the transition 
region from QGP to a hadron gas till kinetic freeze-out. This would 
naturally occur at the same temperature and similar expansion velocity, 
independent of the initial conditions created in the collisions.

Above 2 \gevc, the normalized direct-photon yield becomes \sqsn 
dependent. The \snn{200} \auau data remain consistent with the 
exponential fit until $p_T{\approx}3$~\gevc, where prompt-photon production 
from hard-scattering processes starts to dominate (see 
Fig.~\ref{fig:Norm_pTspectra_all}). In contrast, the \pbpb data from 
\snn{2760} begin to exceed the exponential $p_T{\approx}2$~\gevc, while 
prompt-photon production only becomes the main photon source above 4 to 
5~\gevc, where the \Ncoll-scaled pQCD calculation describes the heavy 
ion data well.


This leaves room for additional contributions to the direct-photon spectrum 
in the range from 2 to 5~\gevc beyond prompt-photon production, which 
are \sqsn dependent. Such contributions could reflect the increasing 
initial temperature that would be expected with increasing collision energy.

\subsection{{\pt} dependence of the scaling variable {$\alpha$}}
\label{sec:alphavspt}

\begin{figure*}[t!]
\begin{minipage}{0.48\linewidth}
 \includegraphics[width=0.99\linewidth]{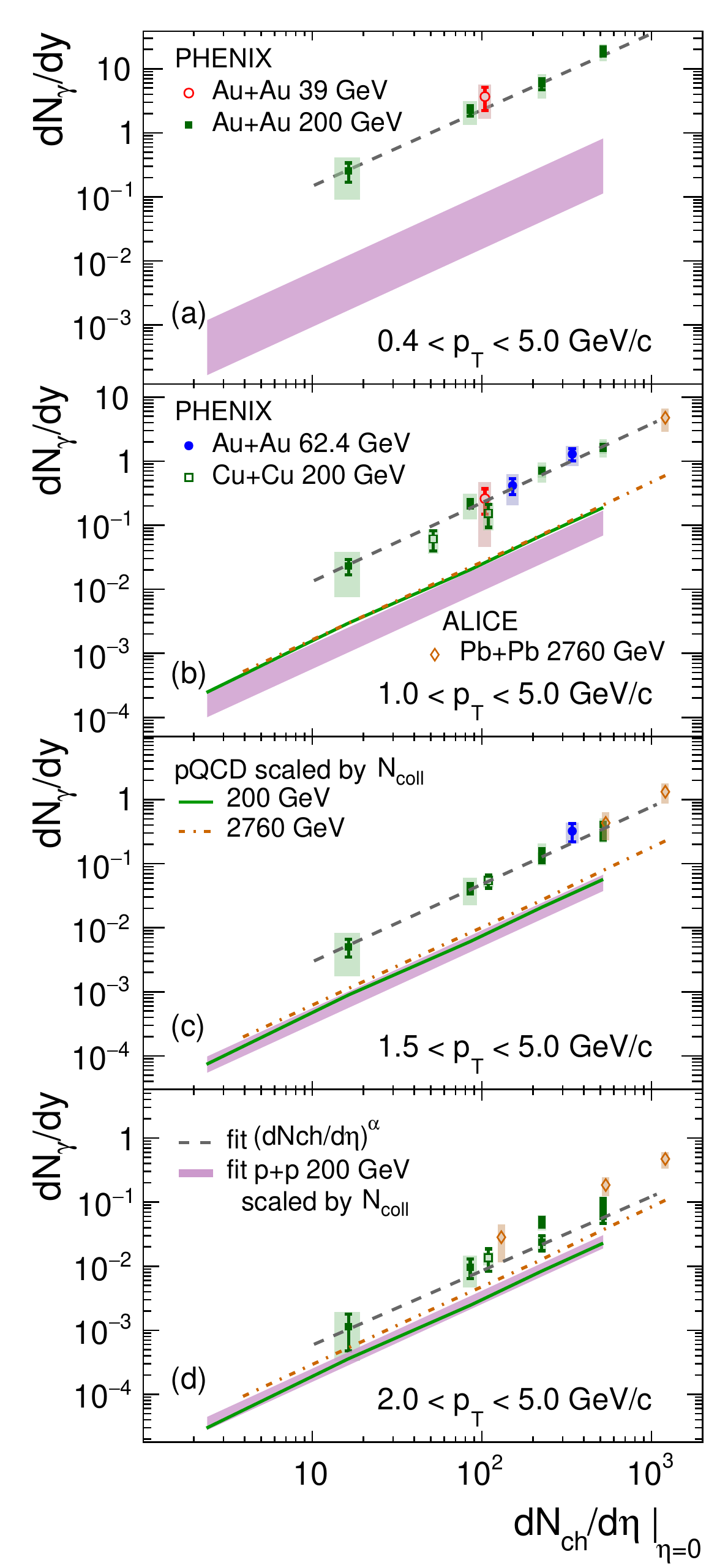}
\caption{Integrated invariant direct-photon yields vs. charged particle 
multiplicity for \pt integrated from (a) 0.4 \gevc , (b) 1.0 \gevc, (c) 
1.5 \gevc, and (d) 2.0 to 5.0 \gevc for all available \AB{A}{A} data 
sets. The band gives the integrated invariant direct-photon yield from 
\pp collisions at \sqs = 200 GeV, scaled by \Ncoll to the corresponding 
\dNch for the \AB{A}{A} data sets. For panels (b) to (d) also the scaled 
and integrated yield from pQCD is given for 200 and 2760 GeV. The dashed 
lines are the result of fitting the PHENIX data with 
$A_{\rm ch}(\dNch)^\alpha$. The fit values for $\alpha$ are consistent with 
a common value of $1.21{\pm}0.04$ independent of \pt. Note that the 
legend for data points, calculations, and fits over panels (a) to (d) 
are valid for all panels.
}
\label{fig:Intlowpt}
\end{minipage}
\hspace{0.2cm}
\begin{minipage}{0.48\linewidth}
\vspace{-1.0cm}
\includegraphics[width=0.99\linewidth]{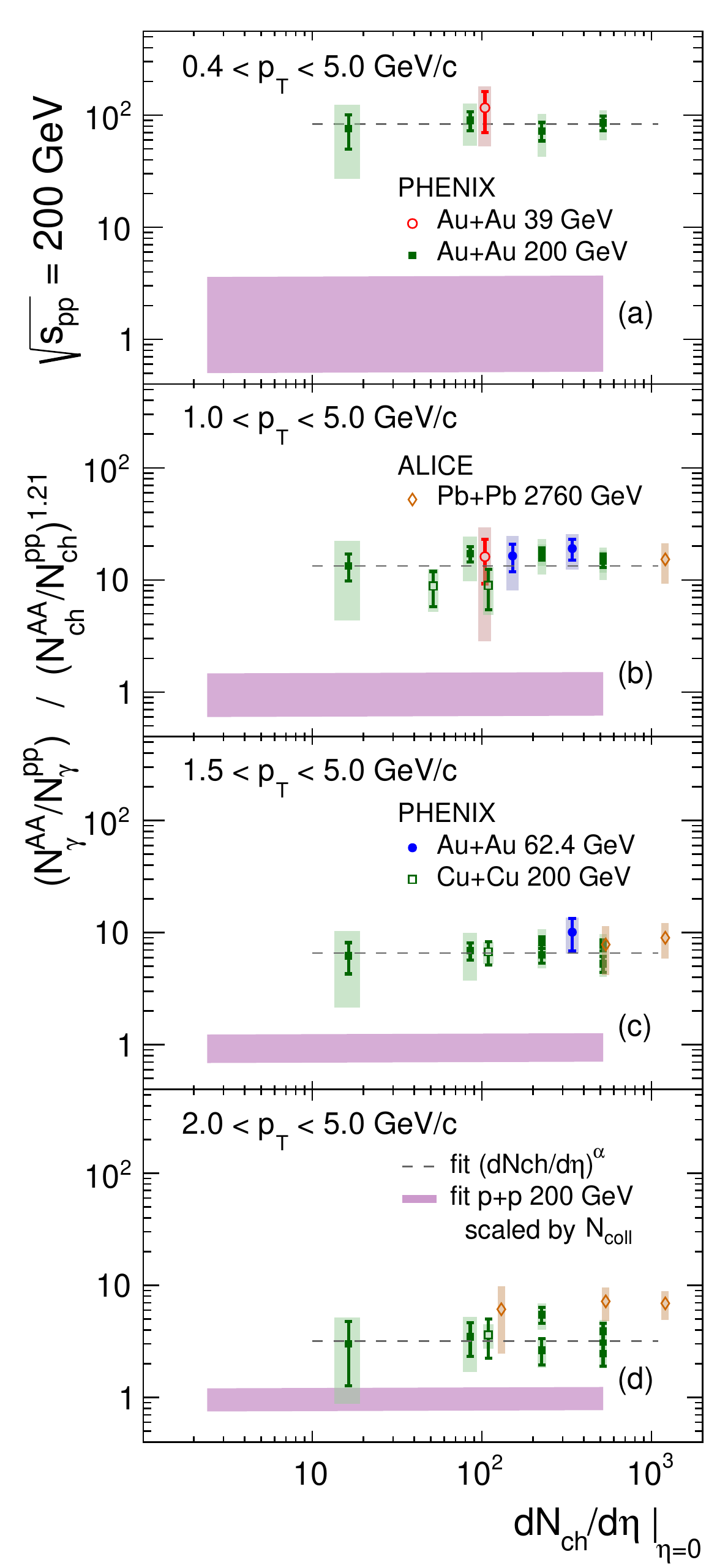}
\caption{The ratio of the integrated direct-photon yields from \AB{A}{A} 
collisions, shown in Fig.~\protect\ref{fig:Intlowpt}, to the integrated 
direct-photon yield from \pp collisions at \snn{200}.  The ratio is 
normalized to the ratio $(\dNch^{AA}/\dNch^{pp})^{\alpha}$, where 
$\alpha=1.21$ is the average value from Tab.~\ref{Tab:alpha}. The four 
panels (a) to (d) show the different integration regions from (a) 0.4, 
(b) 1.0, (c) 1.5, and (d) 2.0, to 5.0~\gevc, respectively. Note that the 
legend for data points, calculations, and fits that are distributed over 
panels (a) to (d) are valid for all panels.}
\label{fig:Ratio_Intlowpt}
\end{minipage}
\end{figure*}

In this final section, the scaling behavior of the direct-photon yield 
with $(\dNch)^\alpha$ will be revisited. So far, a fixed value of 
$\alpha =1.25$ was used to calculate the normalized inclusive 
direct-photon yield. This value was obtained from the scaling relation 
$\Ncoll{\propto}(\dNch)^\alpha$~\cite{Adare:2018wgc}. Here, $\alpha$ 
will be determined from the direct-photon data itself as a function of 
\pt. For this purpose, the direct-photon $p_{T}$ spectra are integrated 
above a minimum \pt value ($p_{T,{\rm min}}$) of 0.4~GeV/$c$, 1.0~GeV/$c$, 
1.5~GeV/$c$, and 2.0~GeV/$c$. Panels (a) to (d) of 
Fig.~\ref{fig:Intlowpt} show the integrated yields as a function of 
\dNch for all data sets shown in Fig.~\ref{fig:Norm_pTspectra_all}. The 
systematic uncertainties, shown as boxes, give the uncertainty on the 
integrated yield and the uncertainty on $\dNch$. The \AB{A}{A} data are 
compared to a band representing the integrated yields obtained from the 
fit to the \pp data at \sqs = 200 GeV, with the functional form given in 
Eq.~\ref{eqn:ppfit}, scaled by \Ncoll. The width of the band is given by 
the uncertainties on the \pp fit and on \Ncoll, combined quadratically. 
Panels (b) to (c) also show the integrated yields from the \Ncoll-scaled 
pQCD calculations for \sqs = 200 and 2760~GeV.


\begin{table}[htb!]
\caption{Fit values obtained from fitting all PHENIX data in panel (a) 
to (d) in Fig.~\protect\ref{fig:Intlowpt} and (a) and (b) in 
Fig.~\protect\ref{fig:IntHighPt} with $A_{\rm ch}(\dNch)^\alpha$. The 
uncertainties on $\alpha$ are quoted separately as statistical and 
systematic uncertainties, with the latter including uncertainties from the 
direct-photon measurements as well as the \dNch. For the normalization, 
$A_{\rm ch}$, total uncertainties are given. }
   \begin{ruledtabular} \begin{tabular}{ccccc}
    $\pt_{\rm min}$  & $\pt_{\rm max}$ & $A_{\rm ch}$ & $\alpha$ & $\chi^{2}/{\rm NDF}$ \\
     \gevc  & \gevc & &   &   \\
\hline
    0.4  & 5.0 & $(1.06\pm0.59)\!\cdot\!10^{-2}$ & $1.19\pm0.09\pm0.18$ & 1.18/3 \\
    1.0  & 5.0 & $(8.16\pm3.46)\!\cdot\!10^{-4}$ & $1.23\pm0.06\pm0.18$ & 5.27/8 \\
    1.5  & 5.0 & $(1.90\pm0.87)\!\cdot\!10^{-4}$ & $1.21\pm0.07\pm0.16$ & 6.50/6 \\ 
    2.0  & 5.0 & $(5.55\pm3.74)\!\cdot\!10^{-5}$ & $1.16\pm0.11\pm0.08$ & 8.85/5 \\
    5.0  & 14.0 & $(5.00\pm1.08)\!\cdot\!10^{-7}$ & $1.21\pm0.02\pm0.07$ & 2.839/7 \\ 
    8.0  & 14.0 & $(7.83\pm1.82)\!\cdot\!10^{-8}$ & $1.17\pm0.02\pm0.06$ & 2.362/7 \\
   \end{tabular} \end{ruledtabular}
\label{Tab:alpha}
\end{table}

It is clear from Fig.~\ref{fig:Norm_pTspectra_all} that all \AB{A}{A} 
data follow a similar common trend. The PHENIX data in each panel of 
Fig.~\ref{fig:Intlowpt} is fitted with the scaling relation:
\begin{equation}
\int_{p_{T, {\rm min}}}^{p_{T,{\rm max}}} \frac{1}{2\pi\pt} \frac{d^2N}{dp_{T}dy}\,d\pt =  A_{\rm ch} \ \left( \frac{dN_{\rm ch}}{d\eta} \right) ^\alpha.
\label{eqn:scaling}
\end{equation}

The fit results for $p_{T,{\rm max}} = 5 \gevc$ are shown as dashed 
lines in Fig.~\ref{fig:Norm_pTspectra_all}; the fit parameters are given 
in Table~\ref{Tab:alpha}. Here the dominant systematic uncertainties are 
due to occupancy dependent differences in the energy scale calibration 
and on \dNch. It is assumed that within a given data set these could be 
anti-correlated and that they are uncorrelated between different data 
sets. The $\alpha$ values are consistent with an average value of 
$\alpha=1.21{\pm}0.04\,({\rm stat})$, with no evident dependence on 
\ptmin. The value is consistent, but slightly lower, than 
$\alpha=1.25\pm0.02$.


Figure~\ref{fig:Ratio_Intlowpt} shows the integrated yield from 
\AB{A}{A} collisions divided by the scaled \pp integrated yield 
normalized by $((\dNch)^{pp}/(\dNch)^{AA})^{1.21}$. In this 
representation, the \pp bands bracket unity with no visible slope. For 
high \pt the vertical scale would be equivalent to the 
nuclear-modification factor of prompt photons. For \ptmin = 0.4, 1.0, 
and 1.5~\gevc all \AB{A}{A} data have the same absolute value, within 
statistical and systematic uncertainties, but are significantly enhanced 
compared to the \pp band. In particular, the \pbpb data at \snn{2760} 
also shows the same value in panels (b) and (c), even though they were 
not included in the fit. The enhancement above \pp drops from nearly two 
orders of magnitude to a factor of $\approx$7 with increasing \ptmin. 
In panel (d) for 
the 2~\gevc threshold the \snn{200} data also have the same 
value, with an enhancement of $\approx$3.  The \pbpb data at \snn{2760}, 
while also being independent of \dNch, have a value roughly 30\% higher 
than the 200 GeV data. This illustrates the breakdown of the scaling 
towards higher \pt, at a \pt for which prompt-photon production is not 
yet expected to be the dominant source. As can be seen from 
Fig.~\ref{fig:Intlowpt}, in this \pt region the \pbpb integrated yield 
exceeds by a factor of 4 to 5 what is calculated by pQCD for prompt-photon 
production. 

\begin{figure}[hbt!]
 \includegraphics[width=1.0\linewidth]{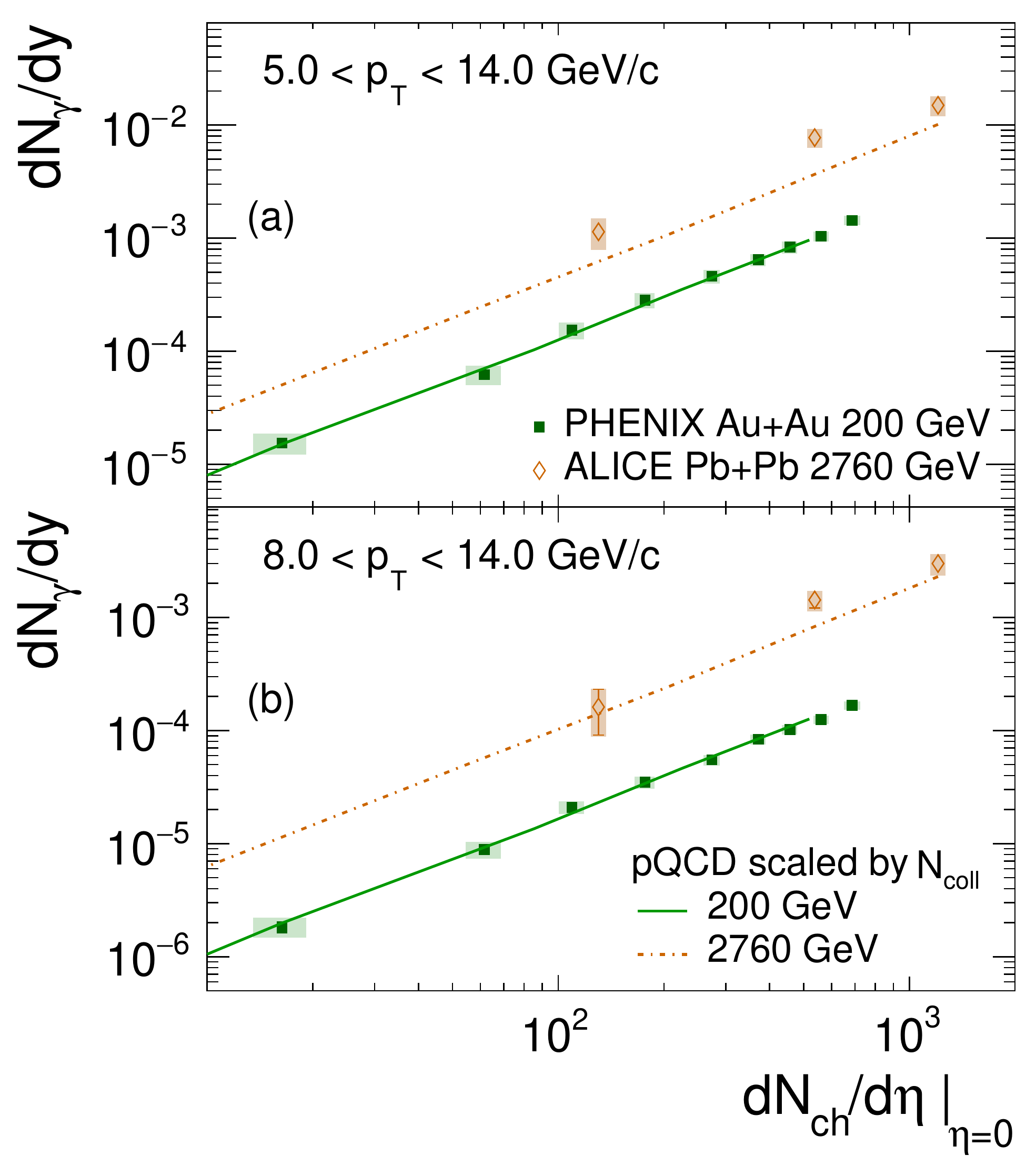}
\caption{Integrated direct-photon yields from \AB{A}{A} collisions for 
\ptmin of 5 \gevc (a) and 8 \gevc. The representation is the same as in 
Fig.~\protect\ref{fig:Intlowpt}. Also shown are the results from pQCD 
calculations scaled by \Ncoll. }
\label{fig:IntHighPt}
\end{figure} 

With increasing \ptmin the integrated yield becomes increasingly 
sensitive to the prompt-photon contribution. Integrated direct-photon 
yields for the ranges $5.0 < \pt < 14$ \gevc and $8.0 < \pt < 14$ \gevc 
are shown in panels (a) and (b) of Fig.~\ref{fig:IntHighPt}, together 
with the corresponding values based on pQCD calculations for the same 
collision energies.  For the integrated yields from \auau at 200~GeV, the 
enhancement compared to \pp has vanished and the measured yield is 
dominated by prompt-photon production, following closely the scaled and 
integrated yield calculated by pQCD. Fitting the data with 
Eq.~\ref{eqn:scaling} results in slope values of $\alpha=1.213 \pm 0.008 
\pm 0.070 $ and $\alpha = 1.172 \pm 0.016 \pm 0.063 $. The full set of 
fit parameters are given in Table.~\ref{Tab:alpha}. Even though the 
direct-photon yield is dominated by prompt-photon production the slope 
values are consistent with those found at lower \ptmin.

The \pbpb data at 2760 GeV continue to be enhanced compared to the pQCD 
calculations even out to \ptmin of 8 \gevc. The enhancement decreases 
with \ptmin and is $\approx$50\% at \ptmin = 5 \gevc and reduces to less 
than 30\% for 8 \gevc.  Given the systematic uncertainties on the data 
and the pQCD calculation these values may already be 
consistent~\cite{Adam:2015lda}. Irrespective of whether in addition to 
prompt-photon production another source is needed to account for the 
data, the \pbpb data can also be well described by a fit with 
Eq.~\ref{eqn:scaling} with $\alpha = 1.12 \pm 0.05$ and $ 1.21 \pm 
0.13$, for $\pt > 5$ \gevc and 8 \gevc, respectively.  These values are 
consistent with values given in Table~\ref{Tab:alpha}, within the quoted 
statistical errors.

\begin{figure}[hbt]
 \includegraphics[width=1.0\linewidth]{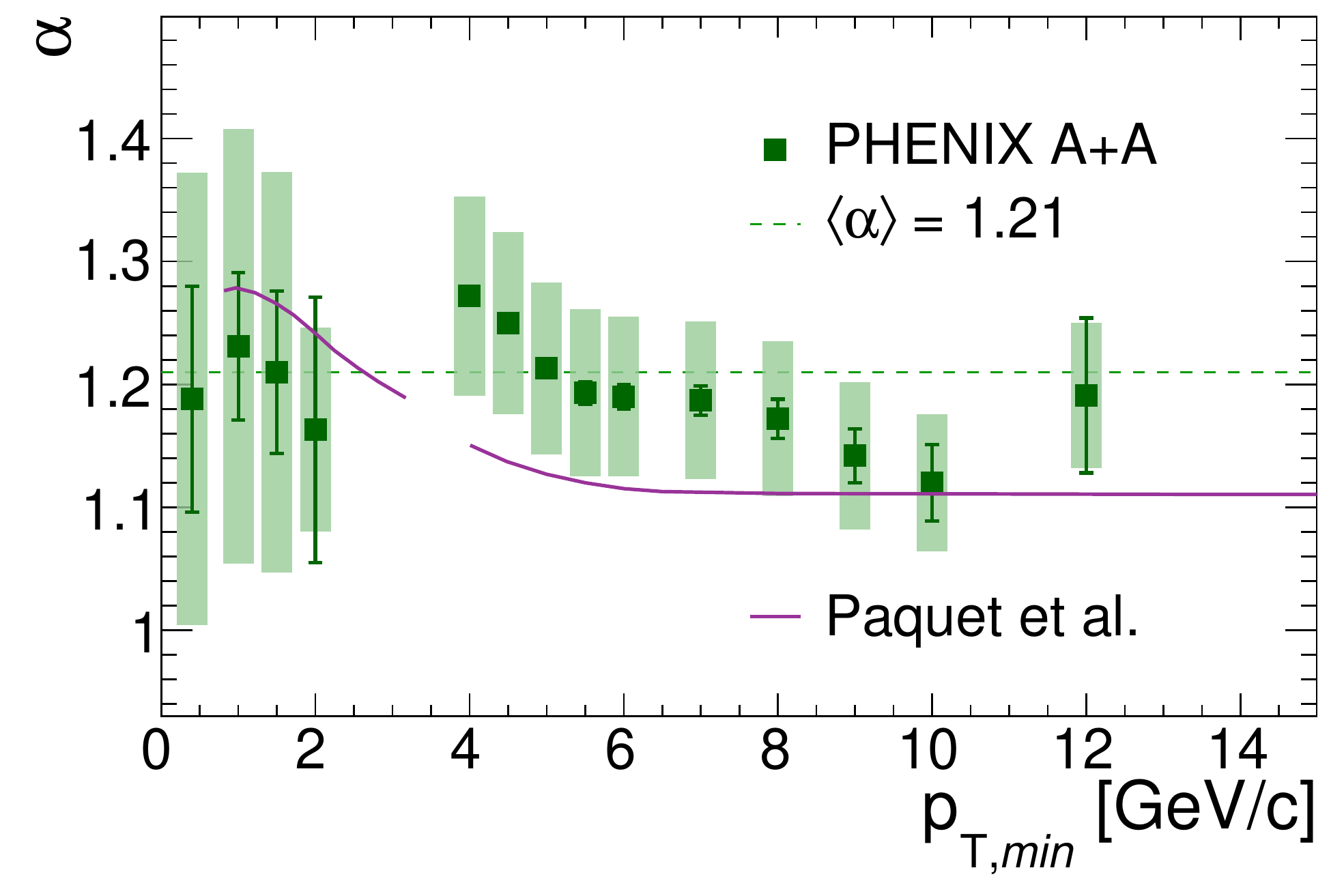}
\caption{The $\alpha$ values extracted using fits to integrated 
direct-photon yields. The dashed line gives the average $\alpha$ value for the 
four lower \ptmin points. Also shown is a model calculation for $\alpha$ 
discussed in the text. }
\label{fig:alpha_vs_ptmin}
\end{figure} 

Figure~\ref{fig:alpha_vs_ptmin} presents the values of $\alpha$ listed 
in Table~\ref{Tab:alpha}, which were obtained from the PHENIX \AB{A}{A} 
data as function of \ptmin.  Also shown in 
Fig.~\ref{fig:alpha_vs_ptmin} are $\alpha$ values from similar fits for 
several other values of $\ptmin>4$ \gevc to integrated direct-photon 
yields from \auau data at \snn{200} published in~\cite{Adare:2014fwh}. 
Within systematic uncertainties, all $\alpha$ values are consistent with 
an average value of 1.21 for the thresholds below 4 \gevc, which is
shown as a dashed line.  

There is no evidence for a dependence of 
$\alpha$ on \ptmin.

Figure~\ref{fig:alpha_vs_ptmin} compares the data to $\alpha$ extracted 
from theoretical model calculations of direct-photon 
radiation~\cite{Paquet:2020,Gale:2020xlg}. 
The model calculation includes prompt-photon production, radiation from 
the pre-equilibrium phase, and thermal photons emitted during the 
evolution from QGP to hadron gas to freeze-out. As discussed in the 
introduction, in general these and similar calculations qualitatively 
reproduce the large direct-photon yield and the large anisotropy with 
respect to the reaction plane observed experimentally, but falls short of 
a simultaneous quantitative description. 
Similarly, the model calculation shown in Fig.~\ref{fig:alpha_vs_ptmin} 
does not fully describe the dependence of $\alpha$ on $p_T$.
In the region where thermal 
radiation is expected to be significant, below \pt = 2 \gevc, the 
calculated $\alpha$ values are consistent with data, but the calculation 
predicts a \pt dependence of $\alpha$ which is not seen in the data.  In 
the model calculation, the thermal-photon contribution from the QGP 
phase depends on \dNch with a higher power of $\alpha \approx 1.8$ than 
the later stage contribution from the hadron gas $\alpha \approx 1.2$. 
The \dNch dependence of the prompt contribution is similar to the one 
from the hadron gas. The dominant sources of direct-photon emission 
change with increasing \pt from hadron gas to QGP to prompt-photon 
production, and therefore $\alpha$ would be expected to depend on \pt. 
While the data do not show such a dependence, the uncertainties, in 
particular systematic uncertainties, are too large to rule out that 
$\alpha$ does change with \pt.

\section{Summary}
\label{sec:Summary}

The PHENIX Collaboration presented the measurement of low \pt 
direct-photon production in MB data samples of \auau collisions at 
39 and 62.4 GeV recorded at RHIC in 2010.  The measurements were 
performed using the PHENIX central arms to detect photon conversions to 
$e^{+}e^{-}$ pairs in the back plane of the HBD, 
following the technique outlined in Ref.~\cite{Adare:2014fwh} for the 
analysis of low-momentum direct photons in \auau collisions at 200~GeV. 
In addition to the MB data samples, the 62.4 \gevc data was 
subdivided into two centrality classes, 0\%--20\% and 20\%--40\%. For all 
samples, the relative direct-photon yield, $R_{\gamma}$, was obtained 
through a double ratio in which many sources of systematic uncertainties 
cancel. In the \pt range from 0.4 to 3 \gevc, a clear direct-photon 
signal is found for all event selections, which significantly exceeds 
the expectations from prompt-photon production.

The direct-photon \pt spectra are not described by one exponential 
function, but are consistent with a local inverse slope increasing with 
\pt.  Comparing the 39 and 62.4 GeV data to direct-photon data from 
Au$+$Au collisions at \snn{200}, also measured by PHENIX, and Pb$+$Pb 
collisions at \snn{2760}, published by ALICE, reveals that the local 
inverse slopes and the shape of the \pt spectra below 2~GeV/$c$ are 
independent of $\sqrt{s_{_{NN}}}$ and centrality of the event sample. 
The combined data for central collisions were fitted with an exponential 
in the \pt range below 1.3~GeV/$c$. The inverse slope value found is 
$T_{\rm eff}=0.174{\pm}0.018$~GeV/$c$. The \pt range from 0.9 to 
2.1~GeV/$c$ was also fitted with an exponential function. The inverse 
slope is significantly larger, with a value of $T_{\rm eff}=0.289 
{\pm}0.024$~GeV/$c$.

Furthermore, the invariant yield of low-\pt direct photons emitted from 
heavy ion collisions shows a common scaling behavior with \dNch that 
takes the form $A_{\rm ch}(\dNch)^\alpha$. Up to \pt of 2 to 2.5 \gevc both 
parameters $A_{\rm ch}$ and $\alpha$ are independent of \sqsn and centrality of 
the event sample. The parameter $A_{\rm ch}$ depends on \pt, but $\alpha$ 
does not. To extend these observations, the \auau data at \snn{200} and 
the \pbpb data at 2760 GeV were analyzed at larger \pt. It was found 
that $A_{\rm ch}$ does depend on \sqsn even in the \pt range from 2 to 
5~\gevc where direct-photon emission is not yet dominated by 
prompt-photon production. However, $\alpha$ remains remarkably 
insensitive to \pt, \sqsn, and centrality.

A possible scenario, consistent with the observations, is that 
direct-photon radiation at low \pt originates from thermal processes while the 
collision system transitions from the QGP phase to a hadron gas. This 
would naturally be at similar temperature and expansion velocity 
independent of \sqsn, collision centrality, and colliding species. In 
the range from 2 to 5~\gevc there might be a contribution from the QGP 
phase earlier in the collision which is more pronounced at higher 
collision energies.  While the data seem qualitatively consistent with 
this conjecture, model calculations suggest that the \dNch dependence of 
the direct-photon yield should vary with \pt, as different photon 
sources are expected to scale differently with \dNch and would 
contribute to different \pt regions. In contrast, within the 
experimental uncertainties, no evidence for such a \pt dependence of 
$\alpha$ was detected.


\begin{acknowledgments}

We thank the staff of the Collider-Accelerator and Physics
Departments at Brookhaven National Laboratory and the staff of
the other PHENIX participating institutions for their vital
contributions.
We also thank J.F. Paquet for many fruitful discussions and 
sharing additional information.
We acknowledge support from the Office of Nuclear Physics in the
Office of Science of the Department of Energy,
the National Science Foundation,
Abilene Christian University Research Council,
Research Foundation of SUNY, and
Dean of the College of Arts and Sciences, Vanderbilt University
(U.S.A),
Ministry of Education, Culture, Sports, Science, and Technology
and the Japan Society for the Promotion of Science (Japan),
Conselho Nacional de Desenvolvimento Cient\'{\i}fico e
Tecnol{\'o}gico and Funda c{\~a}o de Amparo {\`a} Pesquisa do
Estado de S{\~a}o Paulo (Brazil),
Natural Science Foundation of China (People's Republic of China),
Croatian Science Foundation and
Ministry of Science and Education (Croatia),
Ministry of Education, Youth and Sports (Czech Republic),
Centre National de la Recherche Scientifique, Commissariat
{\`a} l'{\'E}nergie Atomique, and Institut National de Physique
Nucl{\'e}aire et de Physique des Particules (France),
J. Bolyai Research Scholarship, EFOP, the New National Excellence
Program ({\'U}NKP), NKFIH, and OTKA (Hungary),
Department of Atomic Energy and Department of Science and Technology
(India),
Israel Science Foundation (Israel),
Basic Science Research and SRC(CENuM) Programs through NRF
funded by the Ministry of Education and the Ministry of
Science and ICT (Korea),
Physics Department, Lahore University of Management Sciences (Pakistan),
Ministry of Education and Science, Russian Academy of Sciences,
Federal Agency of Atomic Energy (Russia),
VR and Wallenberg Foundation (Sweden),
University of Zambia, the Government of the Republic of Zambia (Zambia),
the U.S. Civilian Research and Development Foundation for the
Independent States of the Former Soviet Union,
the Hungarian American Enterprise Scholarship Fund,
the US-Hungarian Fulbright Foundation,
and the US-Israel Binational Science Foundation.

\end{acknowledgments}



\begin{thebibliography}{68}%
\makeatletter
\providecommand \@ifxundefined [1]{%
 \@ifx{#1\undefined}
}%
\providecommand \@ifnum [1]{%
 \ifnum #1\expandafter \@firstoftwo
 \else \expandafter \@secondoftwo
 \fi
}%
\providecommand \@ifx [1]{%
 \ifx #1\expandafter \@firstoftwo
 \else \expandafter \@secondoftwo
 \fi
}%
\providecommand \natexlab [1]{#1}%
\providecommand \enquote  [1]{``#1''}%
\providecommand \bibnamefont  [1]{#1}%
\providecommand \bibfnamefont [1]{#1}%
\providecommand \citenamefont [1]{#1}%
\providecommand \href@noop [0]{\@secondoftwo}%
\providecommand \href [0]{\begingroup \@sanitize@url \@href}%
\providecommand \@href[1]{\@@startlink{#1}\@@href}%
\providecommand \@@href[1]{\endgroup#1\@@endlink}%
\providecommand \@sanitize@url [0]{\catcode `\\12\catcode `\$12\catcode
  `\&12\catcode `\#12\catcode `\^12\catcode `\_12\catcode `\%12\relax}%
\providecommand \@@startlink[1]{}%
\providecommand \@@endlink[0]{}%
\providecommand \url  [0]{\begingroup\@sanitize@url \@url }%
\providecommand \@url [1]{\endgroup\@href {#1}{\urlprefix }}%
\providecommand \urlprefix  [0]{URL }%
\providecommand \Eprint [0]{\href }%
\providecommand \doibase [0]{https://doi.org/}%
\providecommand \selectlanguage [0]{\@gobble}%
\providecommand \bibinfo  [0]{\@secondoftwo}%
\providecommand \bibfield  [0]{\@secondoftwo}%
\providecommand \translation [1]{[#1]}%
\providecommand \BibitemOpen [0]{}%
\providecommand \bibitemStop [0]{}%
\providecommand \bibitemNoStop [0]{.\EOS\space}%
\providecommand \EOS [0]{\spacefactor3000\relax}%
\providecommand \BibitemShut  [1]{\csname bibitem#1\endcsname}%
\let\auto@bib@innerbib\@empty
\bibitem [{\citenamefont {Stankus}(2005)}]{Stankus:2005eq}%
  \BibitemOpen
  \bibfield  {author} {\bibinfo {author} {\bibfnamefont {P.}~\bibnamefont
  {Stankus}},\ }\bibfield  {title} {\bibinfo {title} {{Direct photon production
  in relativistic heavy-ion collisions}},\ }\href {https://doi.org/10.
  1146/annurev.nucl.53.041002.110533} {\bibfield  {journal} {\bibinfo
  {journal} {Ann. Rev. Nucl. Part. Sci.}\ }\textbf {\bibinfo {volume} {55}},\
  \bibinfo {pages} {517} (\bibinfo {year} {2005})}\BibitemShut {NoStop}%
\bibitem [{\citenamefont {David}\ \emph {et~al.}(2008)\citenamefont {David},
  \citenamefont {Rapp},\ and\ \citenamefont {Xu}}]{David:2006sr}%
  \BibitemOpen
  \bibfield  {author} {\bibinfo {author} {\bibfnamefont {G.}~\bibnamefont
  {David}}, \bibinfo {author} {\bibfnamefont {R.}~\bibnamefont {Rapp}},\ and\
  \bibinfo {author} {\bibfnamefont {Z.}~\bibnamefont {Xu}},\ }\bibfield
  {title} {\bibinfo {title} {{Electromagnetic Probes at RHIC-II}},\ }\href
  {https://doi.org/10. 1016/j.physrep.2008.04.003} {\bibfield  {journal}
  {\bibinfo  {journal} {Phys. Rept.}\ }\textbf {\bibinfo {volume} {462}},\
  \bibinfo {pages} {176} (\bibinfo {year} {2008})}\BibitemShut {NoStop}%
\bibitem [{\citenamefont {Linnyk}\ \emph {et~al.}(2016)\citenamefont {Linnyk},
  \citenamefont {Bratkovskaya},\ and\ \citenamefont
  {Cassing}}]{Linnyk:2015rco}%
  \BibitemOpen
  \bibfield  {author} {\bibinfo {author} {\bibfnamefont {O.}~\bibnamefont
  {Linnyk}}, \bibinfo {author} {\bibfnamefont {E.~L.}\ \bibnamefont
  {Bratkovskaya}},\ and\ \bibinfo {author} {\bibfnamefont {W.}~\bibnamefont
  {Cassing}},\ }\bibfield  {title} {\bibinfo {title} {{Effective QCD and
  transport description of dilepton and photon production in heavy-ion
  collisions and elementary processes}},\ }\href {https://doi.org/10.
  1016/j.ppnp.2015.12.003} {\bibfield  {journal} {\bibinfo  {journal} {Prog.
  Part. Nucl. Phys.}\ }\textbf {\bibinfo {volume} {87}},\ \bibinfo {pages} {50}
  (\bibinfo {year} {2016})}\BibitemShut {NoStop}%
\bibitem [{\citenamefont {David}(2020)}]{David:2019wpt}%
  \BibitemOpen
  \bibfield  {author} {\bibinfo {author} {\bibfnamefont {G.}~\bibnamefont
  {David}},\ }\bibfield  {title} {\bibinfo {title} {{Direct real photons in
  relativistic heavy ion collisions}},\ }\href {https://doi.org/10.
  1088/1361-6633/ab6f57} {\bibfield  {journal} {\bibinfo  {journal} {Rept.
  Prog. Phys.}\ }\textbf {\bibinfo {volume} {83}},\ \bibinfo {pages} {046301}
  (\bibinfo {year} {2020})}\BibitemShut {NoStop}%
\bibitem [{\citenamefont {van Hees}\ \emph {et~al.}(2011)\citenamefont {van
  Hees}, \citenamefont {Gale},\ and\ \citenamefont {Rapp}}]{vanHees:2011vb}%
  \BibitemOpen
  \bibfield  {author} {\bibinfo {author} {\bibfnamefont {H.}~\bibnamefont {van
  Hees}}, \bibinfo {author} {\bibfnamefont {C.}~\bibnamefont {Gale}},\ and\
  \bibinfo {author} {\bibfnamefont {R.}~\bibnamefont {Rapp}},\ }\bibfield
  {title} {\bibinfo {title} {{Thermal Photons and Collective Flow at the
  Relativistic Heavy-Ion Collider}},\ }\href
  {https://doi.org/10.1103/PhysRevC.84.054906} {\bibfield  {journal} {\bibinfo
  {journal} {Phys. Rev. C}\ }\textbf {\bibinfo {volume} {84}},\ \bibinfo
  {pages} {054906} (\bibinfo {year} {2011})}\BibitemShut {NoStop}%
\bibitem [{\citenamefont {van Hees}\ \emph {et~al.}(2015)\citenamefont {van
  Hees}, \citenamefont {He},\ and\ \citenamefont {Rapp}}]{vanHees:2014ida}%
  \BibitemOpen
  \bibfield  {author} {\bibinfo {author} {\bibfnamefont {H.}~\bibnamefont {van
  Hees}}, \bibinfo {author} {\bibfnamefont {M.}~\bibnamefont {He}},\ and\
  \bibinfo {author} {\bibfnamefont {R.}~\bibnamefont {Rapp}},\ }\bibfield
  {title} {\bibinfo {title} {{Pseudo-critical enhancement of thermal photons in
  relativistic heavy-ion collisions?}},\ }\href
  {https://doi.org/10.1016/j.nuclphysa.2014.09.009} {\bibfield  {journal}
  {\bibinfo  {journal} {Nucl. Phys. A}\ }\textbf {\bibinfo {volume} {933}},\
  \bibinfo {pages} {256} (\bibinfo {year} {2015})}\BibitemShut {NoStop}%
\bibitem [{\citenamefont {Dion}\ \emph {et~al.}(2011)\citenamefont {Dion},
  \citenamefont {Paquet}, \citenamefont {Schenke}, \citenamefont {Young},
  \citenamefont {Jeon},\ and\ \citenamefont {Gale}}]{Dion:2011pp}%
  \BibitemOpen
  \bibfield  {author} {\bibinfo {author} {\bibfnamefont {M.}~\bibnamefont
  {Dion}}, \bibinfo {author} {\bibfnamefont {J.-F.}\ \bibnamefont {Paquet}},
  \bibinfo {author} {\bibfnamefont {B.}~\bibnamefont {Schenke}}, \bibinfo
  {author} {\bibfnamefont {C.}~\bibnamefont {Young}}, \bibinfo {author}
  {\bibfnamefont {S.}~\bibnamefont {Jeon}},\ and\ \bibinfo {author}
  {\bibfnamefont {C.}~\bibnamefont {Gale}},\ }\bibfield  {title} {\bibinfo
  {title} {{Viscous photons in relativistic heavy ion collisions}},\ }\href
  {https://doi.org/10.1103/PhysRevC.84.064901} {\bibfield  {journal} {\bibinfo
  {journal} {Phys. Rev. C}\ }\textbf {\bibinfo {volume} {84}},\ \bibinfo
  {pages} {064901} (\bibinfo {year} {2011})}\BibitemShut {NoStop}%
\bibitem [{\citenamefont {Shen}\ \emph {et~al.}(2014)\citenamefont {Shen},
  \citenamefont {Heinz}, \citenamefont {Paquet},\ and\ \citenamefont
  {Gale}}]{Shen:2013vja}%
  \BibitemOpen
  \bibfield  {author} {\bibinfo {author} {\bibfnamefont {C.}~\bibnamefont
  {Shen}}, \bibinfo {author} {\bibfnamefont {U.~W.}\ \bibnamefont {Heinz}},
  \bibinfo {author} {\bibfnamefont {J.-F.}\ \bibnamefont {Paquet}},\ and\
  \bibinfo {author} {\bibfnamefont {C.}~\bibnamefont {Gale}},\ }\bibfield
  {title} {\bibinfo {title} {{Thermal photons as a quark-gluon plasma
  thermometer reexamined}},\ }\href
  {https://doi.org/10.1103/PhysRevC.89.044910} {\bibfield  {journal} {\bibinfo
  {journal} {Phys. Rev. C}\ }\textbf {\bibinfo {volume} {89}},\ \bibinfo
  {pages} {044910} (\bibinfo {year} {2014})}\BibitemShut {NoStop}%
\bibitem [{\citenamefont {Shen}\ \emph {et~al.}(2016)\citenamefont {Shen},
  \citenamefont {Paquet}, \citenamefont {Denicol}, \citenamefont {Jeon},\ and\
  \citenamefont {Gale}}]{Shen:2015qba}%
  \BibitemOpen
  \bibfield  {author} {\bibinfo {author} {\bibfnamefont {C.}~\bibnamefont
  {Shen}}, \bibinfo {author} {\bibfnamefont {J.~F.}\ \bibnamefont {Paquet}},
  \bibinfo {author} {\bibfnamefont {G.~S.}\ \bibnamefont {Denicol}}, \bibinfo
  {author} {\bibfnamefont {S.}~\bibnamefont {Jeon}},\ and\ \bibinfo {author}
  {\bibfnamefont {C.}~\bibnamefont {Gale}},\ }\bibfield  {title} {\bibinfo
  {title} {{Thermal photon radiation in high multiplicity $p$$+$Pb collisions
  at the Large Hadron Collider}},\ }\href {https://doi.org/10.
  1103/PhysRevLett.116.072301} {\bibfield  {journal} {\bibinfo  {journal}
  {Phys. Rev. Lett.}\ }\textbf {\bibinfo {volume} {116}},\ \bibinfo {pages}
  {072301} (\bibinfo {year} {2016})}\BibitemShut {NoStop}%
\bibitem [{\citenamefont {Paquet}\ \emph {et~al.}(2016)\citenamefont {Paquet},
  \citenamefont {Shen}, \citenamefont {Denicol}, \citenamefont {Luzum},
  \citenamefont {Schenke}, \citenamefont {Jeon},\ and\ \citenamefont
  {Gale}}]{Paquet:2015lta}%
  \BibitemOpen
  \bibfield  {author} {\bibinfo {author} {\bibfnamefont {J.-F.}\ \bibnamefont
  {Paquet}}, \bibinfo {author} {\bibfnamefont {C.}~\bibnamefont {Shen}},
  \bibinfo {author} {\bibfnamefont {G.~S.}\ \bibnamefont {Denicol}}, \bibinfo
  {author} {\bibfnamefont {M.}~\bibnamefont {Luzum}}, \bibinfo {author}
  {\bibfnamefont {B.}~\bibnamefont {Schenke}}, \bibinfo {author} {\bibfnamefont
  {S.}~\bibnamefont {Jeon}},\ and\ \bibinfo {author} {\bibfnamefont
  {C.}~\bibnamefont {Gale}},\ }\bibfield  {title} {\bibinfo {title}
  {{Production of photons in relativistic heavy-ion collisions}},\ }\href
  {https://doi.org/10.1103/PhysRevC.93.044906} {\bibfield  {journal} {\bibinfo
  {journal} {Phys. Rev. C}\ }\textbf {\bibinfo {volume} {93}},\ \bibinfo
  {pages} {044906} (\bibinfo {year} {2016})}\BibitemShut {NoStop}%
\bibitem [{\citenamefont {Bratkovskaya}\ \emph {et~al.}(2008)\citenamefont
  {Bratkovskaya}, \citenamefont {Kiselev},\ and\ \citenamefont
  {Sharkov}}]{Bratkovskaya:2008iq}%
  \BibitemOpen
  \bibfield  {author} {\bibinfo {author} {\bibfnamefont {E.~L.}\ \bibnamefont
  {Bratkovskaya}}, \bibinfo {author} {\bibfnamefont {S.~M.}\ \bibnamefont
  {Kiselev}},\ and\ \bibinfo {author} {\bibfnamefont {G.~B.}\ \bibnamefont
  {Sharkov}},\ }\bibfield  {title} {\bibinfo {title} {{Direct photon production
  from hadronic sources in high-energy heavy-ion collisions}},\ }\href
  {https://doi.org/10.1103/PhysRevC.78.034905} {\bibfield  {journal} {\bibinfo
  {journal} {Phys. Rev. C}\ }\textbf {\bibinfo {volume} {78}},\ \bibinfo
  {pages} {034905} (\bibinfo {year} {2008})}\BibitemShut {NoStop}%
\bibitem [{\citenamefont {Bratkovskaya}(2014)}]{Bratkovskaya:2014mva}%
  \BibitemOpen
  \bibfield  {author} {\bibinfo {author} {\bibfnamefont {E.~L.}\ \bibnamefont
  {Bratkovskaya}},\ }\bibfield  {title} {\bibinfo {title} {{Phenomenology of
  photon and dilepton production in relativistic nuclear collisions}},\ }\href
  {https://doi.org/10.1016/j.nuclphysa.2014.09.088} {\bibfield  {journal}
  {\bibinfo  {journal} {Nucl. Phys. A}\ }\textbf {\bibinfo {volume} {931}},\
  \bibinfo {pages} {194} (\bibinfo {year} {2014})}\BibitemShut {NoStop}%
\bibitem [{\citenamefont {Linnyk}\ \emph {et~al.}(2014)\citenamefont {Linnyk},
  \citenamefont {Cassing},\ and\ \citenamefont
  {Bratkovskaya}}]{Linnyk:2013wma}%
  \BibitemOpen
  \bibfield  {author} {\bibinfo {author} {\bibfnamefont {O.}~\bibnamefont
  {Linnyk}}, \bibinfo {author} {\bibfnamefont {W.}~\bibnamefont {Cassing}},\
  and\ \bibinfo {author} {\bibfnamefont {E.~L.}\ \bibnamefont {Bratkovskaya}},\
  }\bibfield  {title} {\bibinfo {title} {{Centrality dependence of the direct
  photon yield and elliptic flow in heavy-ion collisions at $\sqrt{s_{NN}}$=200
  GeV}},\ }\href {https://doi.org/10. 1103/PhysRevC.89.034908} {\bibfield
  {journal} {\bibinfo  {journal} {Phys. Rev. C}\ }\textbf {\bibinfo {volume}
  {89}},\ \bibinfo {pages} {034908} (\bibinfo {year} {2014})}\BibitemShut
  {NoStop}%
\bibitem [{\citenamefont {Chiu}\ \emph {et~al.}(2013)\citenamefont {Chiu},
  \citenamefont {Hemmick}, \citenamefont {Khachatryan}, \citenamefont
  {Leonidov}, \citenamefont {Liao},\ and\ \citenamefont
  {McLerran}}]{Chiu:2012ij}%
  \BibitemOpen
  \bibfield  {author} {\bibinfo {author} {\bibfnamefont {M.}~\bibnamefont
  {Chiu}}, \bibinfo {author} {\bibfnamefont {T.~K.}\ \bibnamefont {Hemmick}},
  \bibinfo {author} {\bibfnamefont {V.}~\bibnamefont {Khachatryan}}, \bibinfo
  {author} {\bibfnamefont {A.}~\bibnamefont {Leonidov}}, \bibinfo {author}
  {\bibfnamefont {J.}~\bibnamefont {Liao}},\ and\ \bibinfo {author}
  {\bibfnamefont {L.}~\bibnamefont {McLerran}},\ }\bibfield  {title} {\bibinfo
  {title} {{Production of Photons and Dileptons in the Glasma}},\ }\href
  {https://doi.org/10. 1016/j.nuclphysa.2013.01.014} {\bibfield  {journal}
  {\bibinfo  {journal} {Nucl. Phys. A}\ }\textbf {\bibinfo {volume} {900}},\
  \bibinfo {pages} {16} (\bibinfo {year} {2013})}\BibitemShut {NoStop}%
\bibitem [{\citenamefont {McLerran}\ and\ \citenamefont
  {Schenke}(2014)}]{McLerran:2014hza}%
  \BibitemOpen
  \bibfield  {author} {\bibinfo {author} {\bibfnamefont {L.}~\bibnamefont
  {McLerran}}\ and\ \bibinfo {author} {\bibfnamefont {B.}~\bibnamefont
  {Schenke}},\ }\bibfield  {title} {\bibinfo {title} {{The Glasma, Photons and
  the Implications of Anisotropy}},\ }\href {https://doi.org/10.
  1016/j.nuclphysa.2014.06.004} {\bibfield  {journal} {\bibinfo  {journal}
  {Nucl. Phys. A}\ }\textbf {\bibinfo {volume} {929}},\ \bibinfo {pages} {71}
  (\bibinfo {year} {2014})}\BibitemShut {NoStop}%
\bibitem [{\citenamefont {McLerran}\ and\ \citenamefont
  {Schenke}(2016)}]{McLerran:2015mda}%
  \BibitemOpen
  \bibfield  {author} {\bibinfo {author} {\bibfnamefont {L.}~\bibnamefont
  {McLerran}}\ and\ \bibinfo {author} {\bibfnamefont {B.}~\bibnamefont
  {Schenke}},\ }\bibfield  {title} {\bibinfo {title} {{A Tale of Tails: Photon
  Rates and Flow in Ultra-Relativistic Heavy Ion Collisions}},\ }\href
  {https://doi.org/10. 1016/j.nuclphysa.2015.11.008} {\bibfield  {journal}
  {\bibinfo  {journal} {Nucl. Phys. A}\ }\textbf {\bibinfo {volume} {946}},\
  \bibinfo {pages} {158} (\bibinfo {year} {2016})}\BibitemShut {NoStop}%
\bibitem [{\citenamefont {Klein-B{\"o}sing}\ and\ \citenamefont
  {McLerran}(2014)}]{Klein-Bosing:2014uaa}%
  \BibitemOpen
  \bibfield  {author} {\bibinfo {author} {\bibfnamefont {C.}~\bibnamefont
  {Klein-B{\"o}sing}}\ and\ \bibinfo {author} {\bibfnamefont {L.}~\bibnamefont
  {McLerran}},\ }\bibfield  {title} {\bibinfo {title} {{Geometrical Scaling of
  Direct-Photon Production in Hadron Collisions from RHIC to the LHC}},\ }\href
  {https://doi.org/10.1016/j.physletb.2014.05.063} {\bibfield  {journal}
  {\bibinfo  {journal} {Phys. Lett. B}\ }\textbf {\bibinfo {volume} {734}},\
  \bibinfo {pages} {282} (\bibinfo {year} {2014})}\BibitemShut {NoStop}%
\bibitem [{\citenamefont {Berges}\ \emph {et~al.}(2017)\citenamefont {Berges},
  \citenamefont {Reygers}, \citenamefont {Tanji},\ and\ \citenamefont
  {Venugopalan}}]{Berges:2017eom}%
  \BibitemOpen
  \bibfield  {author} {\bibinfo {author} {\bibfnamefont {J.}~\bibnamefont
  {Berges}}, \bibinfo {author} {\bibfnamefont {K.}~\bibnamefont {Reygers}},
  \bibinfo {author} {\bibfnamefont {N.}~\bibnamefont {Tanji}},\ and\ \bibinfo
  {author} {\bibfnamefont {R.}~\bibnamefont {Venugopalan}},\ }\bibfield
  {title} {\bibinfo {title} {{Parametric estimate of the relative photon yields
  from the glasma and the quark-gluon plasma in heavy-ion collisions}},\ }\href
  {https://doi.org/10.1103/PhysRevC.95.054904} {\bibfield  {journal} {\bibinfo
  {journal} {Phys. Rev. C}\ }\textbf {\bibinfo {volume} {95}},\ \bibinfo
  {pages} {054904} (\bibinfo {year} {2017})}\BibitemShut {NoStop}%
\bibitem [{\citenamefont {Khachatryan}\ \emph {et~al.}(2018)\citenamefont
  {Khachatryan}, \citenamefont {Schenke}, \citenamefont {Chiu}, \citenamefont
  {Drees}, \citenamefont {Hemmick},\ and\ \citenamefont
  {Novitzky}}]{Khachatryan:2018ori}%
  \BibitemOpen
  \bibfield  {author} {\bibinfo {author} {\bibfnamefont {V.}~\bibnamefont
  {Khachatryan}}, \bibinfo {author} {\bibfnamefont {B.}~\bibnamefont
  {Schenke}}, \bibinfo {author} {\bibfnamefont {M.}~\bibnamefont {Chiu}},
  \bibinfo {author} {\bibfnamefont {A.}~\bibnamefont {Drees}}, \bibinfo
  {author} {\bibfnamefont {T.~K.}\ \bibnamefont {Hemmick}},\ and\ \bibinfo
  {author} {\bibfnamefont {N.}~\bibnamefont {Novitzky}},\ }\bibfield  {title}
  {\bibinfo {title} {{Photons from thermalizing matter in heavy ion
  collisions}},\ }\href {https://doi.org/10.1016/j.nuclphysa.2018.07.013}
  {\bibfield  {journal} {\bibinfo  {journal} {Nucl. Phys. A}\ }\textbf
  {\bibinfo {volume} {978}},\ \bibinfo {pages} {123} (\bibinfo {year}
  {2018})}\BibitemShut {NoStop}%
\bibitem [{\citenamefont {Monnai}(2014)}]{monnai:2014kqa}%
  \BibitemOpen
  \bibfield  {author} {\bibinfo {author} {\bibfnamefont {A.}~\bibnamefont
  {Monnai}},\ }\bibfield  {title} {\bibinfo {title} {{Thermal photon $v_2$ with
  slow quark chemical equilibration}},\ }\href
  {https://doi.org/10.1103/PhysRevC.90.021901} {\bibfield  {journal} {\bibinfo
  {journal} {Phys. Rev. C}\ }\textbf {\bibinfo {volume} {90}},\ \bibinfo
  {pages} {021901(R)} (\bibinfo {year} {2014})}\BibitemShut {NoStop}%
\bibitem [{\citenamefont {Lee}\ and\ \citenamefont
  {Zahed}(2014)}]{Lee:2014pwa}%
  \BibitemOpen
  \bibfield  {author} {\bibinfo {author} {\bibfnamefont {C.-H.}\ \bibnamefont
  {Lee}}\ and\ \bibinfo {author} {\bibfnamefont {I.}~\bibnamefont {Zahed}},\
  }\bibfield  {title} {\bibinfo {title} {{Electromagnetic Radiation in Hot QCD
  Matter: Rates, Electric Conductivity, Flavor Susceptibility and Diffusion}},\
  }\href {https://doi.org/10.1103/PhysRevC.90.025204} {\bibfield  {journal}
  {\bibinfo  {journal} {Phys. Rev. C}\ }\textbf {\bibinfo {volume} {90}},\
  \bibinfo {pages} {025204} (\bibinfo {year} {2014})}\BibitemShut {NoStop}%
\bibitem [{\citenamefont {Turbide}\ \emph {et~al.}(2004)\citenamefont
  {Turbide}, \citenamefont {Rapp},\ and\ \citenamefont
  {Gale}}]{Turbide:2003si}%
  \BibitemOpen
  \bibfield  {author} {\bibinfo {author} {\bibfnamefont {S.}~\bibnamefont
  {Turbide}}, \bibinfo {author} {\bibfnamefont {R.}~\bibnamefont {Rapp}},\ and\
  \bibinfo {author} {\bibfnamefont {C.}~\bibnamefont {Gale}},\ }\bibfield
  {title} {\bibinfo {title} {{Hadronic production of thermal photons}},\ }\href
  {https://doi.org/10.1103/PhysRevC.69.014903} {\bibfield  {journal} {\bibinfo
  {journal} {Phys. Rev. C}\ }\textbf {\bibinfo {volume} {69}},\ \bibinfo
  {pages} {014903} (\bibinfo {year} {2004})}\BibitemShut {NoStop}%
\bibitem [{\citenamefont {Dusling}\ and\ \citenamefont
  {Zahed}(2010)}]{Dusling:2009ej}%
  \BibitemOpen
  \bibfield  {author} {\bibinfo {author} {\bibfnamefont {K.}~\bibnamefont
  {Dusling}}\ and\ \bibinfo {author} {\bibfnamefont {I.}~\bibnamefont
  {Zahed}},\ }\bibfield  {title} {\bibinfo {title} {{Thermal photons from heavy
  ion collisions: A spectral function approach}},\ }\href
  {https://doi.org/10.1103/PhysRevC.82.054909} {\bibfield  {journal} {\bibinfo
  {journal} {Phys. Rev. C}\ }\textbf {\bibinfo {volume} {82}},\ \bibinfo
  {pages} {054909} (\bibinfo {year} {2010})}\BibitemShut {NoStop}%
\bibitem [{\citenamefont {Kim}\ \emph {et~al.}(2017)\citenamefont {Kim},
  \citenamefont {Lee}, \citenamefont {Teaney},\ and\ \citenamefont
  {Zahed}}]{Kim:2016ylr}%
  \BibitemOpen
  \bibfield  {author} {\bibinfo {author} {\bibfnamefont {Y.-M.}\ \bibnamefont
  {Kim}}, \bibinfo {author} {\bibfnamefont {C.-H.}\ \bibnamefont {Lee}},
  \bibinfo {author} {\bibfnamefont {D.}~\bibnamefont {Teaney}},\ and\ \bibinfo
  {author} {\bibfnamefont {I.}~\bibnamefont {Zahed}},\ }\bibfield  {title}
  {\bibinfo {title} {{Direct photon elliptic flow at energies available at the
  BNL Relativistic Heavy Ion Collider and the CERN Large Hadron Collider}},\
  }\href {https://doi.org/10.1103/PhysRevC.96.015201} {\bibfield  {journal}
  {\bibinfo  {journal} {Phys. Rev. C}\ }\textbf {\bibinfo {volume} {96}},\
  \bibinfo {pages} {015201} (\bibinfo {year} {2017})}\BibitemShut {NoStop}%
\bibitem [{\citenamefont {Heffernan}\ \emph {et~al.}(2015)\citenamefont
  {Heffernan}, \citenamefont {Hohler},\ and\ \citenamefont
  {Rapp}}]{Heffernan:2014mla}%
  \BibitemOpen
  \bibfield  {author} {\bibinfo {author} {\bibfnamefont {M.}~\bibnamefont
  {Heffernan}}, \bibinfo {author} {\bibfnamefont {P.}~\bibnamefont {Hohler}},\
  and\ \bibinfo {author} {\bibfnamefont {R.}~\bibnamefont {Rapp}},\ }\bibfield
  {title} {\bibinfo {title} {{Universal Parametrization of Thermal Photon Rates
  in Hadronic Matter}},\ }\href {https://doi.org/10.1103/PhysRevC.91.027902}
  {\bibfield  {journal} {\bibinfo  {journal} {Phys. Rev. C}\ }\textbf {\bibinfo
  {volume} {91}},\ \bibinfo {pages} {027902} (\bibinfo {year}
  {2015})}\BibitemShut {NoStop}%
\bibitem [{\citenamefont {Linnyk}\ \emph {et~al.}(2015)\citenamefont {Linnyk},
  \citenamefont {Konchakovski}, \citenamefont {Steinert}, \citenamefont
  {Cassing},\ and\ \citenamefont {Bratkovskaya}}]{Linnyk:2015tha}%
  \BibitemOpen
  \bibfield  {author} {\bibinfo {author} {\bibfnamefont {O.}~\bibnamefont
  {Linnyk}}, \bibinfo {author} {\bibfnamefont {V.}~\bibnamefont
  {Konchakovski}}, \bibinfo {author} {\bibfnamefont {T.}~\bibnamefont
  {Steinert}}, \bibinfo {author} {\bibfnamefont {W.}~\bibnamefont {Cassing}},\
  and\ \bibinfo {author} {\bibfnamefont {E.~L.}\ \bibnamefont {Bratkovskaya}},\
  }\bibfield  {title} {\bibinfo {title} {{Hadronic and partonic sources of
  direct photons in relativistic heavy-ion collisions}},\ }\href
  {https://doi.org/10.1103/PhysRevC.92.054914} {\bibfield  {journal} {\bibinfo
  {journal} {Phys. Rev. C}\ }\textbf {\bibinfo {volume} {92}},\ \bibinfo
  {pages} {054914} (\bibinfo {year} {2015})}\BibitemShut {NoStop}%
\bibitem [{\citenamefont {Basar}\ \emph {et~al.}(2012)\citenamefont {Basar},
  \citenamefont {Kharzeev},\ and\ \citenamefont {Skokov}}]{Basar:2012bp}%
  \BibitemOpen
  \bibfield  {author} {\bibinfo {author} {\bibfnamefont {G.}~\bibnamefont
  {Basar}}, \bibinfo {author} {\bibfnamefont {D.~E.}\ \bibnamefont
  {Kharzeev}},\ and\ \bibinfo {author} {\bibfnamefont {V.}~\bibnamefont
  {Skokov}},\ }\bibfield  {title} {\bibinfo {title} {{Conformal anomaly as a
  source of soft photons in heavy ion collisions}},\ }\href
  {https://doi.org/10. 1103/PhysRevLett.109.202303} {\bibfield  {journal}
  {\bibinfo  {journal} {Phys. Rev. Lett.}\ }\textbf {\bibinfo {volume} {109}},\
  \bibinfo {pages} {202303} (\bibinfo {year} {2012})}\BibitemShut {NoStop}%
\bibitem [{\citenamefont {Basar}\ \emph {et~al.}(2014)\citenamefont {Basar},
  \citenamefont {Kharzeev},\ and\ \citenamefont {Shuryak}}]{Basar:2014swa}%
  \BibitemOpen
  \bibfield  {author} {\bibinfo {author} {\bibfnamefont {G.}~\bibnamefont
  {Basar}}, \bibinfo {author} {\bibfnamefont {D.~E.}\ \bibnamefont
  {Kharzeev}},\ and\ \bibinfo {author} {\bibfnamefont {E.~V.}\ \bibnamefont
  {Shuryak}},\ }\bibfield  {title} {\bibinfo {title} {{Magneto-sonoluminescence
  and its signatures in photon and dilepton production in relativistic heavy
  ion collisions}},\ }\href {https://doi.org/10.1103/PhysRevC.90.014905}
  {\bibfield  {journal} {\bibinfo  {journal} {Phys. Rev. C}\ }\textbf {\bibinfo
  {volume} {90}},\ \bibinfo {pages} {014905} (\bibinfo {year}
  {2014})}\BibitemShut {NoStop}%
\bibitem [{\citenamefont {Muller}\ \emph {et~al.}(2014)\citenamefont {Muller},
  \citenamefont {Wu},\ and\ \citenamefont {Yang}}]{Muller:2013ila}%
  \BibitemOpen
  \bibfield  {author} {\bibinfo {author} {\bibfnamefont {B.}~\bibnamefont
  {Muller}}, \bibinfo {author} {\bibfnamefont {S.-Y.}\ \bibnamefont {Wu}},\
  and\ \bibinfo {author} {\bibfnamefont {D.-L.}\ \bibnamefont {Yang}},\
  }\bibfield  {title} {\bibinfo {title} {{Elliptic flow from thermal photons
  with magnetic field in holography}},\ }\href
  {https://doi.org/10.1103/PhysRevD.89.026013} {\bibfield  {journal} {\bibinfo
  {journal} {Phys. Rev. D}\ }\textbf {\bibinfo {volume} {89}},\ \bibinfo
  {pages} {026013} (\bibinfo {year} {2014})}\BibitemShut {NoStop}%
\bibitem [{\citenamefont {Ayala}\ \emph {et~al.}(2017)\citenamefont {Ayala},
  \citenamefont {Mercado},\ and\ \citenamefont
  {Villavicencio}}]{Ayala:2016awt}%
  \BibitemOpen
  \bibfield  {author} {\bibinfo {author} {\bibfnamefont {A.}~\bibnamefont
  {Ayala}}, \bibinfo {author} {\bibfnamefont {P.}~\bibnamefont {Mercado}},\
  and\ \bibinfo {author} {\bibfnamefont {C.}~\bibnamefont {Villavicencio}},\
  }\bibfield  {title} {\bibinfo {title} {{Magnetic catalysis of a finite size
  pion condensate}},\ }\href {https://doi.org/10.1103/PhysRevC.95.014904}
  {\bibfield  {journal} {\bibinfo  {journal} {Phys. Rev. C}\ }\textbf {\bibinfo
  {volume} {95}},\ \bibinfo {pages} {014904} (\bibinfo {year}
  {2017})}\BibitemShut {NoStop}%
\bibitem [{\citenamefont {Adler}\ \emph {et~al.}(2005)\citenamefont {Adler}
  \emph {et~al.}}]{PHENIX:2005yls}%
  \BibitemOpen
  \bibfield  {author} {\bibinfo {author} {\bibfnamefont {S.~S.}\ \bibnamefont
  {Adler}} \emph {et~al.} (\bibinfo {collaboration} {PHENIX Collaboration}),\
  }\bibfield  {title} {\bibinfo {title} {{Centrality dependence of direct
  photon production in $\sqrt{s_{_{NN}}}=200$~GeV Au$+$Au collisions}},\ }\href
  {https://doi.org/10. 1103/PhysRevLett.94.232301} {\bibfield  {journal}
  {\bibinfo  {journal} {Phys. Rev. Lett.}\ }\textbf {\bibinfo {volume} {94}},\
  \bibinfo {pages} {232301} (\bibinfo {year} {2005})}\BibitemShut {NoStop}%
\bibitem [{\citenamefont {Aggarwal}\ \emph {et~al.}(2000)\citenamefont
  {Aggarwal} \emph {et~al.}}]{WA98:2000vxl}%
  \BibitemOpen
  \bibfield  {author} {\bibinfo {author} {\bibfnamefont {M.~M.}\ \bibnamefont
  {Aggarwal}} \emph {et~al.} (\bibinfo {collaboration} {WA98 Collaboration}),\
  }\bibfield  {title} {\bibinfo {title} {{Observation of direct photons in
  central 158-A-GeV Pb-208 + Pb-208 collisions}},\ }\href
  {https://doi.org/10.1103/PhysRevLett.85.3595} {\bibfield  {journal} {\bibinfo
   {journal} {Phys. Rev. Lett.}\ }\textbf {\bibinfo {volume} {85}},\ \bibinfo
  {pages} {3595} (\bibinfo {year} {2000})}\BibitemShut {NoStop}%
\bibitem [{\citenamefont {Aggarwal}\ \emph {et~al.}(2004)\citenamefont
  {Aggarwal} \emph {et~al.}}]{WA98:2003ukc}%
  \BibitemOpen
  \bibfield  {author} {\bibinfo {author} {\bibfnamefont {M.~M.}\ \bibnamefont
  {Aggarwal}} \emph {et~al.} (\bibinfo {collaboration} {WA98 Collaboration}),\
  }\bibfield  {title} {\bibinfo {title} {{Interferometry of direct photons in
  central Pb-208+Pb-208 collisions at 158-A-GeV}},\ }\href
  {https://doi.org/10.1103/PhysRevLett.93.022301} {\bibfield  {journal}
  {\bibinfo  {journal} {Phys. Rev. Lett.}\ }\textbf {\bibinfo {volume} {93}},\
  \bibinfo {pages} {022301} (\bibinfo {year} {2004})}\BibitemShut {NoStop}%
\bibitem [{\citenamefont {Adare}\ \emph
  {et~al.}(2010{\natexlab{a}})\citenamefont {Adare} \emph
  {et~al.}}]{Adare:2008ab}%
  \BibitemOpen
  \bibfield  {author} {\bibinfo {author} {\bibfnamefont {A.}~\bibnamefont
  {Adare}} \emph {et~al.} (\bibinfo {collaboration} {PHENIX Collaboration}),\
  }\bibfield  {title} {\bibinfo {title} {{Enhanced production of direct photons
  in Au+Au collisions at $\sqrt{s_{NN}}$=200 GeV and implications for the
  initial temperature}},\ }\href {https://doi.org/10.
  1103/PhysRevLett.104.132301} {\bibfield  {journal} {\bibinfo  {journal}
  {Phys. Rev. Lett.}\ }\textbf {\bibinfo {volume} {104}},\ \bibinfo {pages}
  {132301} (\bibinfo {year} {2010}{\natexlab{a}})}\BibitemShut {NoStop}%
\bibitem [{\citenamefont {Adare}\ \emph {et~al.}(2015)\citenamefont {Adare}
  \emph {et~al.}}]{Adare:2014fwh}%
  \BibitemOpen
  \bibfield  {author} {\bibinfo {author} {\bibfnamefont {A.}~\bibnamefont
  {Adare}} \emph {et~al.} (\bibinfo {collaboration} {PHENIX Collaboration}),\
  }\bibfield  {title} {\bibinfo {title} {{Centrality dependence of low-momentum
  direct-photon production in Au$+$Au collisions at $\sqrt{s_{NN}}$=200 GeV}},\
  }\href {https://doi.org/10.1103/PhysRevC.91.064904} {\bibfield  {journal}
  {\bibinfo  {journal} {Phys. Rev. C}\ }\textbf {\bibinfo {volume} {91}},\
  \bibinfo {pages} {064904} (\bibinfo {year} {2015})}\BibitemShut {NoStop}%
\bibitem [{\citenamefont {Adare}\ \emph
  {et~al.}(2012{\natexlab{a}})\citenamefont {Adare} \emph
  {et~al.}}]{PHENIX:2011oxq}%
  \BibitemOpen
  \bibfield  {author} {\bibinfo {author} {\bibfnamefont {A.}~\bibnamefont
  {Adare}} \emph {et~al.} (\bibinfo {collaboration} {PHENIX Collaboration}),\
  }\bibfield  {title} {\bibinfo {title} {{Observation of direct-photon
  collective flow in $\sqrt{s_{NN}}$=200 GeV Au+Au collisions}},\ }\href
  {https://doi.org/10. 1103/PhysRevLett.109.122302} {\bibfield  {journal}
  {\bibinfo  {journal} {Phys. Rev. Lett.}\ }\textbf {\bibinfo {volume} {109}},\
  \bibinfo {pages} {122302} (\bibinfo {year} {2012}{\natexlab{a}})}\BibitemShut
  {NoStop}%
\bibitem [{\citenamefont {Adare}\ \emph
  {et~al.}(2016{\natexlab{a}})\citenamefont {Adare} \emph
  {et~al.}}]{Adare:2015lcd}%
  \BibitemOpen
  \bibfield  {author} {\bibinfo {author} {\bibfnamefont {A.}~\bibnamefont
  {Adare}} \emph {et~al.} (\bibinfo {collaboration} {PHENIX Collaboration}),\
  }\bibfield  {title} {\bibinfo {title} {{Azimuthally anisotropic emission of
  low-momentum direct photons in Au$+$Au collisions at $\sqrt{s_{NN}}$=200
  GeV}},\ }\href {https://doi.org/10.1103/PhysRevC.94.064901} {\bibfield
  {journal} {\bibinfo  {journal} {Phys. Rev. C}\ }\textbf {\bibinfo {volume}
  {94}},\ \bibinfo {pages} {064901} (\bibinfo {year}
  {2016}{\natexlab{a}})}\BibitemShut {NoStop}%
\bibitem [{\citenamefont {Adamczyk}\ \emph
  {et~al.}(2017{\natexlab{a}})\citenamefont {Adamczyk} \emph
  {et~al.}}]{STAR:2016use}%
  \BibitemOpen
  \bibfield  {author} {\bibinfo {author} {\bibfnamefont {L.}~\bibnamefont
  {Adamczyk}} \emph {et~al.} (\bibinfo {collaboration} {STAR Collaboration}),\
  }\bibfield  {title} {\bibinfo {title} {{Direct virtual photon production in
  Au+Au collisions at $\sqrt{s_{NN}}$ = 200 GeV}},\ }\href {https://doi.org/10.
  1016/j.physletb.2017.04.050} {\bibfield  {journal} {\bibinfo  {journal}
  {Phys. Lett. B}\ }\textbf {\bibinfo {volume} {770}},\ \bibinfo {pages} {451}
  (\bibinfo {year} {2017}{\natexlab{a}})}\BibitemShut {NoStop}%
\bibitem [{\citenamefont {Adam}\ \emph {et~al.}(2016)\citenamefont {Adam} \emph
  {et~al.}}]{Adam:2015lda}%
  \BibitemOpen
  \bibfield  {author} {\bibinfo {author} {\bibfnamefont {J.}~\bibnamefont
  {Adam}} \emph {et~al.} (\bibinfo {collaboration} {ALICE Collaboration}),\
  }\bibfield  {title} {\bibinfo {title} {{Direct photon production in Pb-Pb
  collisions at \sqsn = 2.76 TeV}},\ }\href
  {https://doi.org/10.1016/j.physletb.2016.01.020} {\bibfield  {journal}
  {\bibinfo  {journal} {Phys. Lett. B}\ }\textbf {\bibinfo {volume} {754}},\
  \bibinfo {pages} {235} (\bibinfo {year} {2016})}\BibitemShut {NoStop}%
\bibitem [{\citenamefont {Adare}\ \emph {et~al.}(2019)\citenamefont {Adare}
  \emph {et~al.}}]{Adare:2018wgc}%
  \BibitemOpen
  \bibfield  {author} {\bibinfo {author} {\bibfnamefont {A.}~\bibnamefont
  {Adare}} \emph {et~al.} (\bibinfo {collaboration} {PHENIX Collaboration}),\
  }\bibfield  {title} {\bibinfo {title} {{Beam Energy and Centrality Dependence
  of Direct-Photon Emission from Ultrarelativistic Heavy-Ion Collisions}},\
  }\href {https://doi.org/10. 1103/PhysRevLett.123.022301} {\bibfield
  {journal} {\bibinfo  {journal} {Phys. Rev. Lett.}\ }\textbf {\bibinfo
  {volume} {123}},\ \bibinfo {pages} {022301} (\bibinfo {year}
  {2019})}\BibitemShut {NoStop}%
\bibitem [{\citenamefont {Afanasiev}\ \emph {et~al.}(2012)\citenamefont
  {Afanasiev} \emph {et~al.}}]{Afanasiev:2012dg}%
  \BibitemOpen
  \bibfield  {author} {\bibinfo {author} {\bibfnamefont {S.}~\bibnamefont
  {Afanasiev}} \emph {et~al.} (\bibinfo {collaboration} {PHENIX
  Collaboration}),\ }\bibfield  {title} {\bibinfo {title} {{Measurement of
  Direct Photons in Au+Au Collisions at $\sqrt{s_{NN}}$=200 GeV}},\ }\href
  {https://doi.org/10. 1103/PhysRevLett.109.152302} {\bibfield  {journal}
  {\bibinfo  {journal} {Phys. Rev. Lett.}\ }\textbf {\bibinfo {volume} {109}},\
  \bibinfo {pages} {152302} (\bibinfo {year} {2012})}\BibitemShut {NoStop}%
\bibitem [{\citenamefont {Adare}\ \emph {et~al.}(2013)\citenamefont {Adare}
  \emph {et~al.}}]{Adare:2012vn}%
  \BibitemOpen
  \bibfield  {author} {\bibinfo {author} {\bibfnamefont {A.}~\bibnamefont
  {Adare}} \emph {et~al.} (\bibinfo {collaboration} {PHENIX Collaboration}),\
  }\bibfield  {title} {\bibinfo {title} {{Direct photon production in $d$$+$Au
  collisions at $\sqrt{s_{NN}}$=200 GeV}},\ }\href {https://doi.org/10.
  1103/PhysRevC.87.054907} {\bibfield  {journal} {\bibinfo  {journal} {Phys.
  Rev. C}\ }\textbf {\bibinfo {volume} {87}},\ \bibinfo {pages} {054907}
  (\bibinfo {year} {2013})}\BibitemShut {NoStop}%
\bibitem [{\citenamefont {Adare}\ \emph
  {et~al.}(2012{\natexlab{b}})\citenamefont {Adare} \emph
  {et~al.}}]{Adare:2012yt}%
  \BibitemOpen
  \bibfield  {author} {\bibinfo {author} {\bibfnamefont {A.}~\bibnamefont
  {Adare}} \emph {et~al.} (\bibinfo {collaboration} {PHENIX Collaboration}),\
  }\bibfield  {title} {\bibinfo {title} {{Direct-Photon Production in $p$$+$$p$
  Collisions at $\sqrt{s}=200$ GeV at Midrapidity}},\ }\href
  {https://doi.org/10.1103/PhysRevD.86.072008} {\bibfield  {journal} {\bibinfo
  {journal} {Phys. Rev. D}\ }\textbf {\bibinfo {volume} {86}},\ \bibinfo
  {pages} {072008} (\bibinfo {year} {2012}{\natexlab{b}})}\BibitemShut
  {NoStop}%
\bibitem [{\citenamefont {Adler}\ \emph {et~al.}(2007)\citenamefont {Adler}
  \emph {et~al.}}]{Adler:2006yt}%
  \BibitemOpen
  \bibfield  {author} {\bibinfo {author} {\bibfnamefont {S.~S.}\ \bibnamefont
  {Adler}} \emph {et~al.} (\bibinfo {collaboration} {PHENIX Collaboration}),\
  }\bibfield  {title} {\bibinfo {title} {{Measurement of direct photon
  production in $p$$+$$p$ collisions at $\sqrt{s}=200$}},\ }\href
  {https://doi.org/10. 1103/PhysRevLett.98.012002} {\bibfield  {journal}
  {\bibinfo  {journal} {Phys. Rev. Lett.}\ }\textbf {\bibinfo {volume} {98}},\
  \bibinfo {pages} {012002} (\bibinfo {year} {2007})}\BibitemShut {NoStop}%
\bibitem [{\citenamefont {Khachatryan}(2019)}]{Khachatryan:2018evz}%
  \BibitemOpen
  \bibfield  {author} {\bibinfo {author} {\bibfnamefont {V.}~\bibnamefont
  {Khachatryan}} (\bibinfo {collaboration} {PHENIX Collaboration}),\ }\bibfield
   {title} {\bibinfo {title} {{PHENIX measurements of low momentum direct
  photon radiation}},\ }\href {https://doi.org/10.1016/j.nuclphysa.2018.09.069}
  {\bibfield  {journal} {\bibinfo  {journal} {Nucl. Phys. A}\ }\textbf
  {\bibinfo {volume} {982}},\ \bibinfo {pages} {763} (\bibinfo {year}
  {2019})}\BibitemShut {NoStop}%
\bibitem [{\citenamefont {Drees}(2019)}]{Drees:2019ila}%
  \BibitemOpen
  \bibfield  {author} {\bibinfo {author} {\bibfnamefont {A.}~\bibnamefont
  {Drees}} (\bibinfo {collaboration} {PHENIX Collaboration}),\ }\bibfield
  {title} {\bibinfo {title} {{PHENIX Measurements of Beam Energy Dependence of
  Direct Photon Emission}},\ }\href {https://doi.org/10.22323/1.345.0176}
  {\bibfield  {journal} {\bibinfo  {journal} {Proc. Sci.}\ }\textbf {\bibinfo
  {volume} {HardProbes2018}},\ \bibinfo {pages} {176} (\bibinfo {year}
  {2019})}\BibitemShut {NoStop}%
\bibitem [{\citenamefont {Adare}\ \emph {et~al.}(2018)\citenamefont {Adare}
  \emph {et~al.}}]{PHENIX:2018che}%
  \BibitemOpen
  \bibfield  {author} {\bibinfo {author} {\bibfnamefont {A.}~\bibnamefont
  {Adare}} \emph {et~al.} (\bibinfo {collaboration} {PHENIX Collaboration}),\
  }\bibfield  {title} {\bibinfo {title} {{Low-momentum direct photon
  measurement in Cu$+$Cu collisions at $\sqrt{s_{_{NN}}}=200$ GeV}},\ }\href
  {https://doi.org/10.1103/PhysRevC.98.054902} {\bibfield  {journal} {\bibinfo
  {journal} {Phys. Rev. C}\ }\textbf {\bibinfo {volume} {98}},\ \bibinfo
  {pages} {054902} (\bibinfo {year} {2018})}\BibitemShut {NoStop}%
\bibitem [{\citenamefont {Adcox}\ \emph
  {et~al.}(2003{\natexlab{a}})\citenamefont {Adcox} \emph
  {et~al.}}]{PHENIX:overview}%
  \BibitemOpen
  \bibfield  {author} {\bibinfo {author} {\bibfnamefont {K.}~\bibnamefont
  {Adcox}} \emph {et~al.} (\bibinfo {collaboration} {PHENIX Collaboration}),\
  }\bibfield  {title} {\bibinfo {title} {{PHENIX detector overview}},\ }\href
  {https://doi.org/10.1016/S0168-9002(02)01950-2} {\bibfield  {journal}
  {\bibinfo  {journal} {Nucl. Instrum. Methods Phys. Res., Sec. A}\ }\textbf
  {\bibinfo {volume} {499}},\ \bibinfo {pages} {469} (\bibinfo {year}
  {2003}{\natexlab{a}})}\BibitemShut {NoStop}%
\bibitem [{\citenamefont {Khachatryan}(2017)}]{Khachatryan:2017gqp}%
  \BibitemOpen
  \bibfield  {author} {\bibinfo {author} {\bibfnamefont {V.}~\bibnamefont
  {Khachatryan}},\ }\emph {\bibinfo {title} {{The Quark Gluon Plasma probed by
  Low Momentum Direct Photons in Au+Au Collisions at 62.4 GeV and 39 GeV beam
  energies}}},\ \href@noop {} {Ph.D. thesis},\ \bibinfo  {school} {Stony Brook
  University} (\bibinfo {year} {2017}),\ \bibinfo {note}
  {https://inspirehep.net/literature/1804505}\BibitemShut {NoStop}%
\bibitem [{\citenamefont {Adare}\ \emph
  {et~al.}(2010{\natexlab{b}})\citenamefont {Adare} \emph
  {et~al.}}]{PHENIX:2009gyd}%
  \BibitemOpen
  \bibfield  {author} {\bibinfo {author} {\bibfnamefont {A.}~\bibnamefont
  {Adare}} \emph {et~al.} (\bibinfo {collaboration} {PHENIX Collaboration}),\
  }\bibfield  {title} {\bibinfo {title} {{Detailed measurement of the $e^+ e^-$
  pair continuum in $p$$+$$p$ and Au$+$Au collisions at $\sqrt{s_{_{NN}}}=200$
  GeV and implications for direct-photon production}},\ }\href
  {https://doi.org/10.1103/PhysRevC.81.034911} {\bibfield  {journal} {\bibinfo
  {journal} {Phys. Rev. C}\ }\textbf {\bibinfo {volume} {81}},\ \bibinfo
  {pages} {034911} (\bibinfo {year} {2010}{\natexlab{b}})}\BibitemShut
  {NoStop}%
\bibitem [{\citenamefont {Allen}\ \emph {et~al.}(2003)\citenamefont {Allen}
  \emph {et~al.}}]{PHENIX:BBC}%
  \BibitemOpen
  \bibfield  {author} {\bibinfo {author} {\bibfnamefont {M.}~\bibnamefont
  {Allen}} \emph {et~al.} (\bibinfo {collaboration} {PHENIX Collaboration}),\
  }\bibfield  {title} {\bibinfo {title} {{PHENIX inner detectors}},\ }\href
  {https://doi.org/10.1016/S0168-9002(02)01956-3} {\bibfield  {journal}
  {\bibinfo  {journal} {Nucl. Instrum. Methods Phys. Res., Sec. A}\ }\textbf
  {\bibinfo {volume} {499}},\ \bibinfo {pages} {549} (\bibinfo {year}
  {2003})}\BibitemShut {NoStop}%
\bibitem [{\citenamefont {Adcox}\ \emph
  {et~al.}(2003{\natexlab{b}})\citenamefont {Adcox} \emph
  {et~al.}}]{PHENIX:tracking}%
  \BibitemOpen
  \bibfield  {author} {\bibinfo {author} {\bibfnamefont {K.}~\bibnamefont
  {Adcox}} \emph {et~al.} (\bibinfo {collaboration} {PHENIX Collaboration}),\
  }\bibfield  {title} {\bibinfo {title} {{PHENIX central arm tracking
  detectors}},\ }\href {https://doi.org/10.1016/S0168-9002(02)01952-6}
  {\bibfield  {journal} {\bibinfo  {journal} {Nucl. Instrum. Methods Phys.
  Res., Sec. A}\ }\textbf {\bibinfo {volume} {499}},\ \bibinfo {pages} {489}
  (\bibinfo {year} {2003}{\natexlab{b}})}\BibitemShut {NoStop}%
\bibitem [{\citenamefont {Aizawa}\ \emph {et~al.}(2003)\citenamefont {Aizawa}
  \emph {et~al.}}]{PHENIX:PID}%
  \BibitemOpen
  \bibfield  {author} {\bibinfo {author} {\bibfnamefont {M.}~\bibnamefont
  {Aizawa}} \emph {et~al.} (\bibinfo {collaboration} {PHENIX Collaboration}),\
  }\bibfield  {title} {\bibinfo {title} {{PHENIX central arm particle ID
  detectors}},\ }\href {https://doi.org/10.1016/S0168-9002(02)01953-8}
  {\bibfield  {journal} {\bibinfo  {journal} {Nucl. Instrum. Methods Phys.
  Res., Sec. A}\ }\textbf {\bibinfo {volume} {499}},\ \bibinfo {pages} {508}
  (\bibinfo {year} {2003})}\BibitemShut {NoStop}%
\bibitem [{\citenamefont {Aphecetche}\ \emph {et~al.}(2003)\citenamefont
  {Aphecetche} \emph {et~al.}}]{PHENIX:EMCal}%
  \BibitemOpen
  \bibfield  {author} {\bibinfo {author} {\bibfnamefont {L.}~\bibnamefont
  {Aphecetche}} \emph {et~al.} (\bibinfo {collaboration} {PHENIX
  Collaboration}),\ }\bibfield  {title} {\bibinfo {title} {{PHENIX
  calorimeter}},\ }\href {https://doi.org/10.1016/S0168-9002(02)01954-X}
  {\bibfield  {journal} {\bibinfo  {journal} {Nucl. Instrum. Methods Phys.
  Res., Sec. A}\ }\textbf {\bibinfo {volume} {499}},\ \bibinfo {pages} {521}
  (\bibinfo {year} {2003})}\BibitemShut {NoStop}%
\bibitem [{\citenamefont {Anderson}\ \emph {et~al.}(2011)\citenamefont
  {Anderson} \emph {et~al.}}]{PHENIX:HBD}%
  \BibitemOpen
  \bibfield  {author} {\bibinfo {author} {\bibfnamefont {W.}~\bibnamefont
  {Anderson}} \emph {et~al.},\ }\bibfield  {title} {\bibinfo {title} {{Design,
  Construction, Operation and Performance of a Hadron Blind Detector for the
  PHENIX Experiment}},\ }\href {https://doi.org/10.1016/j.nima.2011.04.015}
  {\bibfield  {journal} {\bibinfo  {journal} {Nucl. Instrum. Methods Phys.
  Res., Sec. A}\ }\textbf {\bibinfo {volume} {646}},\ \bibinfo {pages} {35}
  (\bibinfo {year} {2011})}\BibitemShut {NoStop}%
\bibitem [{\citenamefont {Adare}\ \emph
  {et~al.}(2012{\natexlab{c}})\citenamefont {Adare} \emph
  {et~al.}}]{PHENIX:2012oed}%
  \BibitemOpen
  \bibfield  {author} {\bibinfo {author} {\bibfnamefont {A.}~\bibnamefont
  {Adare}} \emph {et~al.} (\bibinfo {collaboration} {PHENIX Collaboration}),\
  }\bibfield  {title} {\bibinfo {title} {{Evolution of $\pi^0$ suppression in
  Au+Au collisions from $\sqrt{s_{NN}} = 39$ to 200 GeV}},\ }\href
  {https://doi.org/10. 1103/PhysRevLett.109.152301} {\bibfield  {journal}
  {\bibinfo  {journal} {Phys. Rev. Lett.}\ }\textbf {\bibinfo {volume} {109}},\
  \bibinfo {pages} {152301} (\bibinfo {year} {2012}{\natexlab{c}})},\ \bibinfo
  {note} {[Phys. Rev.Lett. 125, 049901(E) (2020)]}\BibitemShut {NoStop}%
\bibitem [{\citenamefont {Abelev}\ \emph {et~al.}(2007)\citenamefont {Abelev}
  \emph {et~al.}}]{STAR:2007zea}%
  \BibitemOpen
  \bibfield  {author} {\bibinfo {author} {\bibfnamefont {B.~I.}\ \bibnamefont
  {Abelev}} \emph {et~al.} (\bibinfo {collaboration} {STAR Collaboration}),\
  }\bibfield  {title} {\bibinfo {title} {{Energy dependence of $\pi^{\pm}$, $p$
  and anti-$p$ transverse momentum spectra for Au$+$Au collisions at
  $\sqrt{s_{_{NN}}}=62.4$ and 200 GeV}},\ }\href
  {https://doi.org/10.1016/j.physletb.2007.06.035} {\bibfield  {journal}
  {\bibinfo  {journal} {Phys. Lett. B}\ }\textbf {\bibinfo {volume} {655}},\
  \bibinfo {pages} {104} (\bibinfo {year} {2007})}\BibitemShut {NoStop}%
\bibitem [{\citenamefont {Adamczyk}\ \emph
  {et~al.}(2017{\natexlab{b}})\citenamefont {Adamczyk} \emph
  {et~al.}}]{STAR:2017sal}%
  \BibitemOpen
  \bibfield  {author} {\bibinfo {author} {\bibfnamefont {L.}~\bibnamefont
  {Adamczyk}} \emph {et~al.} (\bibinfo {collaboration} {STAR Collaboration}),\
  }\bibfield  {title} {\bibinfo {title} {{Bulk Properties of the Medium
  Produced in Relativistic Heavy-Ion Collisions from the Beam Energy Scan
  Program}},\ }\href {https://doi.org/10.1103/PhysRevC.96.044904} {\bibfield
  {journal} {\bibinfo  {journal} {Phys. Rev. C}\ }\textbf {\bibinfo {volume}
  {96}},\ \bibinfo {pages} {044904} (\bibinfo {year}
  {2017}{\natexlab{b}})}\BibitemShut {NoStop}%
\bibitem [{\citenamefont {Ren}\ and\ \citenamefont
  {Drees}(2021)}]{Ren:2021xbh}%
  \BibitemOpen
  \bibfield  {author} {\bibinfo {author} {\bibfnamefont {Y.}~\bibnamefont
  {Ren}}\ and\ \bibinfo {author} {\bibfnamefont {A.}~\bibnamefont {Drees}},\
  }\bibfield  {title} {\bibinfo {title} {{Examination of the universal behavior
  of the $\eta$-to-$\pi^0$ ratio in heavy-ion collisions}},\ }\href
  {https://doi.org/10.1103/PhysRevC.104.054902} {\bibfield  {journal} {\bibinfo
   {journal} {Phys. Rev. C}\ }\textbf {\bibinfo {volume} {104}},\ \bibinfo
  {pages} {054902} (\bibinfo {year} {2021})}\BibitemShut {NoStop}%
\bibitem [{\citenamefont {Paquet}(2017)}]{Paquet:2017}%
  \BibitemOpen
  \bibfield  {author} {\bibinfo {author} {\bibfnamefont {J.~F.}\ \bibnamefont
  {Paquet}},\ }\href@noop {} {} (\bibinfo {year} {2017}),\ \bibinfo {note}
  {private communication, uses nuclear PDF nCTEQ15-np and photon fragmentation
  function BFG-II.}\BibitemShut {Stop}%
\bibitem [{\citenamefont {Adare}\ \emph
  {et~al.}(2016{\natexlab{b}})\citenamefont {Adare} \emph
  {et~al.}}]{Adare:2015bua}%
  \BibitemOpen
  \bibfield  {author} {\bibinfo {author} {\bibfnamefont {A.}~\bibnamefont
  {Adare}} \emph {et~al.} (\bibinfo {collaboration} {PHENIX Collaboration}),\
  }\bibfield  {title} {\bibinfo {title} {{Transverse energy production and
  charged-particle multiplicity at midrapidity in various systems from
  $\sqrt{s_{NN}}=7.7$ to 200 GeV}},\ }\href
  {https://doi.org/10.1103/PhysRevC.93.024901} {\bibfield  {journal} {\bibinfo
  {journal} {Phys. Rev. C}\ }\textbf {\bibinfo {volume} {93}},\ \bibinfo
  {pages} {024901} (\bibinfo {year} {2016}{\natexlab{b}})}\BibitemShut
  {NoStop}%
\bibitem [{\citenamefont {Zyla}\ \emph {et~al.}(2020)\citenamefont {Zyla} \emph
  {et~al.}}]{Zyla:2020zbs}%
  \BibitemOpen
  \bibfield  {author} {\bibinfo {author} {\bibfnamefont {P.~A.}\ \bibnamefont
  {Zyla}} \emph {et~al.} (\bibinfo {collaboration} {Particle Data Group}),\
  }\bibfield  {title} {\bibinfo {title} {{Review of Particle Physics}},\ }\href
  {https://doi.org/10.1093/ptep/ptaa104} {\bibfield  {journal} {\bibinfo
  {journal} {Prog. Theor. and Exp. Phys.}\ }\textbf {\bibinfo {volume}
  {2020}},\ \bibinfo {pages} {083C01} (\bibinfo {year} {2020})}\BibitemShut
  {NoStop}%
\bibitem [{\citenamefont {Alpgard}\ \emph {et~al.}(1982)\citenamefont {Alpgard}
  \emph {et~al.}}]{UA5:1982ygd}%
  \BibitemOpen
  \bibfield  {author} {\bibinfo {author} {\bibfnamefont {K.}~\bibnamefont
  {Alpgard}} \emph {et~al.} (\bibinfo {collaboration} {UA5 Collaboration}),\
  }\bibfield  {title} {\bibinfo {title} {{Comparison of $p\bar{p}$ and $pp$
  Interactions at $\sqrt{s}=53$ GeV}},\ }\href
  {https://doi.org/10.1016/0370-2693(82)90325-2} {\bibfield  {journal}
  {\bibinfo  {journal} {Phys. Lett. B}\ }\textbf {\bibinfo {volume} {112}},\
  \bibinfo {pages} {183} (\bibinfo {year} {1982})}\BibitemShut {NoStop}%
\bibitem [{\citenamefont {Alner}\ \emph {et~al.}(1986)\citenamefont {Alner}
  \emph {et~al.}}]{UA5:1986yef}%
  \BibitemOpen
  \bibfield  {author} {\bibinfo {author} {\bibfnamefont {G.~J.}\ \bibnamefont
  {Alner}} \emph {et~al.} (\bibinfo {collaboration} {UA5 Collaboration}),\
  }\bibfield  {title} {\bibinfo {title} {{Scaling of Pseudorapidity
  Distributions at c.m. Energies Up to 0.9 TeV}},\ }\href
  {https://doi.org/10.1007/BF01410446} {\bibfield  {journal} {\bibinfo
  {journal} {Z. Phys. C}\ }\textbf {\bibinfo {volume} {33}},\ \bibinfo {pages}
  {1} (\bibinfo {year} {1986})}\BibitemShut {NoStop}%
\bibitem [{\citenamefont {Adam}\ \emph {et~al.}(2017)\citenamefont {Adam} \emph
  {et~al.}}]{Adam:2015gka}%
  \BibitemOpen
  \bibfield  {author} {\bibinfo {author} {\bibfnamefont {J.}~\bibnamefont
  {Adam}} \emph {et~al.} (\bibinfo {collaboration} {ALICE Collaboration}),\
  }\bibfield  {title} {\bibinfo {title} {{Charged-particle multiplicities in
  proton-proton collisions at $\sqrt{s}=0.9$ to 8 TeV}},\ }\href
  {https://doi.org/10.1140/epjc/s10052-016-4571-1} {\bibfield  {journal}
  {\bibinfo  {journal} {Eur. Phys. J. C}\ }\textbf {\bibinfo {volume} {77}},\
  \bibinfo {pages} {33} (\bibinfo {year} {2017})}\BibitemShut {NoStop}%
\bibitem [{\citenamefont {Aamodt}\ \emph {et~al.}(2011)\citenamefont {Aamodt}
  \emph {et~al.}}]{Aamodt:2010cz}%
  \BibitemOpen
  \bibfield  {author} {\bibinfo {author} {\bibfnamefont {K.}~\bibnamefont
  {Aamodt}} \emph {et~al.} (\bibinfo {collaboration} {ALICE Collaboration}),\
  }\bibfield  {title} {\bibinfo {title} {{Centrality dependence of the
  charged-particle multiplicity density at mid-rapidity in Pb-Pb collisions at
  $\sqrt{s_{NN}}=2. 76$ TeV}},\ }\href {https://doi.org/10.
  1103/PhysRevLett.106.032301} {\bibfield  {journal} {\bibinfo  {journal}
  {Phys. Rev. Lett.}\ }\textbf {\bibinfo {volume} {106}},\ \bibinfo {pages}
  {032301} (\bibinfo {year} {2011})}\BibitemShut {NoStop}%
\bibitem [{\citenamefont {Paquet}(2020)}]{Paquet:2020}%
  \BibitemOpen
  \bibfield  {author} {\bibinfo {author} {\bibfnamefont {J.~F.}\ \bibnamefont
  {Paquet}},\ }\href@noop {} {} (\bibinfo {year} {2020}),\ \bibinfo {note}
  {private communication.}\BibitemShut {Stop}%
\bibitem [{\citenamefont {Gale}\ \emph {et~al.}(2021)\citenamefont {Gale},
  \citenamefont {Paquet}, \citenamefont {Schenke},\ and\ \citenamefont
  {Shen}}]{Gale:2020xlg}%
  \BibitemOpen
  \bibfield  {author} {\bibinfo {author} {\bibfnamefont {C.}~\bibnamefont
  {Gale}}, \bibinfo {author} {\bibfnamefont {J.-F.}\ \bibnamefont {Paquet}},
  \bibinfo {author} {\bibfnamefont {B.}~\bibnamefont {Schenke}},\ and\ \bibinfo
  {author} {\bibfnamefont {C.}~\bibnamefont {Shen}},\ }\bibfield  {title}
  {\bibinfo {title} {{Probing Early-Time Dynamics and Quark-Gluon Plasma
  Transport Properties with Photons and Hadrons}},\ }\href {https://doi.org/10.
  1016/j.nuclphysa.2020.121863} {\bibfield  {journal} {\bibinfo  {journal}
  {Nucl. Phys. A}\ }\textbf {\bibinfo {volume} {1005}},\ \bibinfo {pages}
  {121863} (\bibinfo {year} {2021})}\BibitemShut {NoStop}%
\end{thebibliography}
 
%
 
\end{document}